\documentclass[a4paper, reqno]{amsart}

\usepackage[english]{babel}
\usepackage{graphicx}%
\usepackage{multirow}%
\usepackage{amsmath,amssymb,amsfonts}%
\usepackage{amsthm}%
\usepackage{mathrsfs}%
\usepackage[title]{appendix}%
\usepackage{xcolor}%
\usepackage{textcomp}%
\usepackage{manyfoot}%
\usepackage{booktabs}%
\usepackage{algorithm}%
\usepackage{algorithmicx}%
\usepackage{algpseudocode}%
\usepackage{listings}%
\usepackage{siunitx}
\usepackage{float}  
\usepackage{caption}
\usepackage{subcaption}
\usepackage{placeins}
\usepackage{tabularx}   
\usepackage{makecell} 
\usepackage{multirow}
\usepackage{fancyvrb}
\usepackage{orcidlink}
\usepackage{textcomp}
\usepackage{threeparttable}

\usepackage[backend=biber, sorting=none, doi=true, url=true, style=numeric]{biblatex}
\newcommand{\R}{\mathbb{R}}
\newcommand{\del}{\partial}
\newcommand{\mat}[1]{\begin{pmatrix} #1 \end{pmatrix}}
\renewcommand{\phi}{\varphi}
\newcommand{\norm}[1]{\lVert#1\rVert}

\newcommand{\Wpsb}{W}
\newcommand{\tr}{\mathrm{tr}}
\newcommand{\<}{\left\langle}
\renewcommand{\>}{\right\rangle}
\newcommand{\grad}{\mathrm{grad}}
\newcommand{\m}{O} 
\renewcommand{\div}{\mathrm{div}}
\newcommand{\W}{\tilde{W}}

\newcommand{\pic}[1]{\includegraphics[width=0.11\textwidth]{#1}}

\newcommand{\LossRow}[4]{
   & {\textcolor{gray}{\scalebox{0.8}{OP#1 loss}}} & 
    {\textcolor{gray}{\scalebox{0.8}{\num{#2}}}} & 
    {\textcolor{gray}{\scalebox{0.8}{\num{#3}}}} & 
    {\textcolor{gray}{\scalebox{0.8}{\num{#4}}}} \\}

\hypersetup{
    colorlinks,
    linkcolor={red!50!black},
    citecolor={blue!50!black},
    urlcolor={blue!80!black}
}

\newtheorem{thm}{Theorem}[section]
\theoremstyle{definition}
\newtheorem{example}[thm]{Example}

\title[Data-Based Approach to Hyperelastic Membranes]{Data-Based Approach to Hyperelastic Membranes}

\author{Claudia Grabs}
\address{Claudia Grabs, 
Institut für Mathematik,
Karl-Liebknecht-Str. 24-25,
Universität Potsdam, 
D-14476, Potsdam, Germany
}
\urladdr{\href{http://www.math.uni-potsdam.de/professuren/geometrie/personen/claudia-grabs/}{http://www.math.uni-potsdam.de/professuren/geometrie/personen/claudia-grabs/}}
\email{\href{mailto:claudia.meinel@uni-potsdam.de}{claudia.meinel@uni-potsdam.de}}

\author{Werner Wirges}
\address{Werner Wirges, 
Institut für Physik und Astronomie,
Karl-Liebknecht-Str. 24-25,
Universität Potsdam, 
D-14476, Potsdam, Germany
}
\email{\href{mailto:werner.wirges@uni-potsdam.de}{werner.wirges@uni-potsdam.de}}


\date{\today}

\begin{document}

\begin{abstract}
We study large deformations of hyperelastic membranes using a purely two-dimensional formulation derived from basic balance principles within a modern geometric setting, ensuring a framework that is independent of an underlying three-dimensional formulation. To assess the predictive capabilities of membrane theory, we compare numerical solutions to experimental data from axisymmetric deformations of a silicone rubber film. Five hyperelastic models—Neo-Hookean, Mooney-Rivlin, Gent, Yeoh, and Ogden—are evaluated by fitting their material parameters to our experimental data using TensorFlow. Our results provide a systematic comparison of these models based on their accuracy in capturing observed deformations, establishing a framework for integrating theory, experiment, and data-based parameter identification. 
\end{abstract}

\maketitle

\setcounter{tocdepth}{1}
\tableofcontents

\section{Introduction}

Understanding and modeling hyperelastic membranes is crucial for accurately describing the large deformations of thin structures in engineering and biomechanics. These models enable the prediction of stress distributions and material responses in applications such as soft robotics, biomedical devices, and inflatable structures. 
While the initial motivation for this work arose from discussions in the context of electromechanically active polymers—particularly dielectric elastomers, as surveyed in~\cite{carpi2016}—the present study has remained entirely within the scope of purely mechanical models for hyperelastic membranes.
Reliable simulations require a precise mathematical framework that not only captures the complex material behavior but also provides a robust foundation for numerical implementation.

Over the years, nonlinear membrane theory has been developed from different perspectives. 
Green and Zerna~\cite{GreenZerna1954} introduced membrane theory as part of their broader treatment of elasticity, deriving the governing equations using asymptotic expansions from three-dimensional elasticity and emphasizing geometric interpretations within a continuum mechanics framework. 
Similarly, Green and Adkins~\cite{GreenAdkins1960} devoted a chapter of their book on large elastic deformations to membranes, focusing on material-specific formulations and introducing nonlinear stress-strain relationships for isotropic hyperelastic materials. 
While these works primarily placed membrane theory within the context of general elasticity, later research refined it as an independent discipline. 
Le Dret and Raoult~\cite{LeDretRaoult1995} established a rigorous variational foundation, showing that the nonlinear membrane model emerges as the asymptotic limit of three-dimensional elasticity via $\Gamma$-convergence. 
Expanding on these developments, Libai and Simmonds~\cite{LibaiSimmonds1998} presented a systematic formulation of the nonlinear membrane equations in a dedicated chapter, emphasizing variational principles, kinematic constraints, and stress resultant formulations to bridge continuum mechanics and practical shell models.

In this work, we first derive the governing equations of membrane theory directly from two-dimensional balance equations. 
Following the same conceptual path as in three-dimensional elasticity, we start with purely 2D stress tensors and establish force equilibrium within the deformed surface. 
All quantities are formulated intrinsically in 2D, ensuring a framework that is independent of an underlying three-dimensional formulation. 
Since bending does not induce in-plane strains, the derivation naturally leads to membrane theory. 
The presentation is self-contained, requiring no prior knowledge of elasticity theory, as all necessary concepts are introduced and explained. 
However, the formulation relies on a modern geometric setting, which provides the natural framework for the theory.

Having established the theoretical foundations, we assess how well membrane theory predicts real-world deformations. 
To this end, we conduct a systematic comparison between theory and reality in the special case of axisymmetric deformations, a case we have consistently used as an illustrative example throughout the theoretical developments. 
While such deformations have been extensively studied, most existing works focus on analytical and numerical approaches rather than direct validation against experimental data.

The study of axisymmetric membrane deformations has evolved through different mathematical formulations. 
Yang and Feng~\cite{YangFeng1970} provided an early framework by reformulating the governing equations into a system of three first-order ordinary differential equations, simplifying the analysis of large deformations in axisymmetric structures. 
Later, Fulton and Simmonds~\cite{FultonSimmonds1986} examined large deformations in annular membranes under vertical edge loads, considering various strain energy density functions to explore material-dependent effects. 

To bridge the gap between theory and experiment, we use the experimental data not only for validation but also as training data to calibrate the material parameters of various hyperelastic models. 
The experiments were conducted using Elastosil Film from Wacker, a silicone rubber membrane, which exhibits highly nonlinear elastic behavior and is commonly used in soft robotics and biomedical applications. 
We evaluate five widely used hyperelastic models—Neo-Hookean, Mooney-Rivlin, Gent, Yeoh, and Ogden—which differ in complexity and the number of material parameters. 
Following the ranking on the overall predictive performance of these and other models presented in~\cite{MellyLiuReview}, their accuracy progressively improves as additional nonlinear effects are incorporated. Leveraging state-of-the-art machine learning techniques with TensorFlow, we fit the model parameters to measured deformations, allowing for a highly efficient and automated optimization process. 
This data-based approach enables a systematic evaluation of different constitutive laws and identifies which models best capture the actual material behavior.

Our results show that while simpler models provide reasonable approximations, more complex models such as the Yeoh model offer significantly improved predictive accuracy for the given material. 
The systematic comparison of theory, numerics, and experiments provides new insights into the applicability of hyperelastic models for thin membranes under large deformations.

The paper is structured as follows: Section~\ref{sec:model} presents the theoretical formulation of the membrane equations, derived purely from two-dimensional balance principles. 
Section~\ref{sec:experiments} describes the experimental setup and data acquisition process. 
Section~\ref{sec:simulations} discusses the results, comparing the performance of different hyperelastic models against experimental observations. 
Additional details on the five material models and their mathematical formulations are provided in the Appendix~\ref{secA1}, including a detailed discussion of the corresponding material parameters.

\section{The nonlinear membrane model}\label{sec:model}
In this section, we outline the geometric formulation of elastostatics for hyperelastic membranes, focusing on the main ideas and structures rather than formal derivations.

Throughout this work, we restrict attention to isothermal processes and neglect thermal effects. We also assume the absence of body forces, focusing solely on self-equilibrated configurations.

To follow the theoretical development, some familiarity with differential geometry is helpful—especially with vector and, more generally, tensor fields along a map (often referred to as two-point tensors in continuum mechanics), whose covariant derivatives play a central role in the theory. 
This includes the use of pullbacks and pushforwards of tensor fields and differential forms along maps. 
Once coordinate charts are chosen on the domain and codomain of the map, all relevant objects can also be expressed in components.\footnote{For better readability, we will sometimes omit explicit arguments or be slightly imprecise in notation when the meaning is clear from context.} 
In particular, expressions such as $dr \otimes dr$ denote the tensor product of coordinate 1-forms (sections of the cotangent bundle) and are not to be interpreted as dyadics of infinitesimal displacements. 
All such objects should be understood in the standard coordinate-geometric sense, where, for instance, the Euclidean metric is written as a symmetric $(0,2)$-tensor field using this notation. 
These conventions will be applied consistently throughout the manuscript and should become clear through the step-by-step illustrations in the theoretical discussion. 
For a reference that combines the mathematical foundations with applications to elasticity theory, we recommend the book by Marsden and Hughes~\cite{MH}. 
Standard textbooks on differential geometry, such as~\cite{LeeSM} or~\cite{LeeRM}, provide further background.

As a running example to illustrate the theory and to prepare for the subsequent simulations, we consider rotationally symmetric deformations of initially flat elastic annuli. 
Before we start the purely two-dimensional membrane theory, we want to recall the concept of a membrane as a limiting case of an elastic shell, highlighting its distinct mathematical and physical properties.

\subsection{Shells and membranes}
An elastic body $\overline{M}$ is a \emph{shell} with thickness $d$, if it has the following special geometry in its relaxed, stress-free reference configuration:
There is a regular surface $M\subset \R^3$ with boundary $\del M$
and unit normal vector field $n$, such that 
\begin{equation*}
\overline{M}=\left\{ p+tn(p)\mid p\in M, t\in[-\tfrac{d}{2},\tfrac{d}{2}] \right\}.
\end{equation*}
The surface $M$ is called \emph{midsurface} of the shell $\overline{M}$.
The thickness $d$ is small compared to its other dimensions.
The aim of shell theory is to describe the deformations and motions of a shell by equations that depend only on the two coordinates of the midsurface $M$.
In the introduction of their book~\cite{LibaiSimmonds1998} on shell theory, Libai and Simmonds write: \emph{``An exact two-dimensional theory of shells does not exist. No matter how thin - a shell remains a three-dimensional continuum.''}
There are different shell theories available for different situations.
In shell theory, thickness-integrated stress resultants are introduced, describing membrane and shear forces and bending moments, see for example~\cite{Reddy}.
\emph{Membranes} are special shells that offer no resistance to bending.
For membranes, all shear forces and bending moments are neglected and only tangential stresses are considered.

We found that membrane theory naturally emerges when the concepts and derivations of classical 3D elasticity are transferred to surfaces. 
We introduce everything analogously, from the deformation gradient and the Cauchy-Green deformation tensor, the stress tensors to the equilibrium conditions, leading naturally to membrane theory. 
This intrinsic formulation ensures a purely two-dimensional description, independent of an underlying three-dimensional framework.

\subsection{Deformations of membranes}
Let $M$ now be the midsurface of a membrane.
In the following, we will also call $M$ itself the membrane.

A \emph{deformation} of $M$ is a smooth\footnote{In this work we do not consider lower regularity of the deformation map.} embedding $u:M\to\R^3$. 

\begin{example}[Rotationally symmetric deformation of initially flat elastic annulus]\label{exa:defrotsym}
Let $0<R_1<R_2<\infty$, let $U:=\left[ R_1,R_2 \right]\times (0,2\pi)$.
We consider polar coordinates $(R,\Phi)$,  
\begin{equation} \label{eq:pc}
F_p:U\to\R^3, F_p(R,\Phi)=\mat{R\cos(\Phi) \\ R\sin(\Phi) \\0}. 
\end{equation}
Then $M:=F_p(U)$ is the flat annulus with inner radius $R_1$ and outer radius $R_2$.

On $\R^3$ we consider cylinder coordinates $(r,\phi,z)$ on $\tilde{U}=\R_{\geq0}\times (0,2\pi)
 \times\R$,
$$F_c:\tilde{U} \to\R^3, F_c(r,\phi,z)=\mat{r\cos(\phi)\\ r\sin(\phi) \\z}.$$ 

We introduce two functions $f,h:[R_1,R_2] \to \R$ for the radial and vertical part of the rotationally symmetric deformation $u:M \to \R^3$,  given in coordinates
by
\begin{equation*}
u(R,\Phi)= (u^r(R,\Phi),u^\phi(R,\Phi),u^z(R,\Phi)) =: (f(R), \Phi,h(R)).
\end{equation*}
The deformation map $u$ transforms the flat annulus into a surface of revolution with two circular boundary components.
\end{example}

\subsection{Deformation tensor of membrane deformations}

The differential $du$ of the deformation map $u : M \to \mathbb{R}^3$, commonly referred to as the \emph{deformation gradient}, provides a first local measure of the \emph{strain} induced by the deformation. 
It is usually denoted by $F := du$, and it constitutes a first example of a tensor field \emph{along the deformation} $u$, since it maps tangent vectors of $M$ based at a point $x$ to tangent vectors of the image $u(M)$ based at $u(x)$.

Let $g$ denote the standard Euclidean metric on $\mathbb{R}^3$, and let $G$ be the first fundamental form on $M$, induced by the embedding into $\mathbb{R}^3$. 
These metrics allow us to define the \emph{transpose} of the differential, denoted $du^T$, by the condition that 
\[
g(du(V), w) = G(V, du^T(w))
\]
for all tangent vector fields $V$ on $M$ and all vector fields $w$ along $u$. 
Note that $du^T(w)=0$ if $w$ is normal to $u(M)$.

The \emph{ (right) Cauchy-Green deformation tensor} 
\begin{equation*} \label{eq:defC}
C:=du^T\circ du 
\end{equation*}
is an endomorphism field on the undeformed membrane $M$.

In contrast to the deformation gradient $du$, the Cauchy--Green deformation tensor $C$ remains trivial for deformations that preserve the shape of the membrane but alter only its position in space, such as rigid motions (rotations and translations). 
In this sense, $C$ provides a more intrinsic measure of deformation. 
Note that $C$ is also trivial if $u$ is an isometric embedding of $M$ into $\mathbb{R}^3$, where intrinsic distances are preserved while the membrane may undergo bending and change its curvature.

Since $C$ is symmetric and positive definite, $U:=\sqrt{C}$ is well-defined. 
At any point $x\in M$, for an orthonormal eigenbasis $(e_1,e_2)$ of $T_xM$ for $U(x)$, the corresponding (possibly equal) eigenvalues $(\lambda_1,\lambda_2)$ of $U(x)$ are such that $\left\Vert du(x)(e_\alpha) \right\Vert_g=\left\Vert U(x)e_\alpha \right\Vert_G=\left\Vert \lambda_\alpha e_\alpha \right\Vert_G$ for $\alpha=1,2$. 
This is the reason why the eigenvalues of $U$ are called principal stretches, they describe how much the membrane gets stretched ($\lambda_\alpha>1)$) or compressed ($\lambda_\alpha<1)$) in the direction $e_\alpha$. Then $(\lambda_1^2,\lambda_2^2)$ are the eigenvalues of $C(x)$. 
We consider the dual basis $(e_1^\flat,e_2^\flat)$ with respect to $G$ and define the basic endomorphisms $P_\alpha:=e_\alpha\otimes e_\alpha^\flat$ for $\alpha=1,2$. 
We then have the spectral decomposition

\begin{equation}\label{eq:SpecDecompC}
 C(x)=\sum_{\alpha=1}^2 \lambda_\alpha^2 \, P_\alpha.
\end{equation}

The local \emph{change of area} is described by the function $J:M\to\R$, $$J(x)=\sqrt{\det(C(x))}=\lambda_1\lambda_2.$$  
This is in analogy with the volume change coefficient in three-dimensional elasticity.

\vspace{0.5em}

\begin{example}[Deformation tensor of rotationally symmetric deformation of initially flat annulus]\label{exa:duCrotsym}
Let $u:M\to\R^3$ be as in Example~\ref{exa:defrotsym}.
Then the non-zero components of the deformation gradient with respect to the coordinate bases for polar and cylinder coordinates are
\begin{align*}
du^r_R(R,\Phi) &=\frac{\del u^r}{\del R}(R,\Phi) =\frac{\del f}{\del R}(R)=f'(R), \\
du^\phi_\Phi(R,\Phi) &= \frac{\del u^\phi}{\del \Phi}(R,\Phi) = 1 ,\\
du^z_R (R,\Phi)&=\frac{\del u^z}{\del R}(R,\Phi) =\frac{\del h}{\del R}(R)=h'(R).
\end{align*}

In general, the components of the transpose map $du^T$ are given by

\begin{equation*}
(du^T)^\alpha_i=(g_{ij}\circ u)du^j_\beta G^{\alpha\beta}, 
\end{equation*}
where we use the summation convention and adopt the convention that latin indices take their values in $\left\{ 1,2,3\right\}$ and greek indices take their values in $\left\{ 1,2\right\}$. 
These conventions will henceforth be used without further comment.

The first fundamental form $G$ on $M$ is
\begin{equation*}\label{eq:G}
G(R,\Phi)=dR\otimes dR + R^2 d\Phi\otimes d\Phi,
\end{equation*}
and the standard euclidean metric $g$ on $\R^3$ is
\begin{equation*}\label{eq:g}
g(r,\phi,z)=dr\otimes dr+r^{2}d\phi \otimes d\phi +dz\otimes dz.
\end{equation*}
From this, the non-zero components of the transpose map are

\begin{align*}
(du^T)^R_r(R,\Phi) &=g_{rr}(u(R,\Phi))du^r_R(R,\Phi) G^{RR}(R,\Phi)  =f'(R) \\
(du^T)^\Phi_\phi(R,\Phi) &=g_{\phi\phi}(u(R,\Phi))du^\phi_\Phi (R,\Phi)G^{\Phi\Phi} (R,\Phi)) =\frac{f(R)^2}{R^2} \\
(du^T)^R_z(R,\Phi)&=g_{zz}(u(R,\Phi))du^z_R(R,\Phi) G^{RR}(R,\Phi)  =h'(R),
\end{align*}

giving 
\begin{align*}
C(R,\Phi) &= \left(f'(R)^2+h'(R)^2\right)\frac{\del}{\del R}\otimes dR +   \frac{f(R)^2}{R^2}\frac{\del}{\del\Phi}\otimes d\Phi \text{.} 
\end{align*}

An orthonormal eigenbasis of $C$ is given by $(\frac{\del}{\del R},\frac{1}{R}\frac{\del}{\del \Phi})$ with corresponding dual basis $(dR,R d\Phi)$. 
Defining the projection tensors $P_R:=\frac{\del}{\del R}\otimes dR$ and $P_\Phi:=\frac{1}{R}\frac{\del}{\del \Phi}\otimes R d\Phi =\frac{\del}{\del\Phi}\otimes d\Phi$ the spectral decomposition~\eqref{eq:SpecDecompC} of $C$  takes the form

\begin{align*}\label{eq:Cspecdecomp}
C &=\lambda_R^2 P_R + \lambda_\Phi^2 P_\Phi, 
\end{align*}

with the principal stretches 

\begin{align*}
 \lambda_R(R,\Phi) &=\sqrt{f'(R)^2+h'(R)^2}, & \text{\emph{(``radial stretch'')}} \\
 \lambda_\Phi(R,\Phi) &=\frac{f(R)}{R},  & \text{\emph{(``hoop stretch'').}}
\end{align*}

The radial stretch $\lambda_R$ describes the deformation in the meridional direction (along fixed angles $\Phi$), while the hoop stretch $\lambda_\Phi$ corresponds to the circumferential direction (along circles of constant radius $R$).

\vspace{0.5em}

The area change coefficient is
\begin{equation*}
J =\lambda_R\cdot \lambda_\Phi = \frac{f(R)}{R}\sqrt{f'(R)^2+h'(R)^2}. \label{eq:Jcoords}
\end{equation*}

\end{example}

\subsection{Transformation of integrals}

The metric $g$ restricted to $u(M)\subset \R^3$, i.e. the first fundamental form on $u(M)$, induces the Riemannian volume form $\mathrm{da}$ on $u(M)$. 
On the other hand, $G$ induces the Riemannian volume form $\mathrm{dA}$ on $M$. 
Comparing $\mathrm{dA}$ and the pullback $u^*(\mathrm{da})$ we find exactly the area change coefficient $J$
\begin{equation*}\label{eq:Jforms}
u^*(\mathrm{da})= J\cdot \mathrm{dA}.
\end{equation*}
So by the transformation formula, we have that for $O\subset M$ and $f:u(M)\to \R$ suitable
$$
\int_{u(O)} f\mathrm{da} =\int_O u^*(f\mathrm{da}) = \int_O J\cdot (f\circ u) \mathrm{dA}. 
$$

Moreover, the area change coefficient also appears in the comparison of the induced Riemannian volume forms on 1-dimensional submanifolds, i.e. is needed to transform integrals along curves in the membrane $M$ and their image curves in the deformed membrane $u(M)$. 
Let $c:I\to M$ be a curve in $M$, let $N$ be an unit normal vector field $N$ along $c$ and let $\mathrm{dL}$ be the induced volume form on $c$. 
Consider the image curve $u\circ c$ in the deformed membrane with a unit normal vector field $n$ and induced volume form $\mathrm{dl}$. 
Then for any vector field $v$ along $u\circ c$

$$
u^*(g(v,n)\mathrm{dl}) = J\cdot G(u^*(v),N)\mathrm{dL},
$$
such that again from the transformation formula we have
\begin{equation}\label{eq:trafo}
\int_{u\circ c} g(v,n)\mathrm{dl} = \int_c   u^*(g(v,n)\mathrm{dl}) = \int_c J\cdot G(u^*(v),N) dL.
\end{equation}

\subsection{Stress tensors of membrane deformations}
Following a deformation $ u : M \to \mathbb{R}^3 $ of an elastic membrane, internal stresses arise within the deformed surface $ u(M) $. 
At a given point $ x \in u(M) $, different tangent directions experience different stress levels.

To analyze the resulting forces, we consider a virtual cut through the deformed membrane, represented by a smooth curve $ c : I \to u(M) $, together with a unit normal field $ n $ along $ c $, tangent to $ u(M) $. 
Let $t(x,n)$ be the force vector per unit length at $x$ across the cut, called \emph{Cauchy stress vector}. 
Note that $t(x,n)\in T_x u(M)$, it is tangential to the surface $u(M)$. 
It is the force per unit length engendered by the material outside ( the $n$-direction) on the material inside (the $-n$-direction).

For a fixed spatial direction, given as a fixed vector $e$ on $\R^3$, the total force $F_{c,e}$ along $c$ in the direction $e$ is

\begin{equation} \label{eq:integral}
F_{c,e}=\int_{c} g(t(x,n),e) \mathrm{dl}(x)
\end{equation}
where $dl$ is the induced volume form on the curve. 

By Cauchy's theorem for arbitrary dimensions there exists an endomorphism field $\sigma$ such that $$t(x,n)=\sigma(x)n.$$ 
We call $\sigma$ (membrane) \emph{Cauchy stress tensor field}.

\vspace{0.5em}

For computations, it is convenient to transform the integral~\eqref{eq:integral} using~\eqref{eq:trafo}. 
To this end, we define a new stress tensor field $P$, which is a tensor field along $u$, such that the following relation holds: For any curve $c:I\to M$ in the undeformed membran with unit normal field $N$, and any spatial direction $e$, the total force along $u\circ c$ in direction $e$ satisfies 

\begin{equation} \label{eq:trafo_sigma}
F_{c,e}=\int_{u\circ c} g(\sigma(n),e) \mathrm{dl} = \int_c g(P(N),e)\circ u \, dL.
\end{equation}

It is not hard to see using~\eqref{eq:trafo}, symmetry of $\sigma$ and tangential projections of $e$, that this is equivalent to 

$$
P= J\cdot \sigma \circ (du^{-1})^T.
$$

This is the membrane version of the Piola transformation and $P$ is called \emph{first Piola-Kirchhoff stress tensor field}.

From $P$ we define the \emph{second Piola-Kirchhoff stress tensor field} $S$ to be the endomorphism field on $M$ such that 

\begin{equation} \label{eq:defP}
P=du\circ S. 
\end{equation}

We will discuss in~\ref{sec:stress-strain-rs} how to determine the stress tensor fields $S,P$ and $\sigma$ from the Cauchy-Green deformation tensor $C$ of some membrane deformation $u$ by means of a material law, defined for hyperelastic materials by a membrane energy function $W$.

\vspace{0.5em}

The considerations in~\ref{sec:stress-strain-rs} will prove that for isotropic materials, at every point $x\in M$ of the membrane, the second Piola-Kirchhoff stress tensor $S(x)$ is \emph{coaxial}
\footnote{Two symmetric endomorphisms $A,B$ on an Euclidean vector space are called coaxial, if they admit a common orthonormal basis of eigenvectors or equivalently if they are simultaneously diagonalizable in an orthonormal basis. This condition ensures that $A$ and $B$ share the same principal directions. $A$ and $B$ are coaxial if and only if $AB=BA$.}
to the Cauchy-Green deformation tensor $C(x)$.

Assume a spectral decomposition of $C$ is given by $C=\lambda_1^2P_1+\lambda_2^2P_2$, where $(\lambda_1^2,\lambda_2^2)$ are the eigenvalues of $C$. 
Since $S$ and $C$ are coaxial, the eigenvalues of $S$, called \emph{principal stresses} $(s_1,s_2)$ yield a spectral decomposition $S=s_1P_1+s_2P_2$ with the same projections $P_1,P_2$.

\vspace{0.5em}

\begin{example}[Total vertical force at inner boundary for rotationally symmetric deformation of elastic annulus]\label{exa:Ftot}
We continue the discussion of Examples~\ref{exa:defrotsym} and~\ref{exa:duCrotsym}.
\vspace{0.5em}

The undeformed inner boundary of the annulus $M$ is parametrized by the curve $c:(0,2\pi)\to \R^3$ given as $c(t)=F_p(R_1,t)$ with $F_p$ as in~\eqref{eq:pc}.
The inner unit normal vector field $N(t)\in T_{c(t)}M$ for that curve is the radial coordinate vector field, $N(t)=\frac{\del}{\del R}(c(t))$.

For the total vertical force, the space direction to choose is the euclidean standard basis vector $e_3$, in cylinder coordinates given as the coordinate vector field $\frac{\del}{\del z }$.

\vspace{0.5em}
The total vertical force $F_z$ at the inner boundary for the rotationally symmetric deformation $u$ with first Piola stress tensor $P$ then by~\eqref{eq:trafo_sigma} is
\begin{align*}
 F_z:=F_{u\circ c, \tfrac{\del}{\del z }} 
 &=\int_c g(P(N),\tfrac{\del}{\del z })\circ u \,dL \nonumber \\
 &= \int_0^{2\pi} g(P(\tfrac{\del}{\del R}),\tfrac{\del}{\del z })(u(c(t))) R_1 dt ,
\end{align*}
where $dL(c(t))=\norm{\dot{c}(t)}dt=R_1dt$.

From the second Piola-Kirchhoff stress tensor $S=s_RP_R+s_\Phi P_\Phi$, coaxial to the Cauchy-Green deformation tensor, but with eigenvalues $s_R=S^R_R$ (\emph{radial stress}) and  $s_\Phi=S^\Phi_\Phi$ (\emph{hoop stress}), we obtain the first Piola-Kirchhoff stress tensor $P$ and compute

\begin{align*} 
P(\tfrac{\del}{\del R}) 
&=du(S(\tfrac{\del}{\del R})) = du(s_R \tfrac{\del}{\del R})\\
& = s_R\cdot  du(\tfrac{\del}{\del R}) = s_R \cdot (f' \tfrac{\del}{\del r}\circ u + h' \tfrac{\del}{\del z }\circ u).
\end{align*}

For the total vertical force at the inner boundary we then obtain 

\begin{align}
 F_z  \nonumber
   &=\int_0^{2\pi} g(s_R \cdot (f' \tfrac{\del}{\del r} + h' \tfrac{\del}{\del z }),\tfrac{\del}{\del z }))(u(c(t))) R_1 dt  \\ \nonumber
 &= R_1\cdot s_R(R_1) \int_0^{2\pi} \left( f' g_{rz}\circ u+h' g_{zz}\circ u \right)(c(t)) dt  \\ \nonumber
  &= R_1\cdot s_R(R_1) \int_0^{2\pi} h'(c(t)) dt  \\ \nonumber
 &=  2\pi\cdot R_1\cdot s_R(R_1) \cdot h'(R_1)   ,\\ \label{eq:Fz}
\end{align}

using that $g_{rz}=0$ and where $s_R(c(t))$ and $h'(c(t))$ do not depend on $t$.
\end{example}

\subsection{Energy of membrane deformations}

The elastic energy stored in $u(M)$ following a deformation $u : M \to \mathbb{R}^3$ depends on the material properties of the membrane.  
Analogous to the stored energy density functions in three-dimensional hyperelasticity, we define \emph{membrane energy functions}:  
A membrane energy function $w$ assigns to each positive definite symmetric deformation tensor an associated energy density (energy per area), such that the integral
\begin{equation}\label{eq:Etot}
E[u] = \int_U w(C(x))\, dA(x),
\end{equation}
where $C$ is the Cauchy--Green deformation tensor field associated with a membrane deformation $u : M \to \mathbb{R}^3$ and $U \subset M$ is a suitable subset, represents the internal energy stored in the deformed region $u(U) \subset u(M)$.

Classically, the energy function is defined on a matrix space—specifically, the space of symmetric positive definite matrices.
Note that the matrix representation of $C(x)$ in an arbitrary basis of the tangent space is not necessarily symmetric; although $C$ is a positive definite, symmetric tensor. However, $C(x)$ is always similar to a positive definite, symmetric matrix, and therefore it is diagonalizable with real, strictly positive eigenvalues.

Let $\mathbb{D}_{>0} \subset \mathbb{R}^{2 \times 2}$ denote the set of all diagonalizable $2 \times 2$ matrices with strictly positive eigenvalues.

We consider energy functions $w : \mathbb{D}_{>0} \to \mathbb{R}$ that are invariant under similarity transformations\footnote{This invariance corresponds to material isotropy; see Corollary~5.11 in Chapter~5 of~\cite{MH}.}.  
Such invariance ensures that $w(C(x))$ is well-defined and independent of coordinates, since a change of coordinates yields a similar matrix representation for $C(x)$.

\vspace{0.5em}

Any such invariant membrane energy function $w : \mathbb{D}_{>0} \to \mathbb{R}$ can be expressed in terms of a symmetric function $\Wpsb : \mathbb{R}^2 \to \mathbb{R}$ of the \emph{principal stretches}:
\[
w(C) = \Wpsb(\lambda_1, \lambda_2),
\]
where $C = \lambda_1^2 P_1 + \lambda_2^2 P_2$ is the spectral decomposition of $C$.

According to~\cite[Chapter~VII, Section~L]{LibaiSimmonds1998}, there are two common approaches to determining membrane energies:  
the \emph{direct approach}, which relies on rational analysis and experimentation, and the \emph{descent approach}, which derives membrane energy functions from the three-dimensional or shell formulation of the stored energy density.

In this work, we consider membrane energy functions of the second type, derived\footnote{One defines  
\[
W^{2D}(\lambda_1, \lambda_2) := d \cdot W^{3D}\left( \lambda_1, \lambda_2, \tfrac{1}{\lambda_1 \lambda_2} \right),
\]  
where $W^{3D}$ is an incompressible 3D stored energy density function and $d$ denotes the membrane thickness. The plane stress condition is used to eliminate the hydrostatic pressure that appears in incompressible elasticity.}
from incompressible three-dimensional stored energy functions.

\vspace{0.5em}
We consider the following membrane versions of well-known material models\footnote{In Appendix~\ref{secA1} a review on these material models can be found.}, depending on different sets of constant material parameters $C_i,a_i$.

\begin{align}
\text{Neo-Hookean}\quad  & W=C_1(I_1-3), \label{eq:defWN} \\
\text{Mooney}\quad & W=C_1(I_1-3)+C_2(I_2-3), \label{eq:defWM} \\
\text{Gent}\quad & W=-C_1C_2\ln(1-C_2^{-1}(I_1-3)), \label{eq:defWG} \\
\text{Yeoh, 3rd order}\quad & W=\sum_{i=1}^3C_i(I_1-3)^i,\label{eq:defWY} \\
\text{Ogden, 3rd order}\quad & W=\sum_{i=1}^3 \frac{C_i}{a_i}(I_1(a_i)-3), \label{eq:defWO} 
\end{align}

where\footnote{ $I_1,I_2$ can be interpreted as the first two principal invariants of the deformation tensor $\overline{C}$ of a particular isochor extension $\overline{u}:\overline{M}\to\R^3$ of the membrane deformation $u$ to the three-dimensional shell $\overline{M}$. The right Cauchy-Green deformation tensor $\overline{C}$ of this extension then has the principal stretches $\lambda_1,\lambda_2,\lambda_3$, with $\lambda_3$ being the stretch in the direction normal to the midsurface. 
Due to the incompressibility constraint, $\det(\overline{C})=\lambda_1\lambda_2\lambda_3=1$, and thus $I_1=\lambda_1^2+\lambda_2^2+\lambda_3^2=\lambda_1^2+\lambda_2^2+1/(\lambda_1^2\lambda_2^2)$. }
\begin{align*}
I_1 & =\lambda_1^2+\lambda_2^2+\tfrac{1}{\lambda_1^2\cdot \lambda_2^2} ,\\
I_2 & =\lambda_1^2\cdot \lambda_2^2 + \frac{\lambda_1^2+\lambda_2^2}{\lambda_1^2\cdot \lambda_2^2} , \\
I_1(a) &= \lambda_1^a+\lambda_2^a+\left(\lambda_1\cdot \lambda_2\right)^{-a}.
\end{align*}

\subsection{Stress-strain relationship for hyperelastic membranes} \label{sec:stress-strain-rs}

The stresses arising in a deformed membrane $u(M)$ depend on the strains following $u$ by means of a material law. 
Different materials respond with different stresses for the very same deformation, depending on how soft or stiff the material is.

It can be shown, that for a time dependend deformation (a motion) with time dependend Cauchy-Green deformation tensor field $C(t,x)$ the stress power per unit area at a point $x\in M$ is given by $\tfrac12\langle S(x,t),\dot{C}(x,t) \rangle$ where $\dot{C}(x,t)$ is the time derivative of the Cauchy-Green deformation tensor and $\<\cdot,\cdot\>$ is the induced scalar product on endomorphisms\footnote{This coincides in coordinates with the standard scalar product $\<A,B\>=\tr(A^TB)$ for matrices.} of the tangent space at $x$.  
Using basic balance principles, including balance of energy, this equals the rate of increase of internal elastic energy. 
For hyperelastic membrane materials with membrane energy function $w$ we find

\begin{equation}\label{eq:balanceofenergylocal}
\frac{\del}{\del t} w(C(t,x)) = \tfrac12 \<S(t,x),\dot{C}(x,t)\>.
\end{equation}

On the other hand we have

\begin{align}
 \frac{\del}{\del t} w(C(t,x)) 
 &=(\del_{\dot{C}(x,t)} w ) (C(t,x)) \nonumber \\
 &=Dw(C(t,x))(\dot{C}(x,t)) \nonumber \\
 &=\<\grad w(C(t,x)), \dot{C}(x,t) .\label{eq:grad}\>
\end{align}

So, since~\eqref{eq:balanceofenergylocal} and~\eqref{eq:grad} hold for any $\dot{C}(x,t)$ we see that the second Piola-Kirchhoff stress tensor can be computed from the membrane energy function $w:\mathbb{D}_{>0}\to\R$ by taking the \emph{gradient} of $w$ with respect to the standard scalar product on matrices and multiplying with a factor of $2$. 
This can now be stated dropping time-dependency:

\begin{equation*}\label{eq:SSR}
S(x)=2\grad(w)(C(x))
\end{equation*}

This is the stress-strain relationship for hyperelastic materials.

If the membrane energy function is given as a symmetric function of the principal stretches $\lambda_1,\lambda_2$, (spectral decomposition $C=\lambda_1^2P_1+\lambda_2^2P_2$, $\lambda_\alpha^2=\<C,P_\alpha\>$) it can be shown\footnote{This can be achieved using the fact that $\<P_\alpha,\dot{P_\beta}\>=0$ for any $\alpha,\beta$. Some subtleties have to be adressed in the case of multiplicities of the eigenvalues larger than 1 in order to compute the derivative.}
that for a time dependend deformation

\begin{align*}
  \frac{\del}{\del t} w(C(t,x)) 
  &= \frac{\del}{\del t} \left( \Wpsb(\lambda_1(t,x),\lambda_2(t,x) \right) \\
  &= \<\frac{1}{2}\sum_{\alpha=1}^2 \frac{1}{\lambda_\alpha(x,t)}\del_\alpha \Wpsb(\lambda_1(x,t),\lambda_2(x,t)))P_\alpha(x,t), \dot{C}(x,t). \>
\end{align*}

Comparing with~\eqref{eq:balanceofenergylocal} this yields the stress-strain relationship

\begin{equation*}
S(x)=\sum_{\alpha=1}^2 \frac{1}{\lambda_\alpha(x)}\del_\alpha \Wpsb(\lambda_1(x),\lambda_2(x)))P_\alpha(x)\text{.}
\end{equation*}

In particular, this proves that the second Piola-Kirchhoff stress tensor is everywhere coaxial to the Cauchy-Green deformation tensor, the principal stresses are given by

$$
s_1 = \frac{1}{\lambda_1}\del_1 \Wpsb(\lambda_1,\lambda_2), \qquad s_2 = \frac{1}{\lambda_2}\del_2 \Wpsb(\lambda_1,\lambda_2).
$$

For the membrane energies~\eqref{eq:defWN} to~\eqref{eq:defWO} this yields the principal stresses

\begin{align*}
\text{Neo-Hookean}\quad  & s_\alpha  = 2 C_{1}\left(1 - (\lambda_{\alpha} J)^{-2}\right),  \\
\text{Mooney}\quad &  s_\alpha = 2C_{1}\left(1-(\lambda_{\alpha}J)^{-2}\right)+2C_{2}\left(J^{2}\lambda_{\alpha}^{-2}-\lambda_{\alpha}^{-4}\right), \\
\text{Gent}\quad & s_\alpha = 2C_{1}\left((\lambda_{\alpha}J)^{-2}-1\right)\left(1-C_{2}^{-1}(I_{1}-3)\right)^{-1} ,\\
\text{Yeoh, 3rd order}\quad & s_{\alpha}= 2\left(1-\left(\lambda_{\alpha}J\right){}^{-2}\right)(C_{1}+2C_{2}(I_{1}-3)+3C_{3}(I_{1}-3)^{2}), \\
\text{Ogden, 3rd order}\quad & s_\alpha = \sum_{i=1}^{3}C_{i}\lambda_{\alpha}^{-2}\left(\lambda_{\alpha}^{a_{i}}-J^{a_{i}}\right).    
\end{align*}

\begin{example}[Principal stresses in Neo-Hookean membrane under rotationally symmetric deformation]
\label{exa:stresses}

For the rotationally symmetric deformation $u:M\to \R^3$ as in Examples~\ref{exa:defrotsym},\ref{exa:duCrotsym} with radial stretch $\lambda_R=\sqrt{f'(R)^2+h'(R)^2}$ and hoop stretch $\lambda_\Phi=\tfrac{f(R)}{R}$ we thus obtain the corresponding radial stress $s_R$ and  hoop stress $s_\Phi$ for a Neo-Hookean membrane with membrane energy function~\eqref{eq:defWN}:

\begin{align*}
 s_R(R,\Phi) &= C_1 \left(1-\frac{1}{\lambda_R(R,\Phi)^4\lambda_\Phi(R,\Phi)^2}\right)= C_1 \left(1-\frac{R^2}{f(R)^2(f'(R)^2+h'(R)^2)^2}\right),\\
 s_\Phi(R,\Phi) &= C_1 \left(1-\frac{1}{\lambda_R(R,\Phi)^2\lambda_\Phi(R,\Phi)^4}\right) = C_1 \left(1-\frac{R^4}{f(R)^4(f'(R)^2+h'(R)^2)}\right),
\end{align*}

where we see the dependence of the principal stresses on $R,f(R),h(R),f'(R),h'(R)$.

\end{example}

\subsection{Equilibrium equations for membrane deformations}

The deformed membrane is in a static equilibrium, if all inner forces are balanced.

Mathematically, a membrane deformation $u : M \to \mathbb{R}^3$ is said to be in equilibrium if, for every spatial direction $e$ and every suitable subset $\m \subset M$, the total force along the boundary of the deformed region $\partial u(\m)$ in direction $e$ vanishes. Here, $\partial u(\m)$ is equipped with the outward unit normal field $n$ and the induced Riemannian line element $\mathrm{d}l$.

With equation~\eqref{eq:integral} this equilibrium condition can be expressed as

\begin{equation*}
\forall e,\m: \quad \int_{\del u(\m)} g(\sigma(n),e) \mathrm{dl} =0.
\end{equation*}

By~\eqref{eq:trafo_sigma} this can equivalently be formulated with the first Piola-Kirchhoff stress tensor field $P$ of the deformation 

$$
\forall e,\m: \int_{\del \m} g(P(N),e)\circ u \, dL =0
$$

By transposing $P$ onto $e$ and applying the divergence theorem, the boundary integral is transformed into an integral over~$\m$.

Using a general product rule for the divergence, defined as usual via tracing the covariant derivative of the tensor field along $u$, the integrand becomes

\begin{equation*}
\div(P^T(e)))=\<\nabla e,P\>+g(\div (P),e) \overset{\nabla e = 0}{=} g(\div (P),e).
\end{equation*}

Since the integral vanishes for arbitrary subsets $\m$, the integrand must vanish pointwise. 
As this holds for any spatial direction $e$, we obtain the local form of the static membrane equilibrium equations:

\begin{equation}\label{eq:divP}
 \div(P)=0 .
\end{equation}

In local coordinates $(y^1,y^2,y^3)$ on $\R^3$ and local coordinates $(X^1,X^2)$ on $M$ the  vector field $\div(P)$ along the deformation map $u$ has three components $\div(P)^i$, it is

\begin{align}
\div(P) = \div(P)^i \frac{\del}{\del y^i}\circ u  =G^{\alpha\beta}\left(\frac{\del P_{\beta}^{i}}{\del X^{\alpha}}+P_{\beta}^{k}\frac{\del u^{j}}{\del X^{\alpha}}\left(\gamma_{jk}^{i}\circ u\right)-\Gamma_{\alpha\beta}^{\delta}P_{\delta}^{i}\right) \frac{\del}{\del y^i}\circ u. \label{eq:divPcoords}
\end{align}

The $\gamma_{ij}^{k}$ are the Christoffel symbols of the standard derivative for vector fields on $\R^3$ and $\Gamma_{\alpha\beta}^{\delta}$ are the Christoffel symbols of the Levi-Civita connection of $G$ on $M$.

Taking into account the definition of the first Piola-Kirchhoff stress tensor field~\eqref{eq:defP}, written in coordinates as 

\[
P_{\beta}^{i}=\frac{\del u^{i}}{\del X^{\varepsilon}}S_{\beta}^{\varepsilon},
\]

we obtain the components 

\begin{align*}
\div(P)^i =  G^{\alpha\beta}&\left(\frac{\del^{2}u^{i}}{\del X^{\alpha}\del X^{\varepsilon}}S_{\beta}^{\varepsilon}+\frac{\del u^{i}}{\del X^{\varepsilon}}\frac{\del S_{\beta}^{\varepsilon}}{\del X^{\alpha}}+\frac{\del u^{i}}{\del X^{\varepsilon}}\frac{\del u^{j}}{\del X^{\alpha}}S_{\beta}^{\varepsilon}\left(\gamma_{jk}^{i}\circ u\right)-\Gamma_{\alpha\beta}^{\delta}\frac{\del u^{i}}{\del X^{\varepsilon}}S_{\delta}^{\varepsilon}\right),
\end{align*}

that have to vanish for a static equilibrium.

From the previous section we know, that the components of the second Piola-Kirchhoff stress tensor field $S^\alpha_\beta$ depend on the deformation $u$ only up to first derivatives, since they are all expressions in the principal stretches. 

Therefore, the static membrane equilibrium equations are a coupled system of three second order quasi-linear partial differential equations for the unknown deformation $u$. 
Assuming  a boundary condition of place for the deformation $u$ on $\del M$ turns this into a boundary value problem.

The analytical questions of existence, uniqueness, and stability of solutions to general boundary value problems in nonlinear membrane theory will not be addressed in this work. 
These questions are closely tied to suitable convexity and growth conditions on the membrane energy density. 
A comprehensive treatment of the corresponding analytical framework in the context of three-dimensional elasticity can be found in the literature:
For instance, Part B of Ciarlet’s monograph on three-dimensional elasticity~\cite{CiarletI} provides both an existence theory based on linearization and the implicit function theorem (Chapter~6), as well as an approach via direct minimization of the energy functional under appropriate conditions (Chapter~7). 
The book by Marsden and Hughes~\cite{MH} covers related aspects from a variational and functional-analytic perspective, including linearization techniques (Chapter~4), variational principles (Chapter~5), and the use of functional analytic tools in elasticity (Chapter~6).

\vspace{1em}

The vector field $\div(P)$ splits into a part tangential and a part normal to the deformed surface $u(M)$. 
Let us write $\div(P)=\div(P)^{\parallel}+\div(P)^{\perp}$.
Then it turns out, that the tangential part is $$\div(P)^{\parallel}=J\div^{\mathrm{int}}(\sigma)\circ u \text{,}$$ whereas the normal part is given by $$\div(P)^{\perp}=J\tr(\mathrm{II}\circ \sigma )\circ u.$$
Here $\div^{\mathrm{int}}(\sigma)$ denotes the intrinsic divergence of the Cauchy stress vector field on the deformed membrane $u(M)$, computed with the intrinsic connection associated with the first fundamental form of the deformed membrane $u(M)$, $\mathrm{II}$ denotes the vector valued second fundamental form on $u(M)$ and the trace is $\tr(\mathrm{II}\circ \sigma )=\sum_{i=1}^2\mathrm{II}(\sigma(e_i),e_i)$ for a local orthonormal frame $(e_1,e_2)$ of the tangent space of the deformed surface. 

This splitting of the equilibrium equations involves computing the first and second fundamental form of the deformed membrane, this itself depending on the deformation. 
Therefore it is not convenient for computations, but the splitting sheds light on a difference of these equilibrium equations and their classical three-dimensional counterpart where no such splitting appears. Moreover, we computed the splitting in order to compare with the usual equilibrium equations as stated in Green and Zerna~\cite{GreenZerna1954}.

\vspace{1em}

As was to be expected, one can also check that the Euler-Lagrange equations derived from the energy functional~\eqref{eq:Etot} are exactly the equilibrium equations~\eqref{eq:divP}.

\vspace{0.5em}

\begin{example}[Equilibrium equations for rotationally symmetric deformation of elastic annulus]
We recall from Example~\ref{exa:defrotsym} and~\ref{exa:duCrotsym} the deformation given in polar coordinates $(R,\Phi)$ on $M$  and cylinder coordinates $(r,\phi,z)$ on $\R^3$ as
$$
u(R,\Phi)=(f(R), \Phi,h(R))
$$
with deformation gradient 
$$du=f'dR\otimes \frac{\del}{\del r}\circ u +d\Phi\otimes\frac{\del}{\del\phi}\circ u + h'dR\otimes \frac{\del}{\del z}\circ u,$$ 
Cauchy-Green deformation tensor 
$$
C = \lambda_R^2\frac{\del}{\del R}\otimes dR +  \lambda_\Phi^2 \frac{\del}{\del\Phi}\otimes d\Phi
$$
and principal stretches 
$$
 \lambda_R =\sqrt{f'(R)^2+h'(R)^2}, \qquad \lambda_\Phi =\frac{f(R)}{R}. 
$$

For any choice of a membrane energy function $\Wpsb$ we obtain the second Piola-Kirchhoff stress tensor $S$ with the principal stresses $(s_R,s_\Phi)$ as functions of the principal stretches 

$$ s_R  =\frac{1}{\lambda_R}(\del_{1}\Wpsb)(\lambda_R,\lambda_\Phi), \qquad s_\Phi =\frac{1}{\lambda_{\Phi}}(\del_{2}\Wpsb)(\lambda_{R},\lambda_{\Phi}).$$

The first Piola-Kirchhoff stress tensor field is 
\begin{align*}
P &=du\circ S \\
 &=f's_RdR\otimes \frac{\del}{\del r}\circ u +s_\Phi d\Phi\otimes\frac{\del}{\del\phi}\circ u + h's_RdR\otimes \frac{\del}{\del z}\circ u.
\end{align*}
With the Christoffel symbols for polar and cylinder coordinates and equation~\eqref{eq:divPcoords}, the divergence of~$P$ evaluates to

\begin{align*}
\div(P) &= \left(s_{R}'\cdot f'+s_{R}\cdot f''+\frac{s_{R}\cdot f'}{R}-\frac{f\cdot s_{\Phi}}{R^{2}}\right)\frac{\del}{\del r}\circ u \\
&\phantom{=} \,\, + \left(s_{R}'\cdot h'+s_{R}\cdot h''+\frac{s_{R}\cdot h'}{R}\right)\frac{\del}{\del z}\circ u.
\end{align*}

As expected for reasons of symmetry the component $\div(P)^\phi$ of the vector field $\div(P)$ along~$u$ vanishes. 

The membrane equilibrium equations $\div(P)=0$ are thus equivalent to 
\begin{align}
 0 =\div(P)^r  = & \,\, s_{R}'\cdot f'+s_{R}\cdot f''+\frac{s_{R}\cdot f'}{R}-\frac{f\cdot s_{\Phi}}{R^{2}},  \nonumber \\  
 0 =\div(P)^z  = & \,\, s_{R}'\cdot h'+s_{R}\cdot h''+\frac{s_{R}\cdot h'}{R}  \text{,} \label{eq:ODEs},
\end{align}
a system of two non-linear second order ordinary differential equations for the unknown functions $f,h$ that determine the deformation $u$.\footnote{They coincide with the equilibrium equations for this deformation given in~\cite{FultonSimmonds1986}}

In the following we want to reduce the order of this system to prepare for numerical routines.
To this end, we first have to identify the coefficients of the second derivatives of $f,h$.
The derivative of the radial stress is $s_R'=\del_1 s_R \lambda_R'+\del_2 s_R\lambda_\Phi'$. 
Using 
$$
\lambda_R'=\frac{f'}{\lambda_R}f'' + \frac{h'}{\lambda_R}h'' \text{ and } \lambda_\Phi'=\frac{Rf'-f}{R^2}
$$
we obtain the decomposition $$s_R'=\underbrace{\del_1 s_R \frac{f'}{\lambda_R}}_{=:A}\cdot f'' + \underbrace{\del_1 s_R \frac{h'}{\lambda_R}}_{=:B}\cdot h'' +\underbrace{\del_2 s_R\frac{Rf'-f}{R^2}}_{=:C}$$ where the terms $A,B$ and $C$ are of lower order in $f$ and $h$.

Using this decomposition in the equilibrium equations~\eqref{eq:ODEs} we find\footnote{The quasi-linearity of the system becomes apparent again here.}

\begin{align*}
\div(P)^r &= \underset{=:F_{r}}{\underbrace{\left(Af'+s_R\right)}}f'' + \underset{=:H_{r}}{\underbrace{Bf'}}h''+\underset{=:R_{r}}{\underbrace{Cf'+\frac{f s_{\Phi}}{R^{2}}}},  \\
\div(P)^z &= \underset{=:F_{z}}{\underbrace{Ah'}}f''+\underset{=:H_{z}}{\underbrace{\left(Ah'+s_R\right)}}h''+\underset{=:R_{z}}{\underbrace{Ch'+\frac{s_{R}\cdot h'}{R}}}. 
\end{align*}

We can now rewrite~\eqref{eq:ODEs} as

\begin{equation} \label{eq:sys2}
\begin{pmatrix}F_{r} & H_{r}\\
F_{z} & H_{z}
\end{pmatrix}\begin{pmatrix}f''\\
h''
\end{pmatrix}=\begin{pmatrix}-R_{r}\\
-R_{z}
\end{pmatrix}.
\end{equation}

We can reduce~\eqref{eq:sys2} to a first order system by introducing
\begin{equation*}
M:=\begin{pmatrix}1 & 0 & 0 & 0\\
0 & 1 & 0 & \\
0 & 0 & F_{r} & H_{r}\\
0 & 0 & F_{z} & H_{z}
\end{pmatrix}, \quad y:=\begin{pmatrix} f\\ h\\ f'\\h' \end{pmatrix} \quad \text{ and }
b:=\begin{pmatrix}f'\\
h'\\
-R_{r}\\
-R_{z}
\end{pmatrix},
\end{equation*}
then~\eqref{eq:sys2} is equivalent to $$My'=b,$$
hence to
\begin{align*}
y' & = M^{-1}\cdot b \\
 &= \frac{1}{\det\left(\begin{pmatrix}F_{r} & H_{r}\\
F_{z} & H_{z}
\end{pmatrix}\right)}\begin{pmatrix}1 & 0 & 0 & 0\\
0 & 1 & 0 & 0\\
0 & 0 & H_{z} & -H_{r}\\
0 & 0 & -F_{z} & F_{r}
\end{pmatrix}\begin{pmatrix}f'\\
h'\\
-R_{r}\\
-R_{z}
\end{pmatrix}. \\
\end{align*}

With $D:=\det\left(\begin{pmatrix}F_{r} & H_{r}\\
F_{z} & H_{z}
\end{pmatrix}\right)=F_rH_z-F_zH_r$ we obtain

\begin{equation} \label{eq:ode_for_numerics}
y'=\begin{pmatrix}f'\\
h'\\
D^{-1}\left(H_{r}R_{z}-R_{r}H_{z}\right)\\
D^{-1}\left(R_{r}F_{z}-F_{r}R_{z}\right)
\end{pmatrix}.
\end{equation}

The problem is now of the standard form $$y'(R)=A(R,y(R))$$ with $A:\R\times \R^4\to \R^4$ given by the right hand side of~\eqref{eq:ode_for_numerics}. 
By adding for example initial values or boundary values this system can be given to numerical routines to obtain an approximate solution. 

The terms $F_r,H_r,R_r$ and $F_z,H_z,R_z$ depend via the expressions $s_R,s_\Phi,\del_1 s_R,\del_2 s_R$ on the material.

For a Neo-Hookean membrane, using the principal stretches from Example~\ref{exa:stresses} it is

\begin{align*}
 F_r &= \frac{8 C_{1} R^{2} {f'}^{2}}{f^{2} \left({f'}^{2} + {h'}^{2}\right)^{3}} - \frac{2 C_{1} R^{2}}{f^{2} \left({f'}^{2} + {h'}^{2}\right)^{2}} + 2 C_{1} ,\\
 H_r &= \frac{8 C_{1} R^{2} f' h'}{f^{2} \left({f'}^{2} + {h'}^{2}\right)^{3}}, \\
 R_r &= \frac{4 C_{1} R {f'} \left(R {f'} - f\right)}{f^{3} \left({f'}^{2} + {h'}^{2}\right)^{2}} + \frac{{f'} \left(- \frac{2 C_{1} R^{2}}{f^{2} \left({f'}^{2} + {h'}^{2}\right)^{2}} + 2 C_{1}\right)}{R} - \frac{f \left(- \frac{2 C_{1} R^{4}}{f^{4} \left({f'}^{2} + {h'}^{2}\right)} + 2 C_{1}\right)}{R^{2}}, \\
 F_z &=\frac{8 C_{1} R^{2} {f'} {h'}}{f^{2} \left({f'}^{2} + {h'}^{2}\right)^{3}}, \\
 H_z &= \frac{8 C_{1} R^{2} {h'}^{2}}{f^{2} \left({f'}^{2} + {h'}^{2}\right)^{3}} - \frac{2 C_{1} R^{2}}{f^{2} \left({f'}^{2} + {h'}^{2}\right)^{2}} + 2 C_{1}, \\
 R_z &= \frac{4 C_{1} R {h'} \left(R {f'} - f\right)}{f^{3} \left({f'}^{2} + {h'}^{2}\right)^{2}} + \frac{{h'} \left(- \frac{2 C_{1} R^{2}}{f^{2} \left({f'}^{2} + {h'}^{2}\right)^{2}} + 2 C_{1}\right)}{R}.
\end{align*}

\end{example}

\section{Experiments}\label{sec:experiments}

\subsection{Material and Preparation}
\begin{figure}[H]
\centering
\includegraphics[width=.3\textwidth]{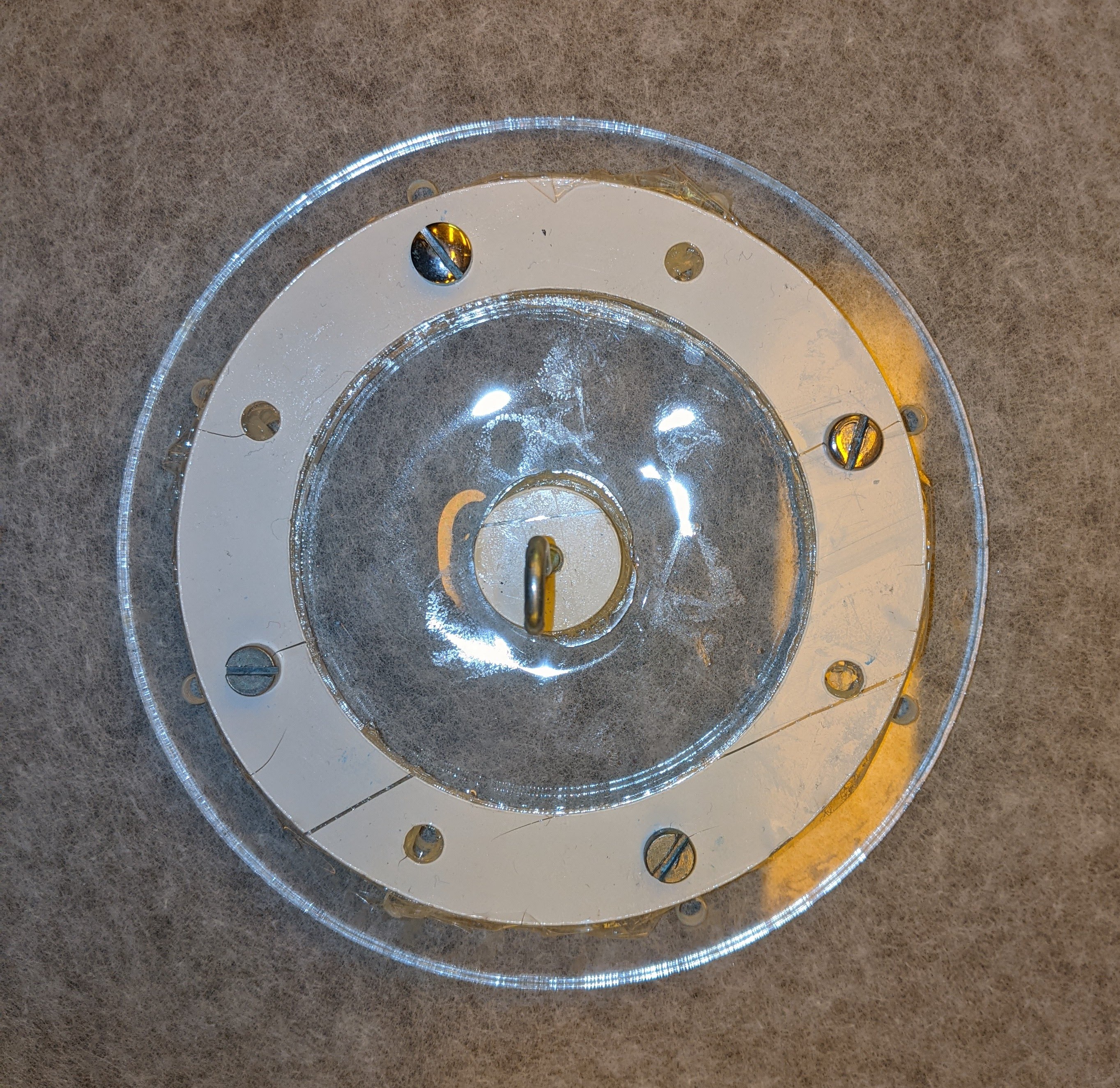}\hfill
\includegraphics[width=.3\textwidth]{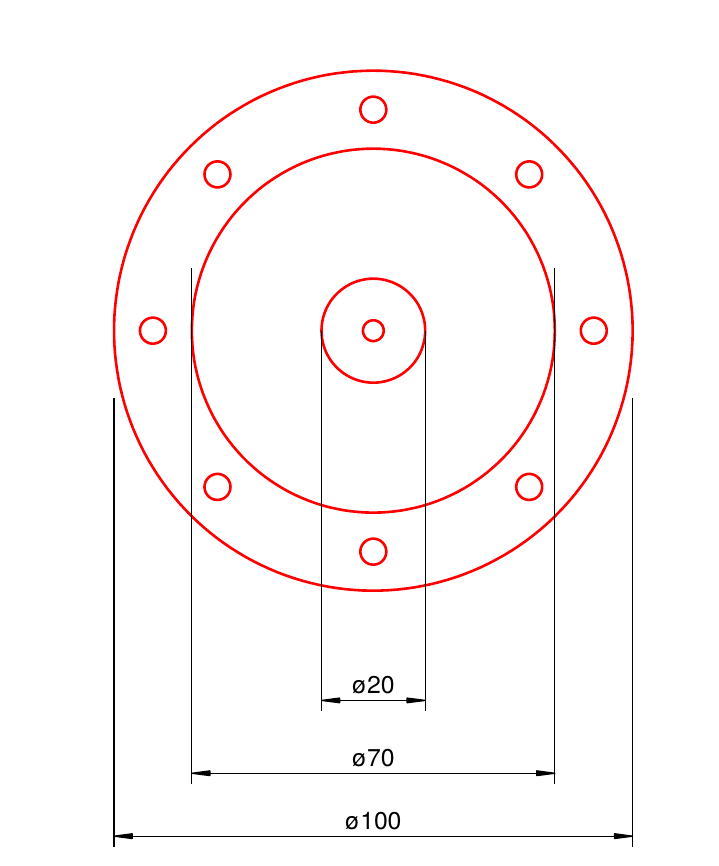}\hfill
\includegraphics[width=.3\textwidth]{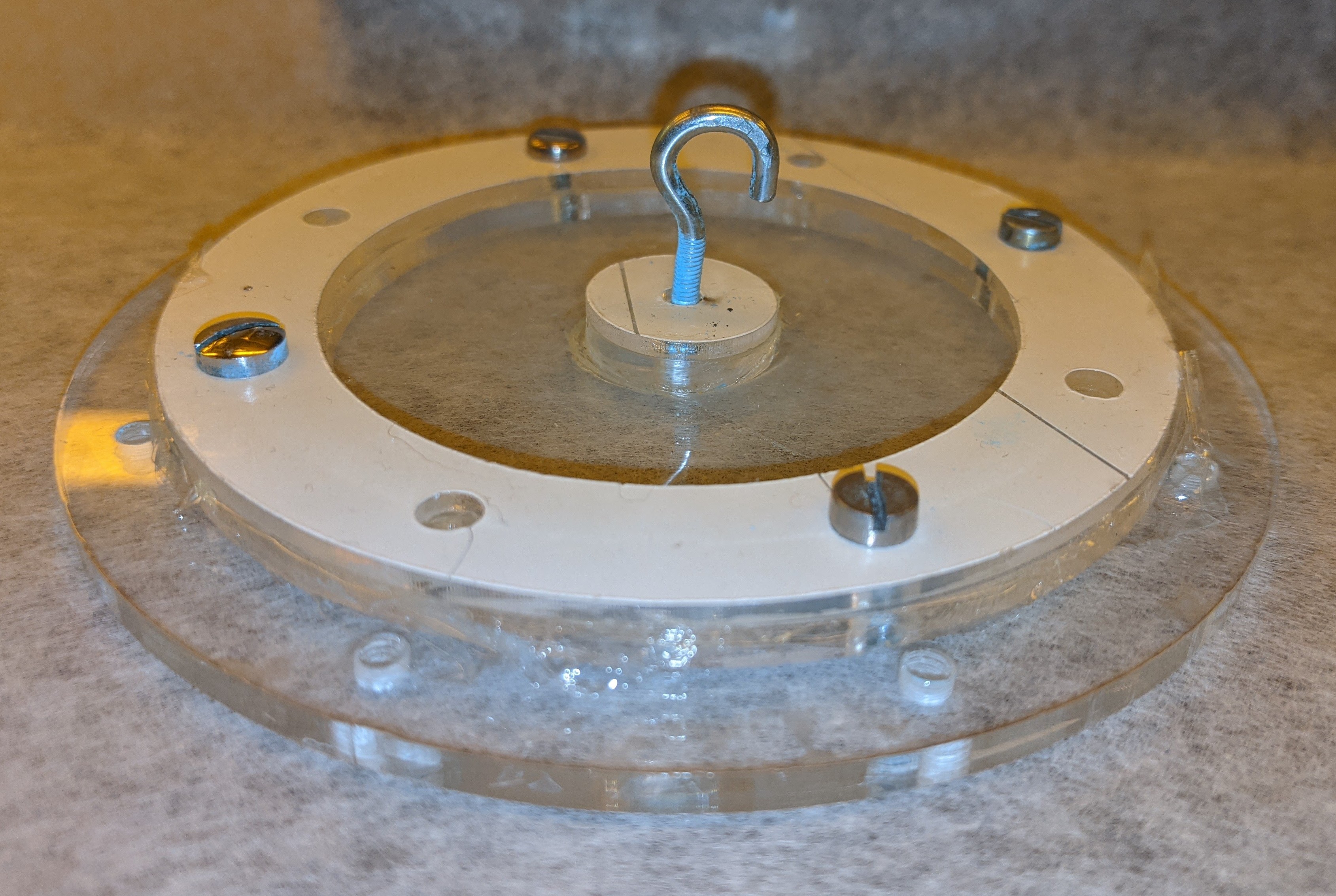}
\caption{Experimental setup: clamped film}
\label{fig:exp:setup}
\end{figure}

A relaxed silicone rubber film was clamped flat between an interior disc with radius $R_1=$\SI{10}{\milli\meter} and an exterior ring with radius $R_2=$\SI{35}{\milli\meter}, see Figure~\ref{fig:exp:setup}.
The film was cut from sheets of Elastosil\textregistered Film 2030 from Wacker~\cite{WA}, a silicone rubber film from cross-linked silicone rubber in three different thicknesses: \SI{50}{\micro\metre}, \SI{100}{\micro\metre} and \SI{200}{\micro\metre}.
The film was glued on the frame with standard silicone rubber adhesive.
Elastosil film is known to have little bending resistance, using the membrane model should be adequat. 
The limitations of the model will be explored when comparing with the simulations.
We prepared 6 specimen in total, two specimen for each thickness. 
To visually distinguish the two specimen with the same thickness, the second film was covered with blue color.
\subsection{Procedure}
\begin{figure}[H]
\centering
\includegraphics[width=.4\textwidth]{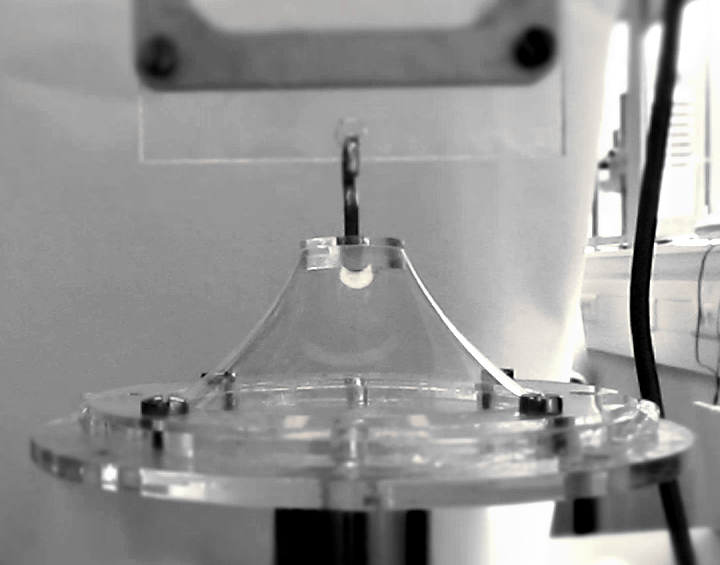}
\caption{Deflection of clamped film}
\label{fig:exp:deflection}
\end{figure}

The specimens were subjected to a vertical tensile force applied at the central hook, lifting the interior disc normal to the initially flat surface, as illustrated in Figure~\ref{fig:exp:deflection}.

The specimen was attached in a tensile tester, the exterior circle was in a fixed position, whereas the interior disc was movable.
The speed of the crosshead was \SI{1}{\milli\meter\per\second}. 
This was performed by a materials testing machine developed for the performance of standardized tests on materials and components, namely a Zwick/Roell Z005 tensile tester~\cite{ZR}.

The amount of tensile force and the elongation were recorded throughout the experiment.
The elongation was controlled by the testing software, the stepsize was \SI{0.25}{\mm}.
In addition to recording the forces and elongations, all performed tests were also video-recorded, such that the shape of the resulting surface of revolution for all elongations was chronicled.

\subsection{Results}

In Table~\ref{tab:exp:shapes} screenshots of the deformed specimen for different deflections can be seen, to catch a first impression of the deformed shapes. 
Below the last picture in each column, the maximal deflection at break, together with the maximal force at break, is given.

\begin{table}[h]
\caption{Shapes of the deformed membranes}
\label{tab:exp:shapes}
\centering
    \begin{tabular}{@{}c|c|c|c|c|c|c}
        specimen   & 1                & 2                & 3                & 4                & 5                & 6                \\
        thickness  & \SI{50}{\um}      & \SI{50}{\um}     & \SI{100}{\um}    & \SI{100}{\um}    & \SI{200}{\um}    & \SI{200}{\um}    \\
        deflection &                  &                  &                  &                  &                  &                  \\
        40mm       & \pic{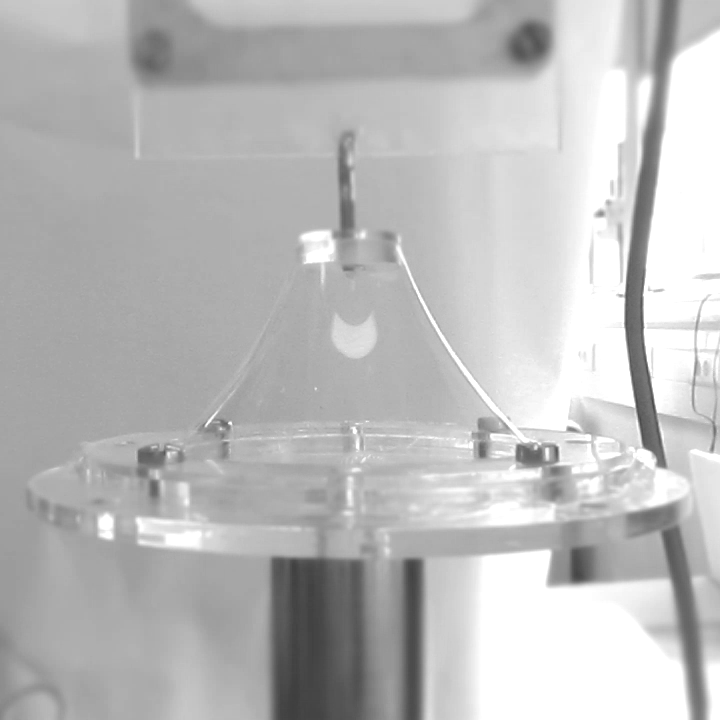}  
                   & \pic{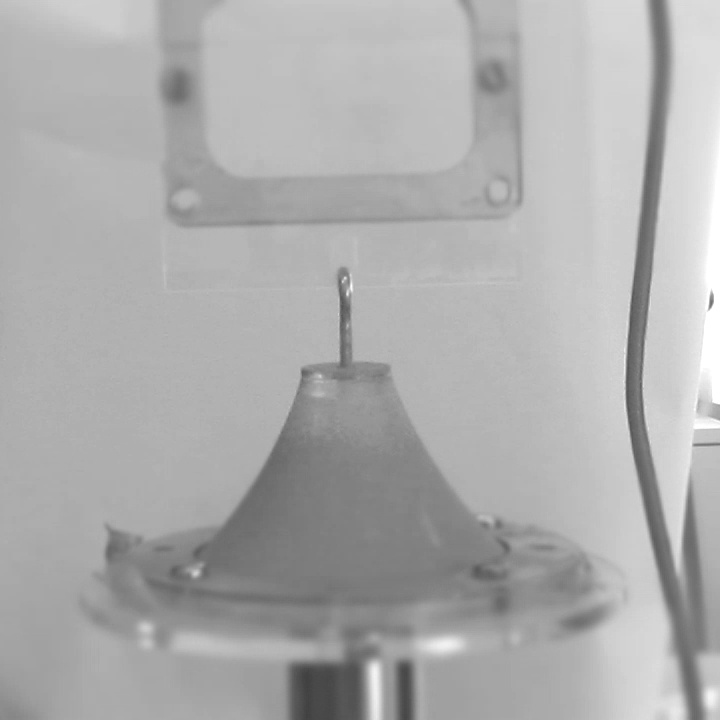}  
                   & \pic{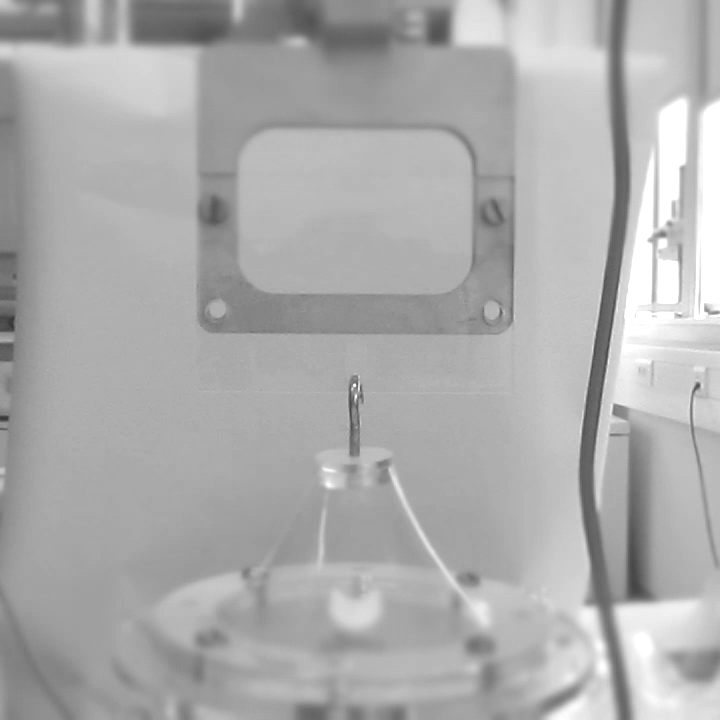}  
                   & \pic{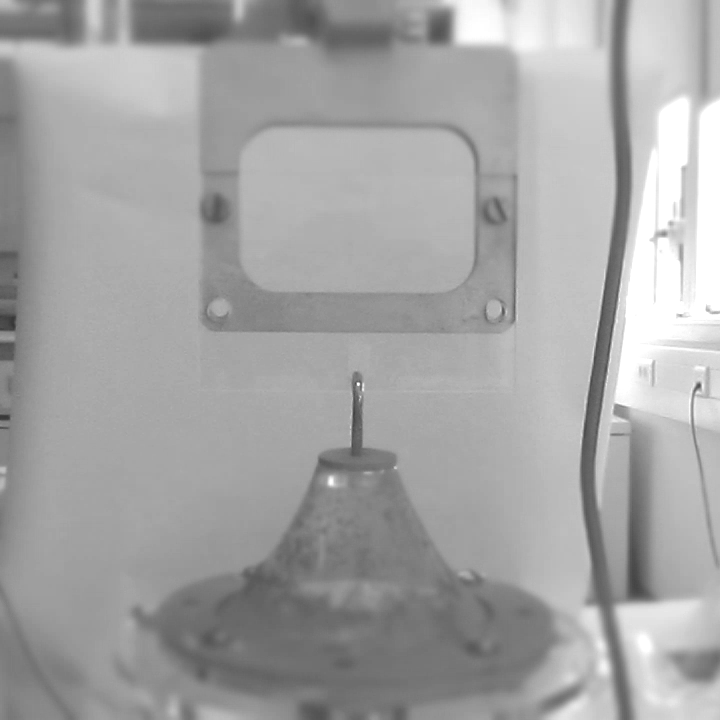}  
                   & \pic{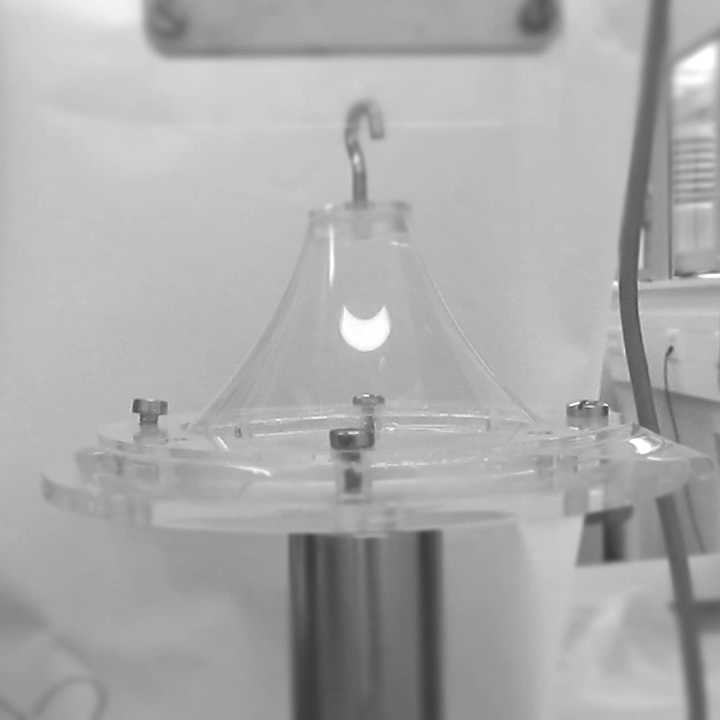}  
                   & \pic{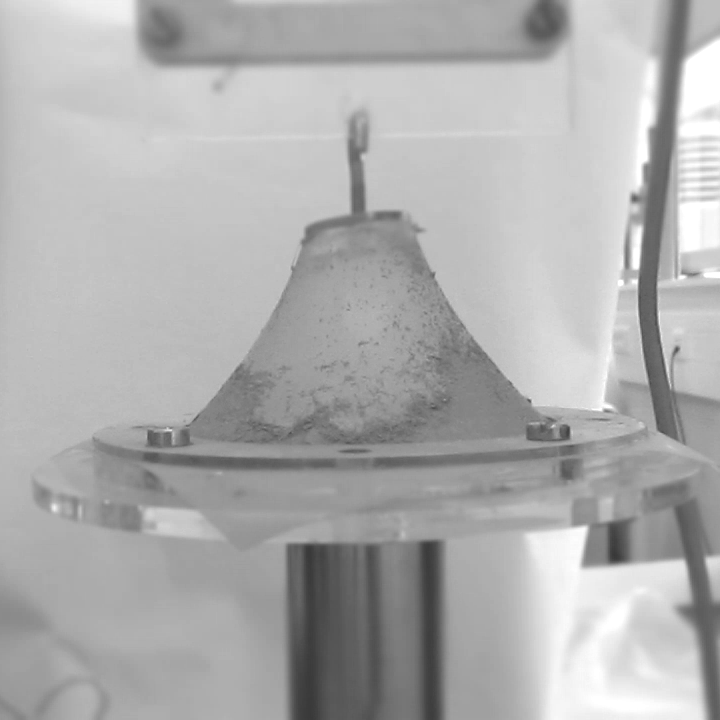}  \\
        60mm       & \pic{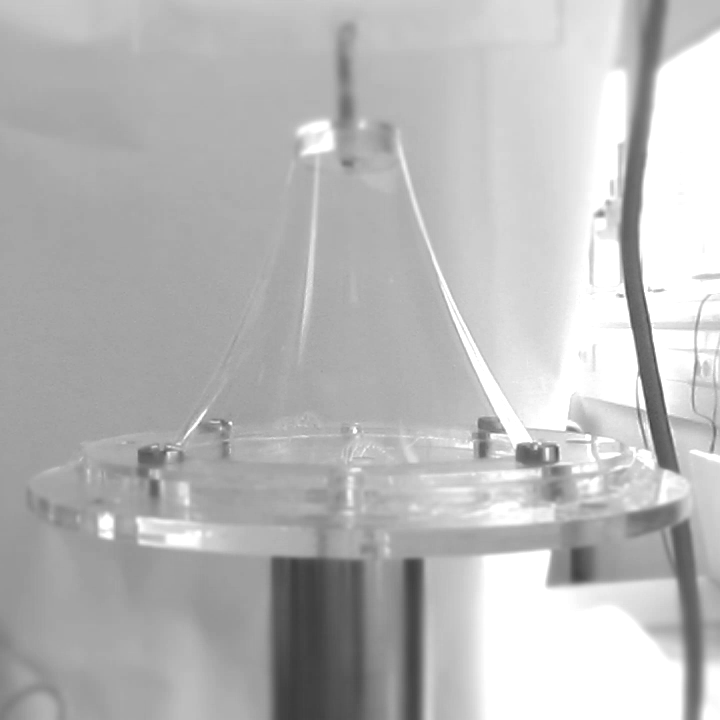}  
                   & \pic{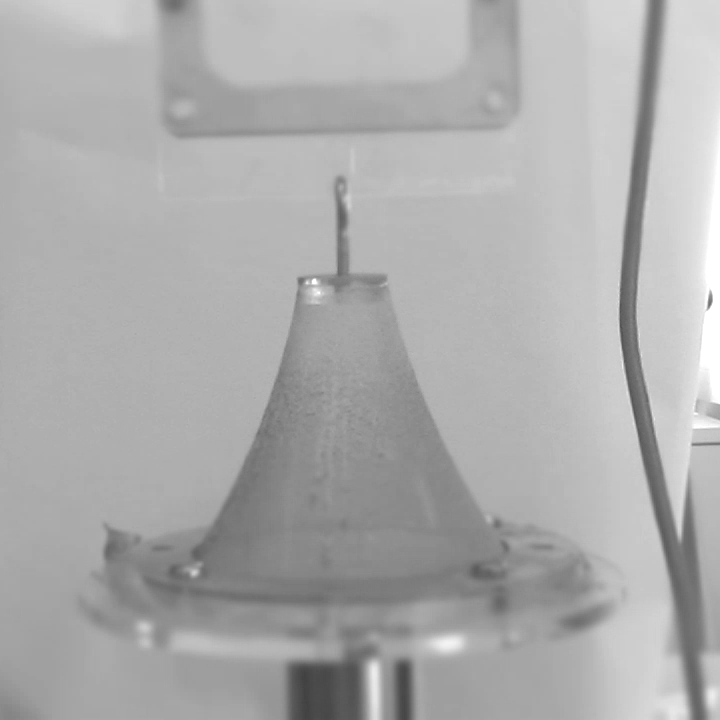}  
                   & \pic{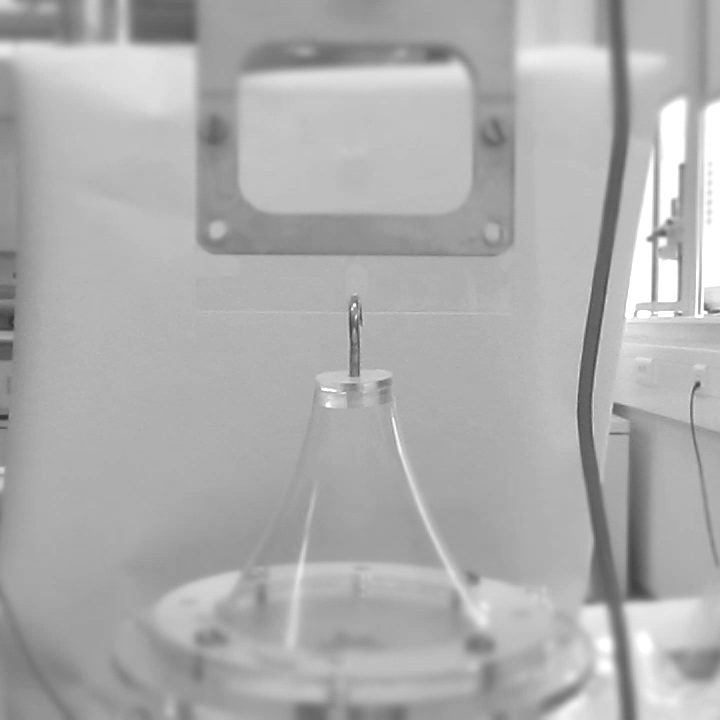}  
                   & \pic{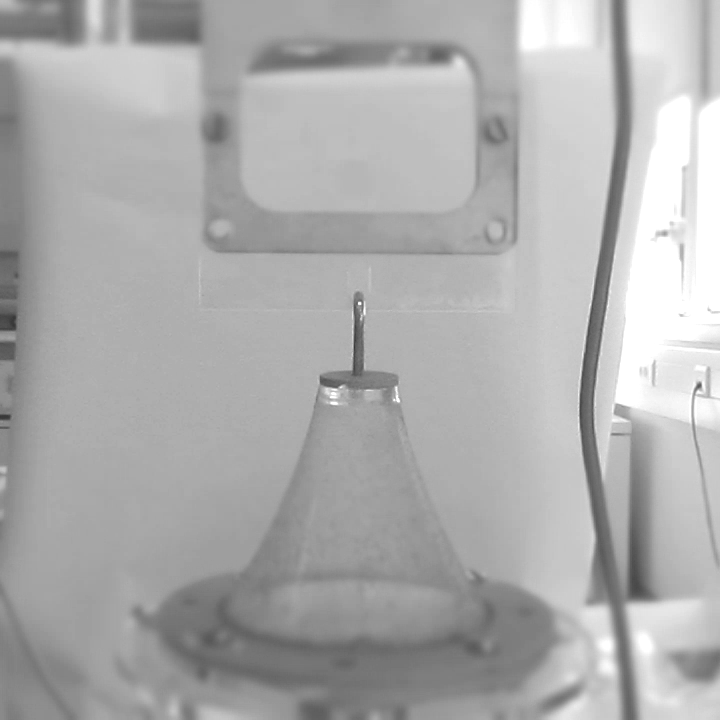}  
                   & \pic{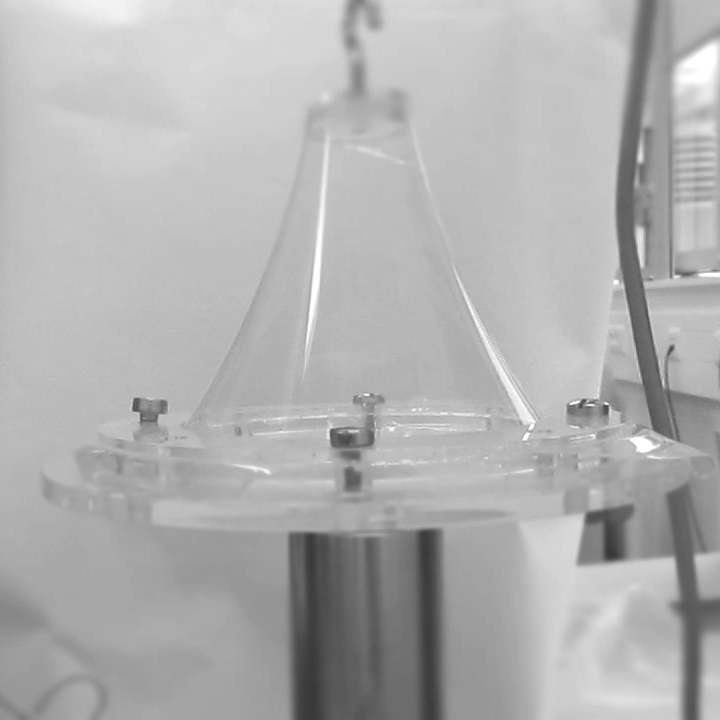}  
                   & \pic{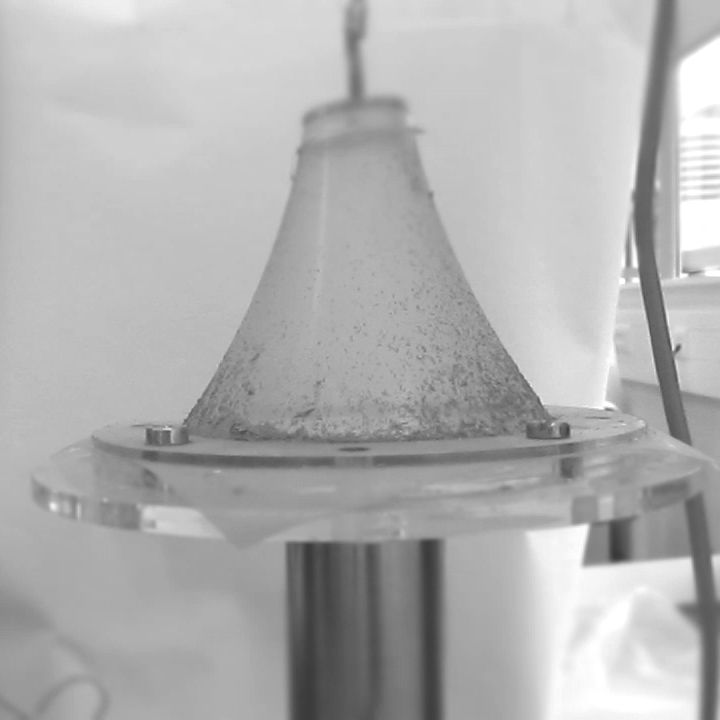}  \\
        80mm       & \pic{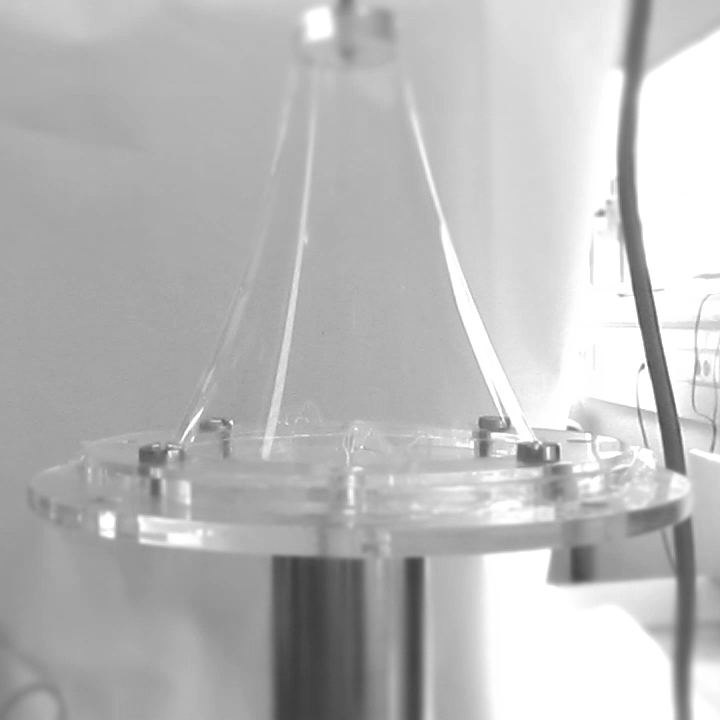}  
                   & \pic{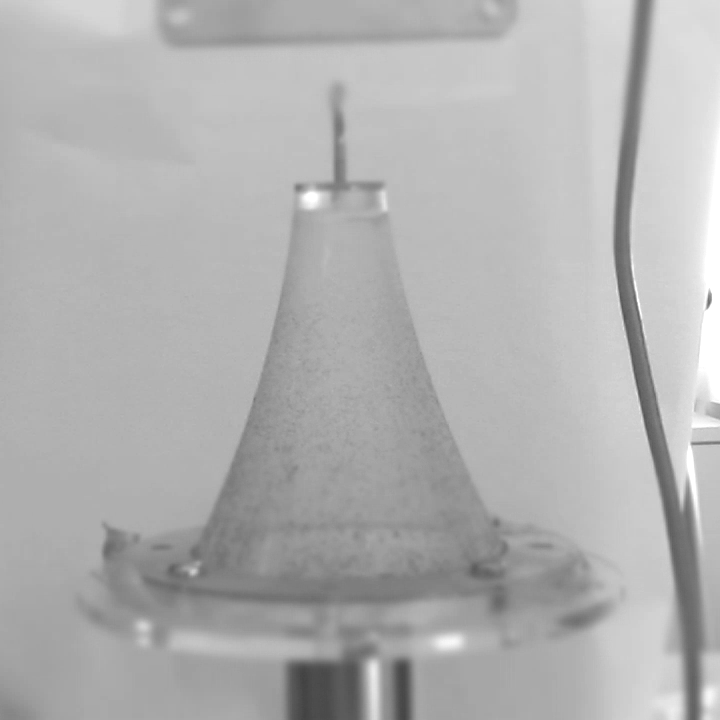}  
                   & \pic{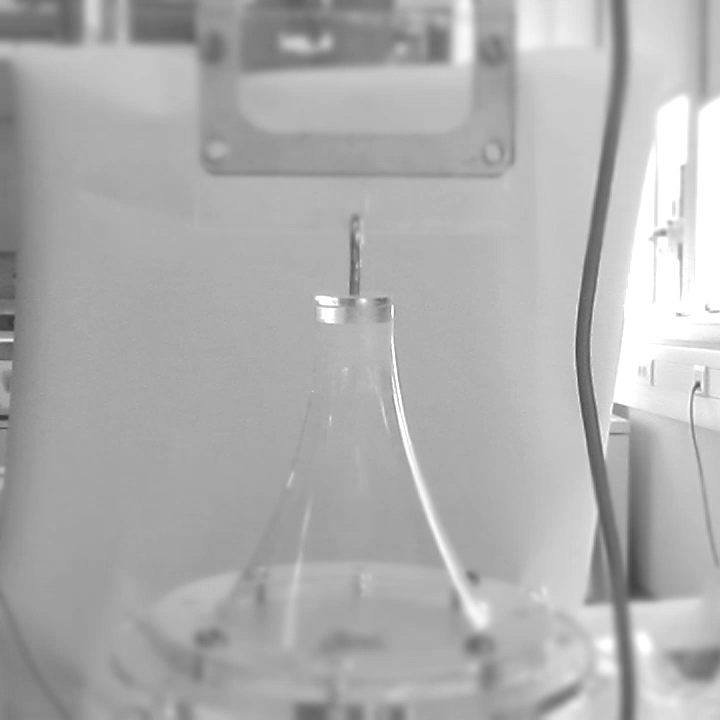}  
                   & \begin{minipage}{0.11\textwidth} 
                         \vspace{-1.5cm} \begin{center} broke at  \\ 74 mm \\ 19.8 N \end{center} 
                     \end{minipage}  
                   & \begin{minipage}{0.11\textwidth} 
                         \vspace{-1.5cm} \begin{center}  broke at \\ 69 mm \\ 26.5 N \end{center}  
                     \end{minipage}  
                   & \begin{minipage}{0.11\textwidth} 
                         \vspace{-1.5cm} \begin{center} broke at \\  68 mm \\ 29.4 N \end{center}  
                     \end{minipage}  \\
        100mm      & \pic{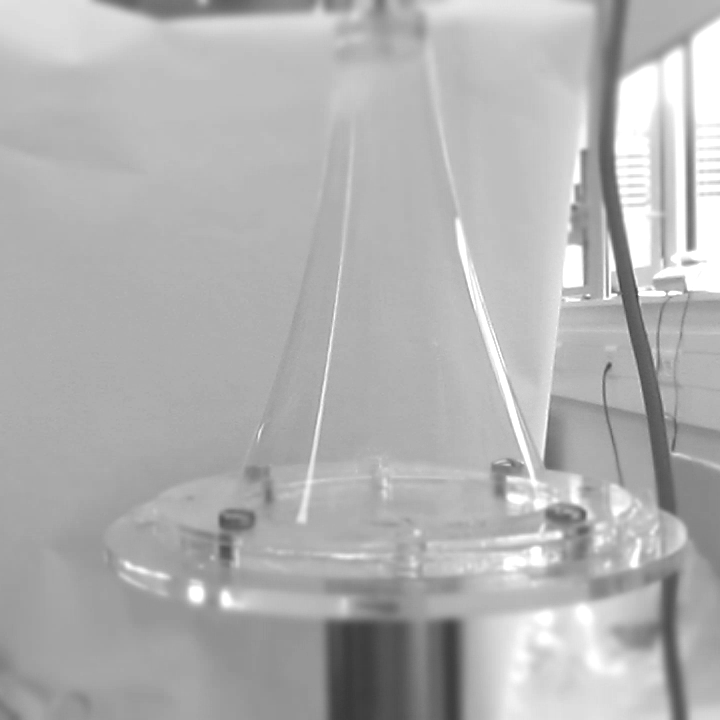}  
                   & \pic{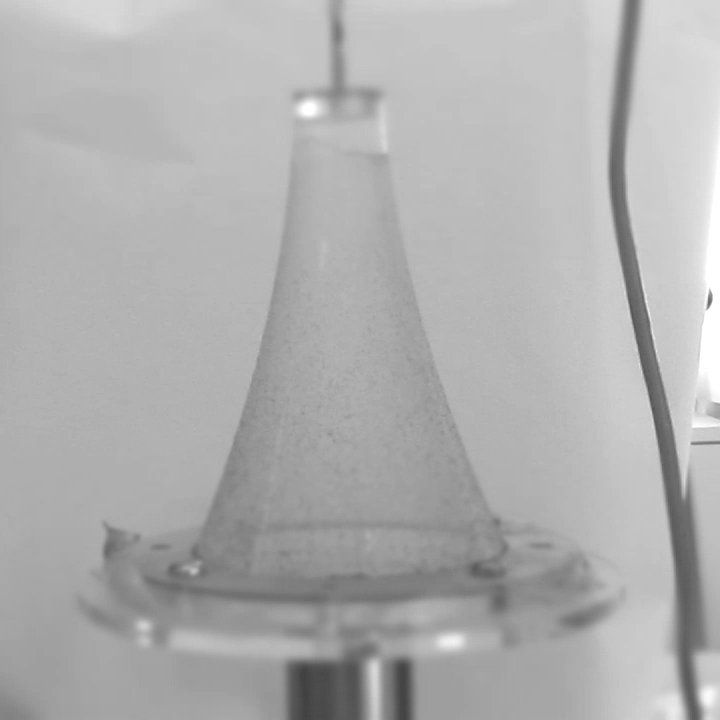}  
                   & \begin{minipage}{0.11\textwidth} 
                         \vspace{-1.5cm} \begin{center} broke at\\  83 mm \\ 18.8 N \end{center}  
                     \end{minipage}  
                   &                  &                  &                  \\
        120mm      & \begin{minipage}{0.11\textwidth} 
                         \vspace{-1.5cm} \begin{center} broke at\\  112 mm \\ 21.5 N \end{center}  
                     \end{minipage}  
                   & \pic{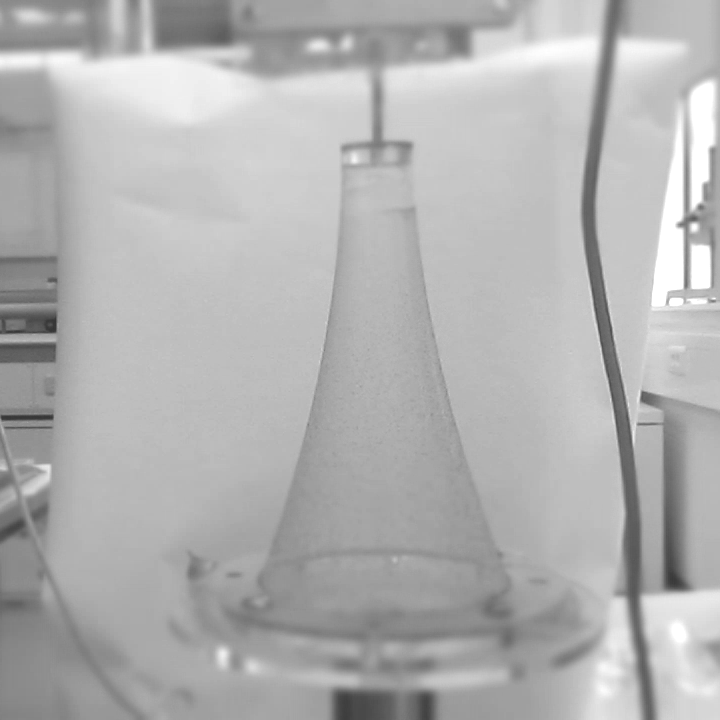}  
                   &                  &                  &                  &                  \\
        140mm      & \pic{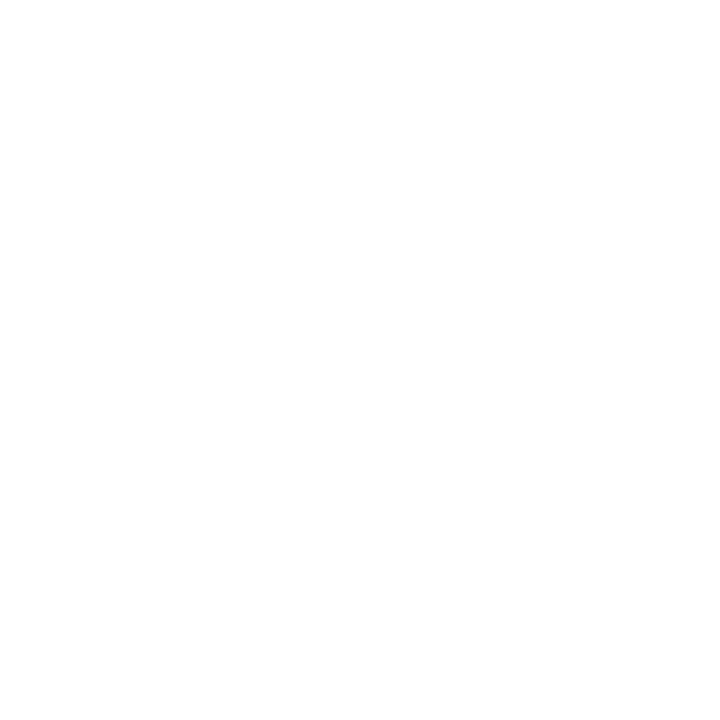}  
                   & \begin{minipage}{0.11\textwidth} 
                         \vspace{-1.5cm} \begin{center} broke at \\ 133 mm \\ 27.8 N \end{center}  
                     \end{minipage}  
                   &                  &                  &                  &                  \\
    \end{tabular}
\end{table}

\begin{figure}[H]
\centering
\includegraphics[width=1\textwidth]{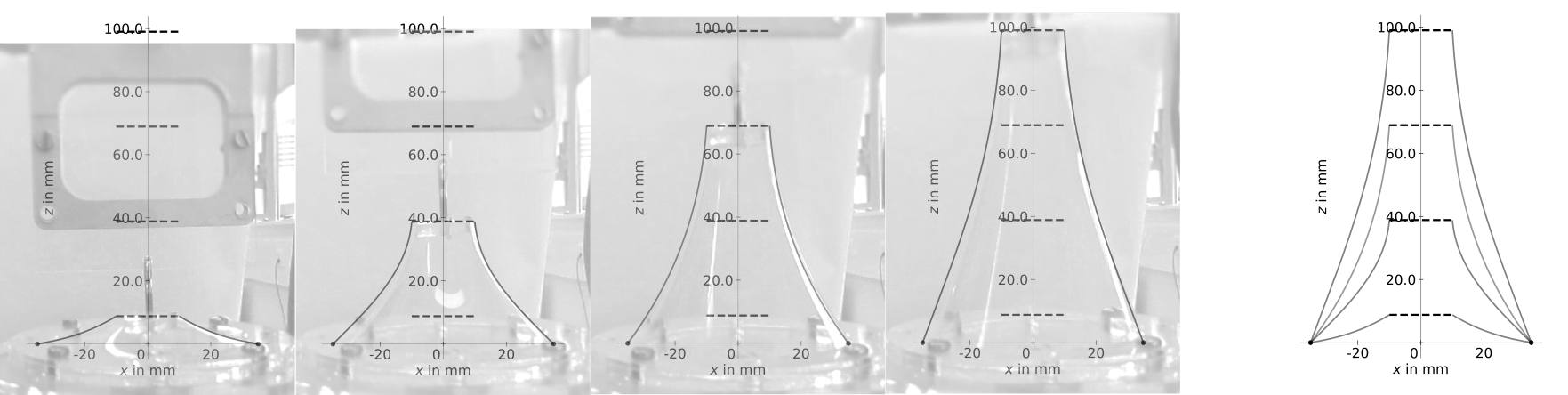}
\caption{Traced contours of deformed profiles for increasing deflections}
\label{fig:exp:profilesEXP}
\end{figure}

To visualize the deformation behavior, we have graphically reconstructed the shape of the deformed membrane for four different vertical displacements of the inner edge of the elastic annulus (Figure~\ref{fig:exp:profilesEXP}). 
This was done exclusively for specimen 1, a \SI{50}{\um} film, which is also the subject of our later simulations. 
The contours were traced using a Bézier curve-based approach in Inkscape to ensure smooth transitions in the graphical representation. 

\begin{figure}[H]
\centering
\includegraphics[width=.4\textwidth]{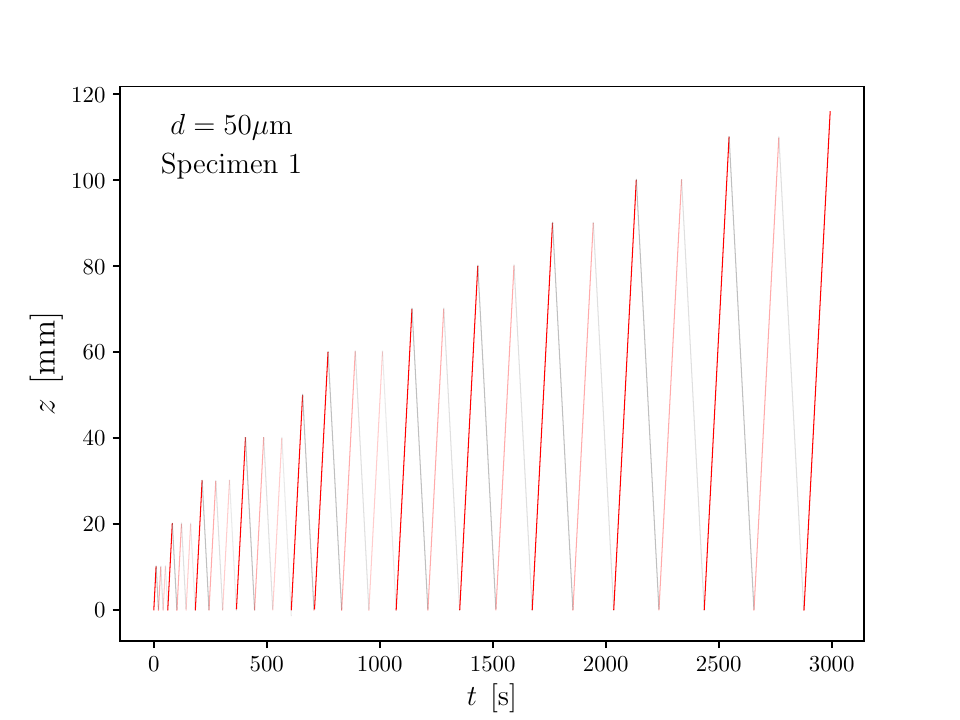}
\includegraphics[width=.4\textwidth]{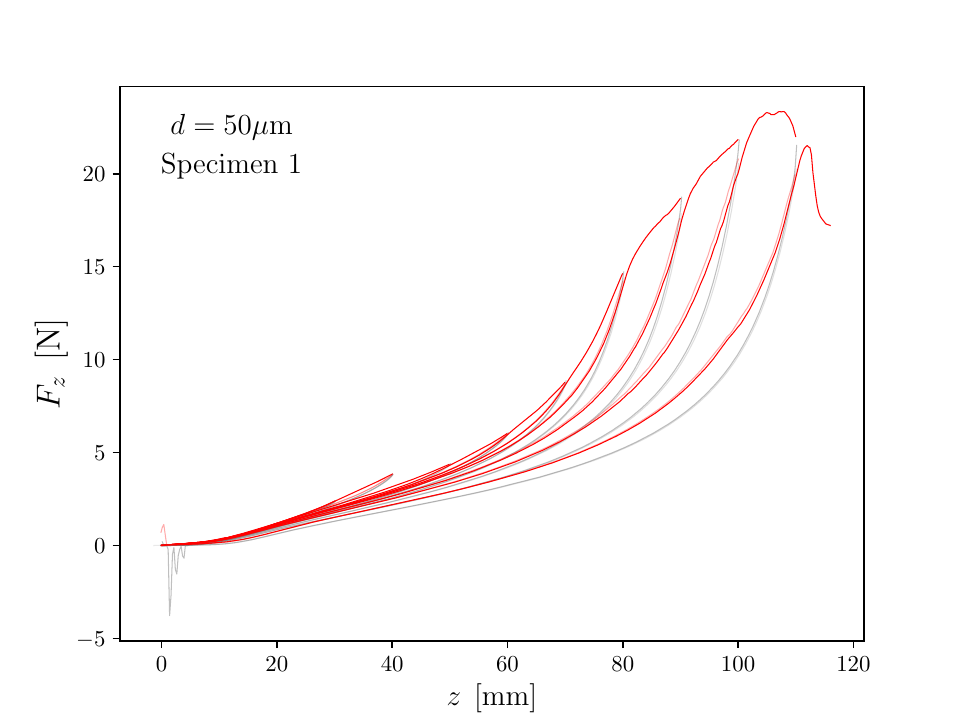}
\caption{Deflection-time diagram first specimen ($d=$\SI{50}{\um}) and corresponding load-deflection curves}
\label{fig:exp:elong-time-diag}
\end{figure}

\begin{figure}[h]
\centering
\includegraphics[width=.49\textwidth]{pdf/50umP1.pdf}\hfill
\includegraphics[width=.49\textwidth]{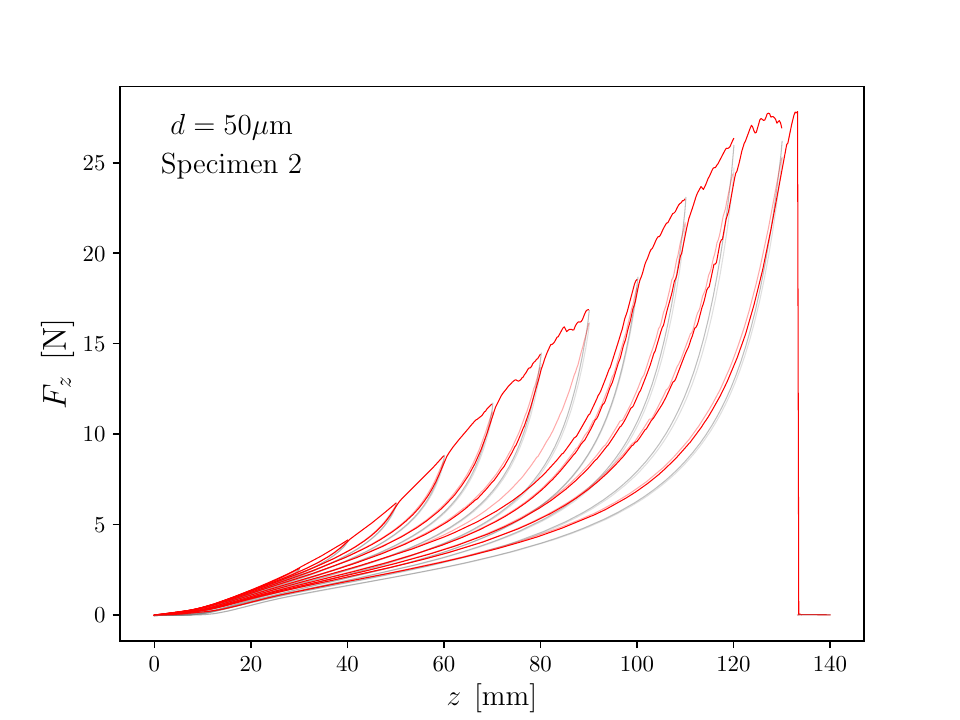}
\\ 
\includegraphics[width=.49\textwidth]{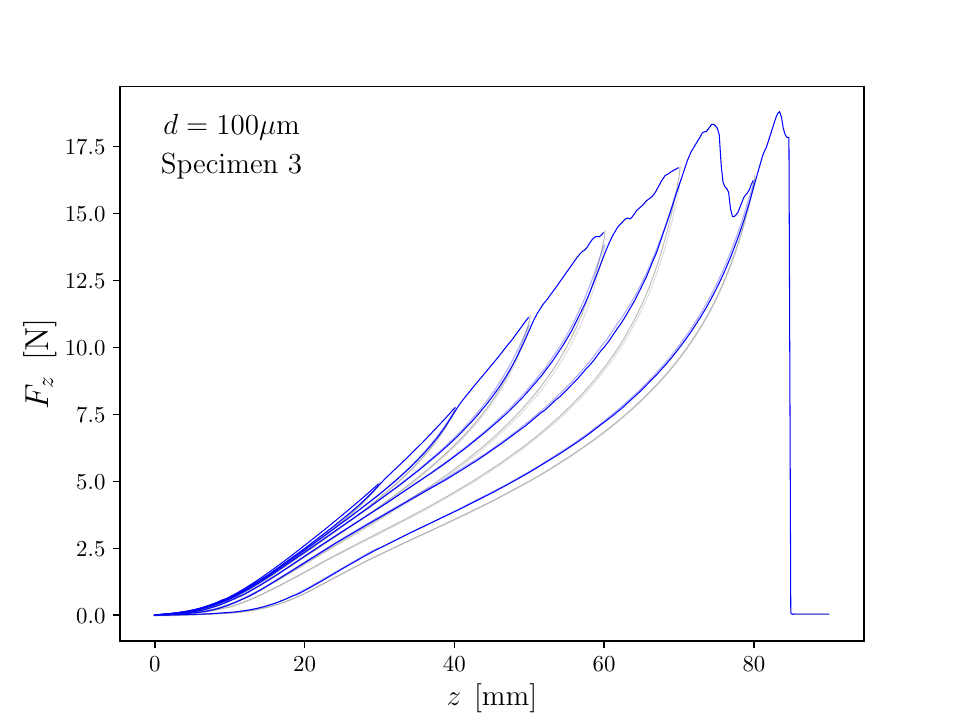}\hfill
\includegraphics[width=.49\textwidth]{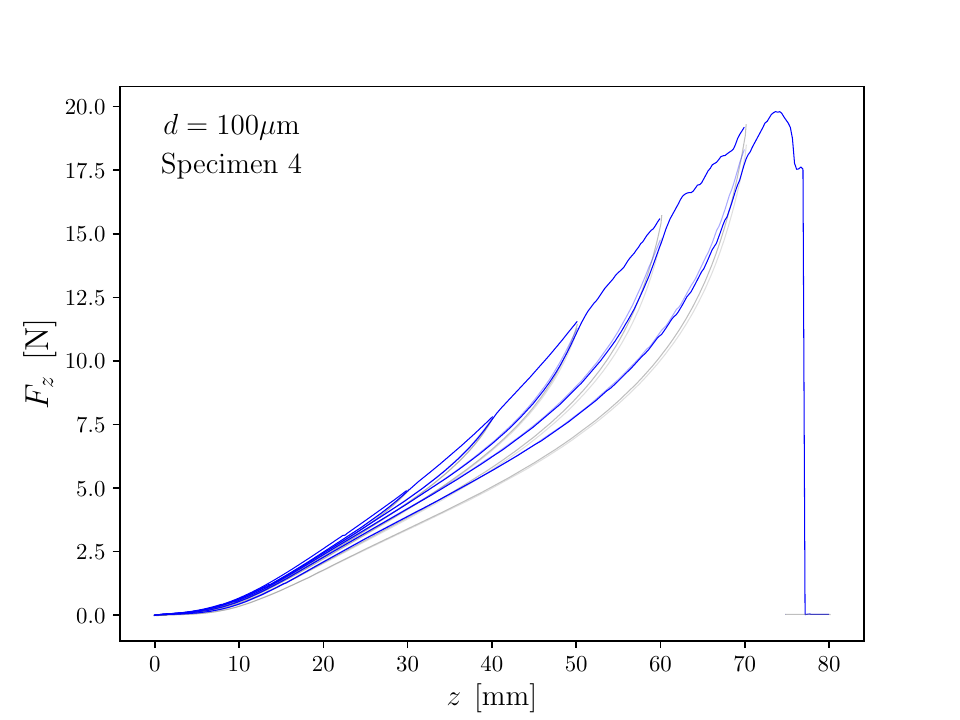}
\\
\includegraphics[width=.49\textwidth]{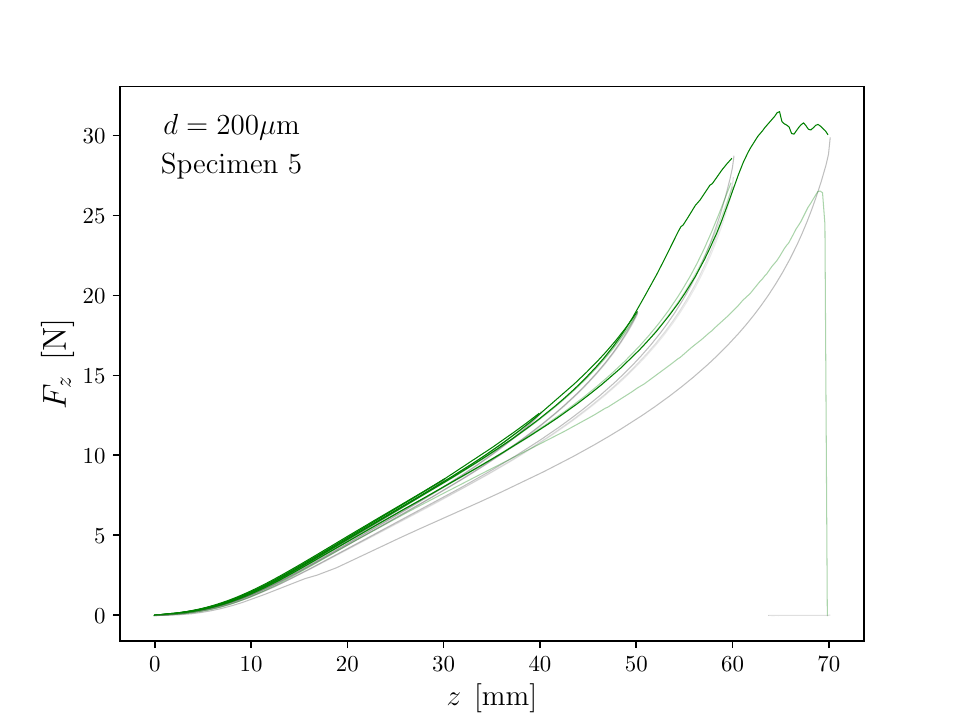}\hfill
\includegraphics[width=.49\textwidth]{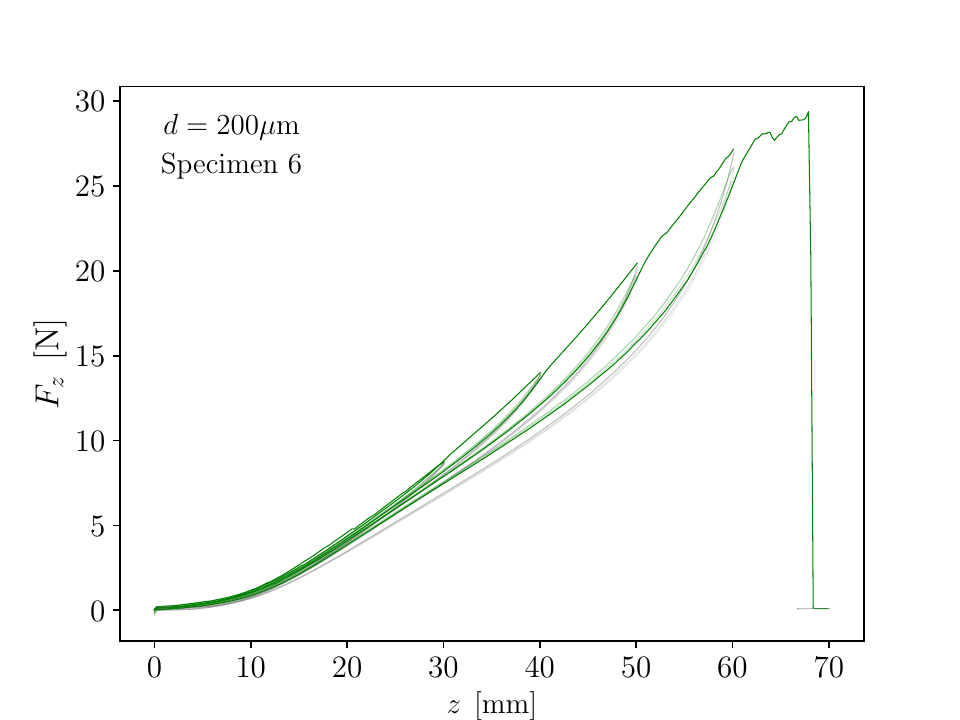}

\caption{Experimental load-deflection curves for all specimen}
\label{fig:exp:plots_proben_all}
\end{figure}

Starting with some initial maximal elongation we performed cycles of lifting and lowering up to the current maximal elongation. 
We then increased the maximal elongation by \SI{10}{\mm} and again perform multiple cycles of lifting and lowering. 
We continued like this, increasing the maximal elongation until break. 
This was done for each of the 6 specimen.
In Figure~\ref{fig:exp:elong-time-diag}, the first diagram shows the cyles of lifting and lowering for the first specimen ($d=\SI{50}{\um}$) in a deflection-time diagram and the corresponding load-deflection curves are in the second diagram. 
Red curves depict lifting and gray curves depict the lowering. 
Different shades of red and grey were used depending on number of cycle.  
In Figure~\ref{fig:exp:plots_proben_all} we plot all load-deflection curves. 
For each specimen there is one diagram, showing all cycles of going upwards and downwards. 

From Figure~\ref{fig:exp:plots_proben_all} we observe hysteresis, tearing effects and material property changes over multiple cycles. 
More precisely, the unloading curves do not follow the same path as the loading curves, leading to a difference in forces for the same displacement, indicating energy dissipation during each loading-unloading cycle. 
The curves show abrupt changes in slope or small force drops when a new maximum displacement is reached. 
This suggests micro-tearing or structural rearrangement in the elastic film, potentially altering its mechanical response in subsequent cycles. 
Repeated stretching cycles lead to progressive changes in the force-displacement behavior, indicating material fatigue. 
The reduction in stiffness over successive cycles suggests softening effects, potentially due to molecular rearrangement or microstructural damage.

Measurements with the \SI{50}{\um} film moreover indicated that the force required to reach a displacement of \SI{100}{\mm} also depends on the crosshead speed. 
At a speed of \SI{0.1}{\milli\meter\per\second}, the maximum force recorded was \SI{16.8}{\newton}, while at \SI{1}{\milli\meter\per\second}, it increased slightly to \SI{17.0}{\newton}. 
At \SI{10}{\milli\meter\per\second}, the maximum force reached \SI{19.5}{\newton}. 
Although a trend of increasing force with speed is observed, the overall change remains relatively small.

\FloatBarrier

\section{Simulations}\label{sec:simulations}

The simulations in this study are not designed to capture all experimentally observed effects, such as hysteresis, tearing, rate dependence, or fatigue-related changes in material properties. 
These phenomena arise from complex viscoelastic and plastic deformation mechanisms, which cannot be adequately represented within the framework of our model.

\begin{figure}[H]
\centering
\includegraphics[width=.6\textwidth]{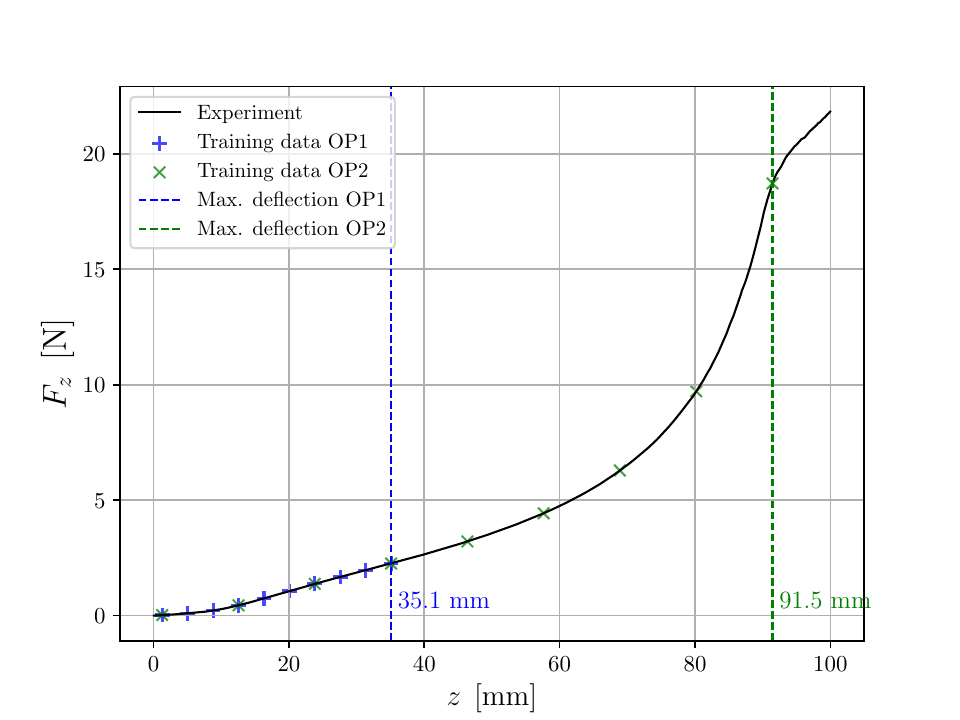}
\caption{Upward-loading curve with markers of training data in both optimizations}
\label{fig:exp:force-elong-diag}
\end{figure}

Instead, we focus on optimally reproducing a single force-displacement curve under well-defined conditions. 
Specifically, we have chosen the film with thickness \SI{50}{\micro\metre}, with a maximum displacement of \SI{100}{\mm}, considering only the first loading cycle and modeling exclusively the upward-loading curve, see Figure~\ref{fig:exp:force-elong-diag}. 

To further ensure that our model remains within its intended scope, we restrict the analysis to the displacement range before the slope change occurs, which is observed around \SI{90}{\mm}. 
Beyond this point, the force-displacement response may be influenced by micro-tearing or structural rearrangements, which our model does not account for. 
By focusing on the initial part of the curve, we aim to capture the fundamental elastic response without introducing complexities related to material damage. 
This approach allows for a clear evaluation of how well the fundamental mechanical behavior can be described within the given constraints while maintaining computational feasibility.

To systematically investigate the material behavior within different strain regimes, we conduct two separate optimized simulations with maximal displacements \SI{35.1}{\mm} (OP1) and  \SI{91.5}{\mm} (OP2). 
The first case corresponds to strains in the simulations below 300\%, which can still be classified as large strain, but remains within a range where nonlinearity is moderate. 
The material response is expected to be dominated by elastic effects without significant structural rearrangements. 
In the second case, strains of the simulations reach up to 700-800\%, entering an extreme strain regime where strong material nonlinearities, molecular reorientation, and potential micro-tearing effects become relevant.  
Although our model does not explicitly account for these effects, it allows us to assess how well a simplified representation can approximate the force-displacement behavior in this range. 

By performing simulations at both strain levels, we can evaluate the validity of the simulations across different deformation regimes while maintaining computational feasibility.

\subsection{Procedure}

For each of the five hyperelastic membrane energy densities~\eqref{eq:defWN} to~\eqref{eq:defWO}, we aim to determine optimal material parameters,  first in the large strain regime (OP1) and then in the extreme strain regime (OP2), such that the simulated loading curve closely matches the experimentally observed loading curve. 
To achieve this, we compare the experimental training data with simulated data points by computing the Euclidean distance between the experimental and simulated forces for the same finite set of vertical deflections. 
By iteratively updating the material model parameters, we seek to minimize this discrepancy and improve the predictive accuracy of the simulation.

Given the complexity of this optimization problem, where gradients must be computed efficiently and parameter updates systematically applied, we employ TensorFlow~\cite{TensorFlow}, an open-source framework developed by Google that is well-suited for numerical computation and optimization tasks. 
TensorFlow operates efficiently with tensors (multi-dimensional arrays) and provides automatic differentiation, enabling efficient gradient-based optimization.

At the core of the optimization process lies the system~\eqref{eq:ode_for_numerics} for the corresponding material model, the system of first order ordinary differential equations (ODEs) that governs the mechanical response of the hyperelastic membrane. 
To determine the unknown deformation corresponding to a given displacement, we must solve a boundary value problem (BVP). 
However, TensorFlow’s built-in ODE solvers, provided via TensorFlow Probability, are designed for initial value problems (IVPs) and do not natively handle boundary conditions.

To overcome this limitation, we employ a shooting method, which reformulates the BVP as an iterative search for suitable initial conditions at the inner edge of the annulus that approximate a valid solution to the boundary conditions at the outer edge. 
This is achieved using a quasi-Newton method, which iteratively refines the initial conditions by minimizing the discrepancy between the numerically integrated solution (starting from the inner boundary) and the desired values at the outer boundary. 
Specifically, we search for a root of a function that measures the deviation between the computed and expected boundary values, such that the numerical solution converges toward a valid approximation of the BVP. 

However, certain difficulties arise in the implementation of this quasi-Newton shooting method. 
First, a good initial guess for the initial conditions is essential for convergence, as the method relies on local linearization and may fail if the starting point is too far from a valid solution. 
Additionally, the numerical inversion of the Jacobian matrix can become problematic if the system is poorly conditioned or close to singular, leading to large or unstable updates. 
Furthermore, if the shooting method enters regions where the boundary conditions become highly sensitive to small perturbations in the initial conditions, the iterative updates may oscillate or fail to converge within a reasonable number of iterations. 
To address these issues, step size adjustments, regularization of the Jacobian inversion, and fallback strategies for ill-conditioned cases are necessary to improve robustness.

Once a solution to the BVP is found, we can compute the total vertical force acting on the system by~\eqref{eq:Fz} from our considerations in Example ~\ref{exa:Ftot}. 
Note that for this, we do not need the full solution, but only the suitable initial conditions to compute the radial stress at the inner edge and from this the total vertical force. 
This stress response depends again on the chosen material model.

With the force-displacement response computed, we proceed with optimizing the material parameters. 
A loss function quantifies the discrepancy between the simulated and experimental force data. 
The optimization algorithm then iteratively updates the parameters, leveraging TensorFlow’s automatic differentiation to compute gradients efficiently.

At each iteration, the simulation is evaluated, the loss is computed, and parameters are adjusted accordingly. 
A convergence check determines whether the optimization has reached a stable state. 
If the loss continues to decrease, further iterations are performed; otherwise, the process terminates, yielding the final optimized material parameters.

This structured approach ensures that the estimated parameters align with experimental data while maintaining computational efficiency, despite the additional complexity introduced by the shooting method for solving the BVP.

The outcome of the optimization process depends on several factors, with the choice of initial parameters (INI) playing a particularly crucial role. 
Since the optimization landscape can contain multiple local minima, different initial conditions may lead to different final parameter sets. 
The selection of initial values for each material model is discussed in Appendix~\ref{secA1}.

Moreover, the loss function may exhibit not only isolated local minima but also entire minimal curves (valleys), where the gradient flow can meander unpredictably. 
This behavior is particularly evident in the two-parameter models, for which we generated loss function plots that reveal such valleys. 
Consequently, the obtained optimal parameters should be interpreted as good parameters rather than the optimal parameters in an absolute sense.

In addition to the initial parameters, other key factors influencing the results include the optimization algorithm, the learning rate and the convergence criteria.

Beyond the issue of multiple possible optimization outcomes, an even more fundamental challenge arises during the optimization process: as the parameters are iteratively updated, the algorithm may enter regions of the parameter space where the governing equations become ill-posed or where physically meaningless solutions emerge. 
This can occur if the energy function no longer satisfies the necessary convexity conditions, leading to non-physical material behavior such as loss of stability or singularities in the stress-strain response.

Additionally, certain parameter configurations may cause numerical instabilities in the equation solver, making it difficult or even impossible to compute the simulated force-displacement curves. 
These issues lead to abrupt failures in the optimization or resulted in the algorithm converging to a mathematically valid but physically unrealistic solution.

Finding optimal parameters that not only minimize the loss function but also ensure a physically consistent and numerically stable solution proved to be a challenge.

\clearpage 

\subsection{Results}
Table~\ref{tab:optimization_results} summarizes the initial and optimized material parameters for all considered models, along with the corresponding loss values.

\begin{table}[h]
\centering
\sisetup{
    table-format=3.5,
    table-number-alignment = center,
    round-mode=places,
    round-precision=3,
    output-decimal-marker = {.},
    group-separator = {},
    group-minimum-digits = 100
}
\begin{threeparttable}
\caption{Model parameters and loss values}
\label{tab:optimization_results}
\renewcommand{\arraystretch}{1.2}
\begin{tabular}{l l S S S}
    \toprule
    Model &  & {INI} & {OP1} & {OP2} \\
    \midrule
    NH    & $C_1$ [\si{\newton\per\metre}]  & 7.5   & 7.6191   & 13.245621  \\
          \LossRow{1}{0.24048306}{0.23184398}{2.9224505} 
          \LossRow{2}{12.545496}{12.405822}{8.4766445}  
    \midrule
    M     & $C_1$ [\si{\newton\per\metre}]  & 7.0   & 6.21436   & 13.459625  \\  
          & $C_2$ [\si{\newton\per\metre}] & 0.5  & 1.4703426   & -0.1179734  \\
          \LossRow{1}{0.3044866}{0.22636904}{2.972474}  
          \LossRow{2}{12.049761}{12.967404}{8.474841}  
    \midrule
    G     & $C_1$ [\si{\newton\per\metre}]  & 7.5   & 7.6059127   & 5.312   \\ 
          & $C_2$ [--]  & 50.00 & {523.010*}   & 17.42  \\
          \LossRow{1}{0.27449587}{0.23496124148368835}{1.0299202}  
          \LossRow{2}{9.206874}{12.129879}{0.8092156}  
    \midrule
    Y     & $C_1$ [\si{\newton\per\metre}]  & 7.5   & 8.821764   & 8.840838   \\
          & $C_2$ [\si{\newton\per\metre}] & -0.075 & -0.65316385  &  -0.6303258  \\
          & $C_3$ [\si{\newton\per\metre}]  & 0.00075 & 0.06163691   & 0.05632747   \\
          \LossRow{1}{0.25876385}{0.10043436}{0.10185724}  
          \LossRow{2}{14.43088}{1.9970075}{0.53057563}  
    \midrule
    O     & $C_1$ [\si{\newton\per\metre}]  & 30.9  & 27.999382  & 11.060008   \\  
          & $C_2$ [\si{\newton\per\metre}]  & 0.06  & 0.03160439   & 0.1959017   \\
          & $C_3$ [\si{\newton\per\metre}]  & -0.5  & -0.48299527  & -0.3679535  \\
          & $a_1$ [--]  & 1.30  & 1.1806412   & 1.5196142   \\
          & $a_2$ [--]  & 5.00  & 4.9906154   & 5.805068   \\
          & $a_3$ [--]  & -2.00 & -1.9481684  & -1.1529809  \\
          \LossRow{1}{0.96127945}{0.12466633}{1.4499232}  
          \LossRow{2}{11.611522}{14.417482}{1.9558717}  
    \bottomrule
\end{tabular}
\begin{tablenotes}
\item[*]{This value is not an optimum. Increasing $C_2$, while optimizing $C_1$ still very slowly reduces the loss, while bringing the solution closer to the Neo-Hookean response.}
\end{tablenotes}
\end{threeparttable}
\end{table}

\begin{figure}[htb]
\centering
\begin{tabular}{>{\centering\arraybackslash}m{1.0cm} 
                >{\centering\arraybackslash}m{5.5cm} 
                >{\centering\arraybackslash}m{5.5cm} }
     model & large strains & extreme strains \\ 
{NH}  & \includegraphics[width=0.4\textwidth]{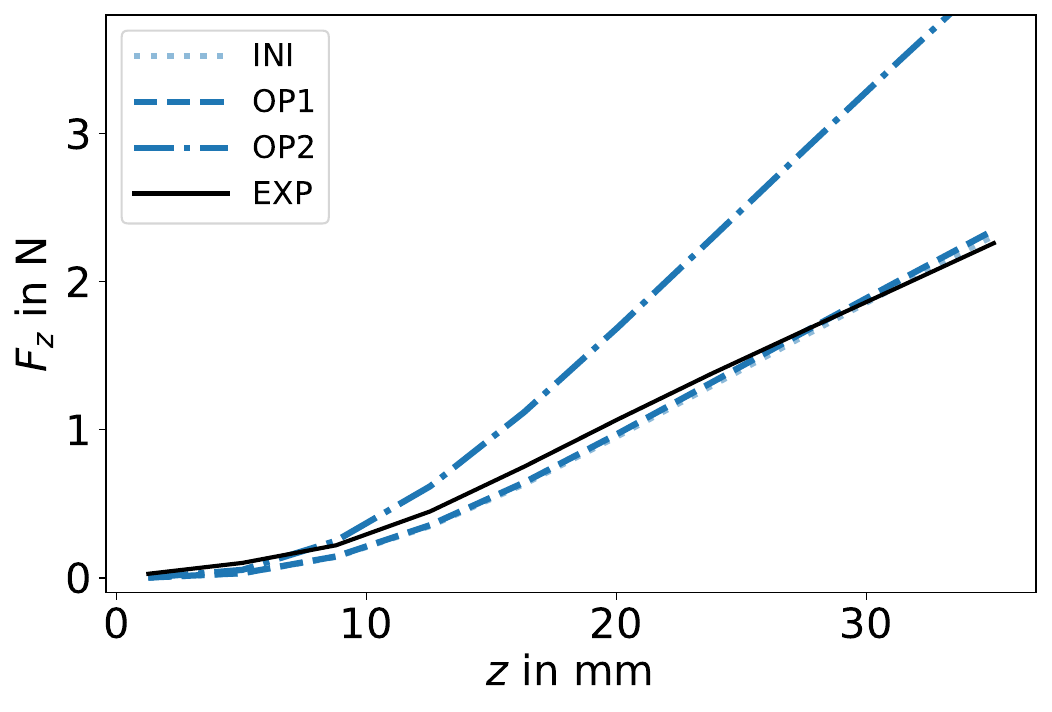}
              & \includegraphics[width=0.4\textwidth]{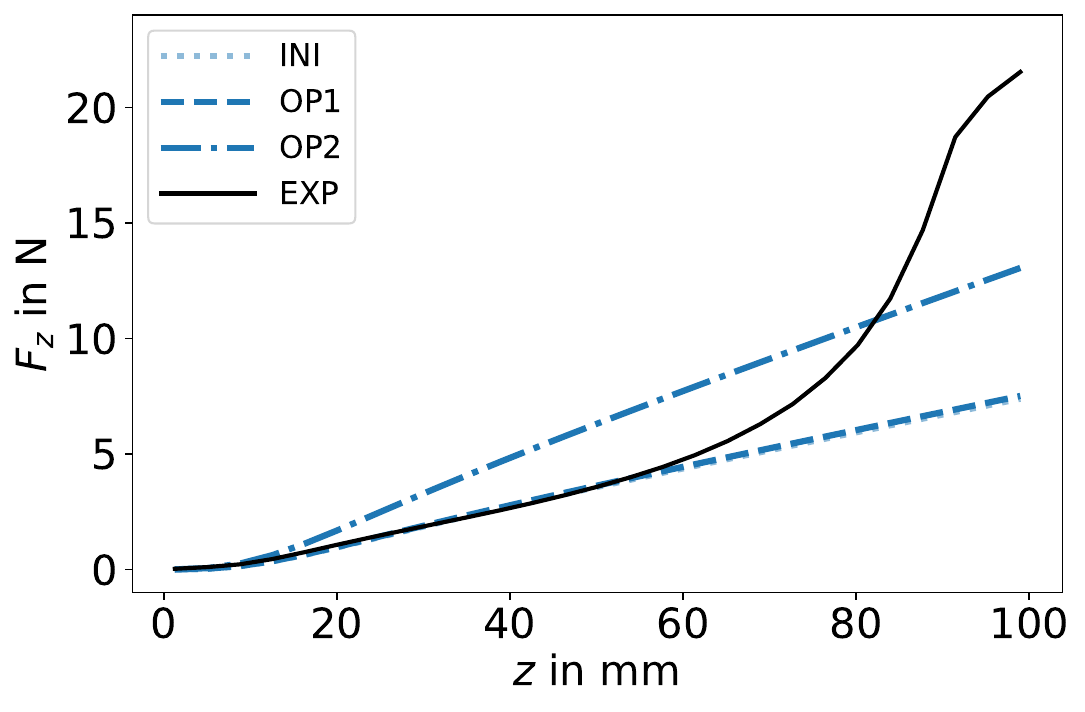} \\
{M}      & \includegraphics[width=0.4\textwidth]{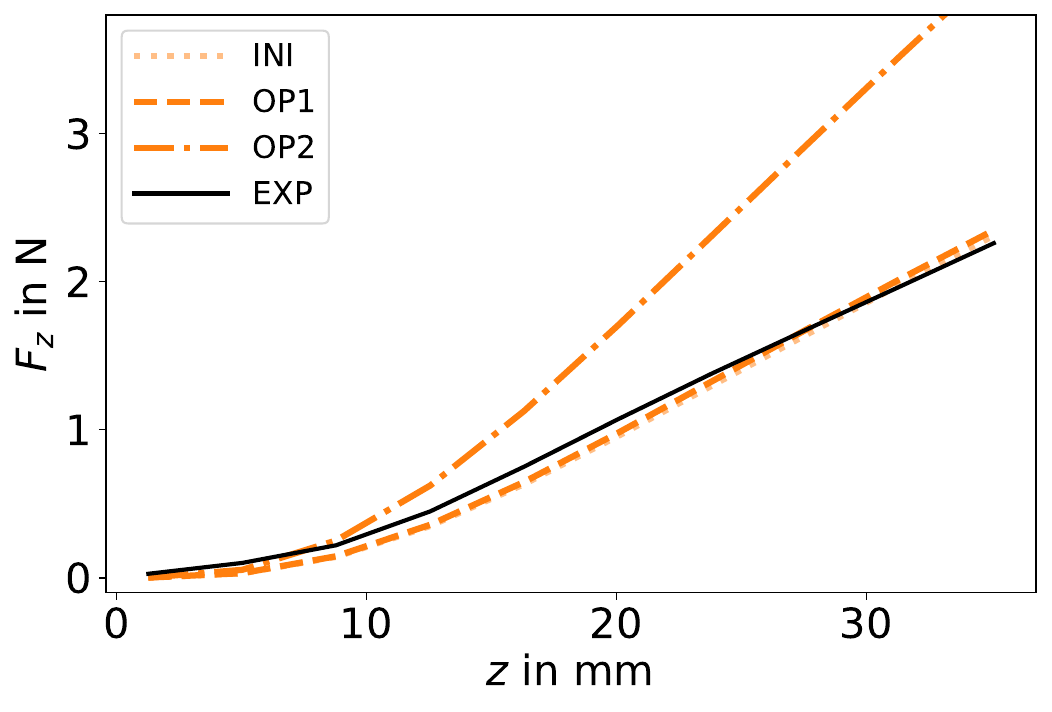}
              & \includegraphics[width=0.4\textwidth]{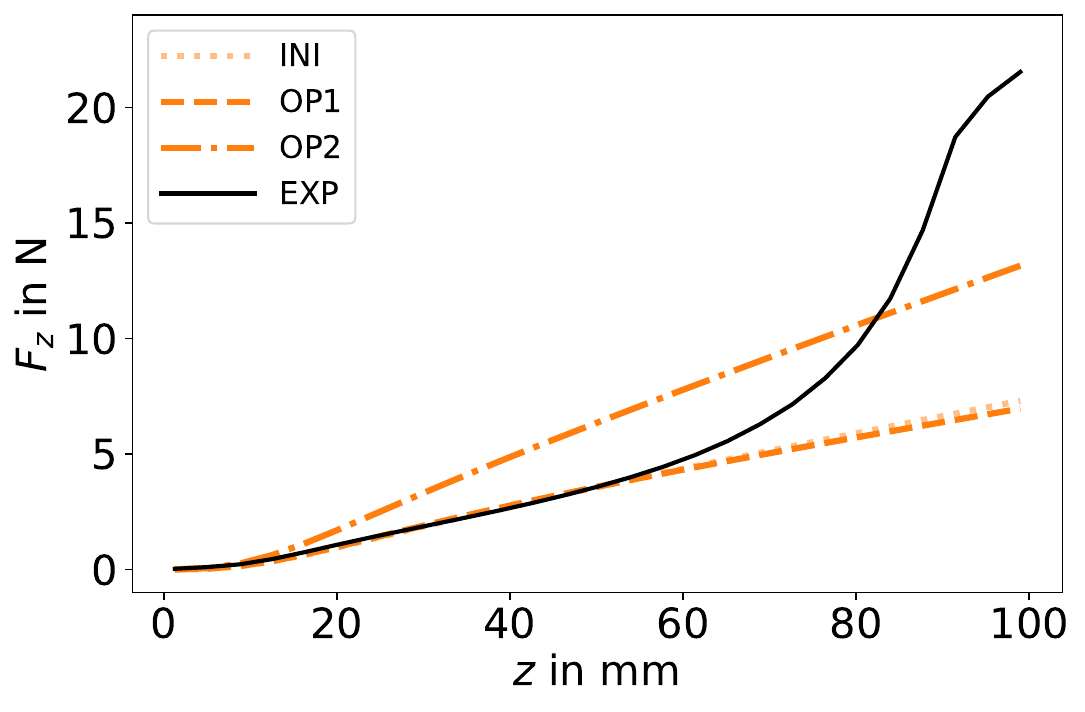} \\
{G}     & \includegraphics[width=0.4\textwidth]{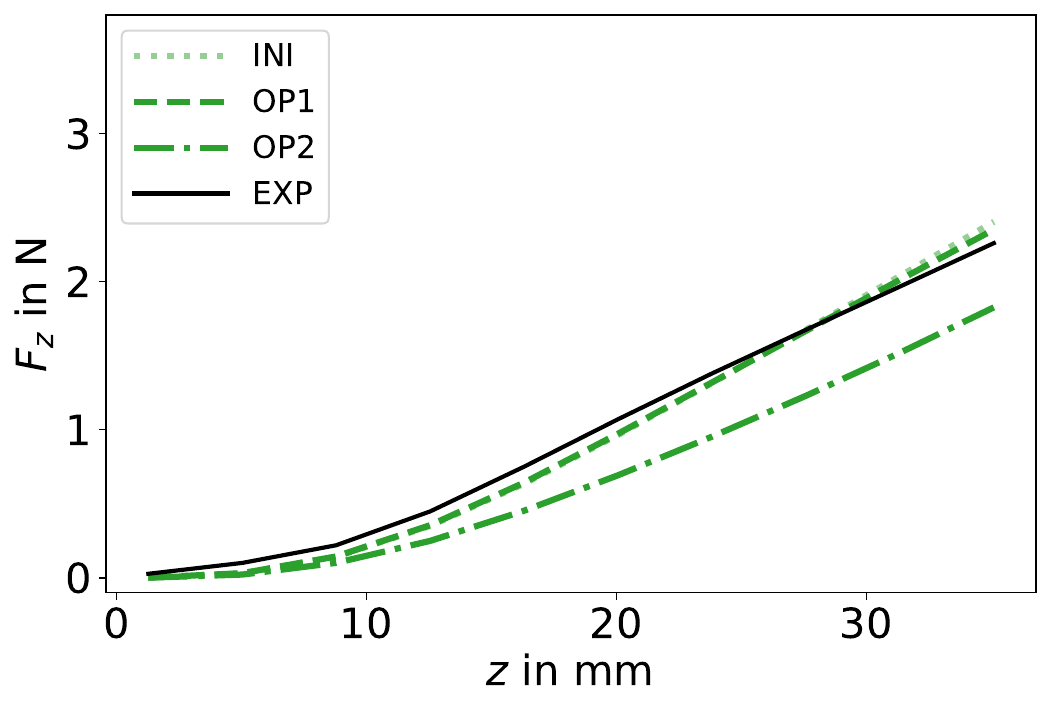}
           & \includegraphics[width=0.4\textwidth]{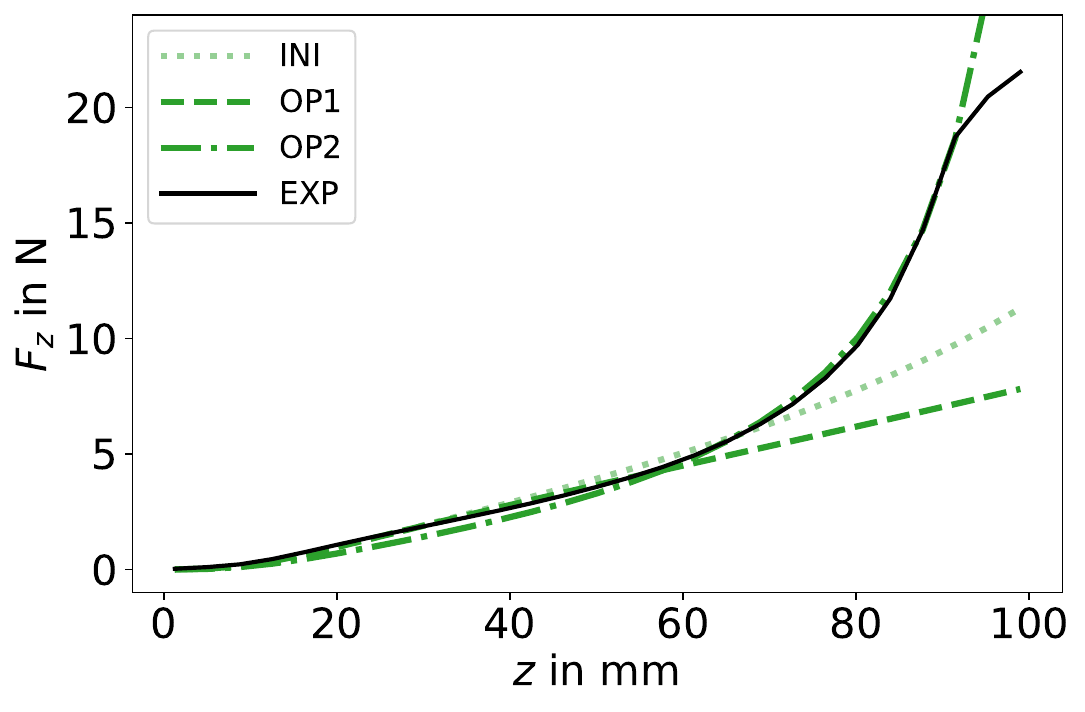} \\
{Y}     & \includegraphics[width=0.4\textwidth]{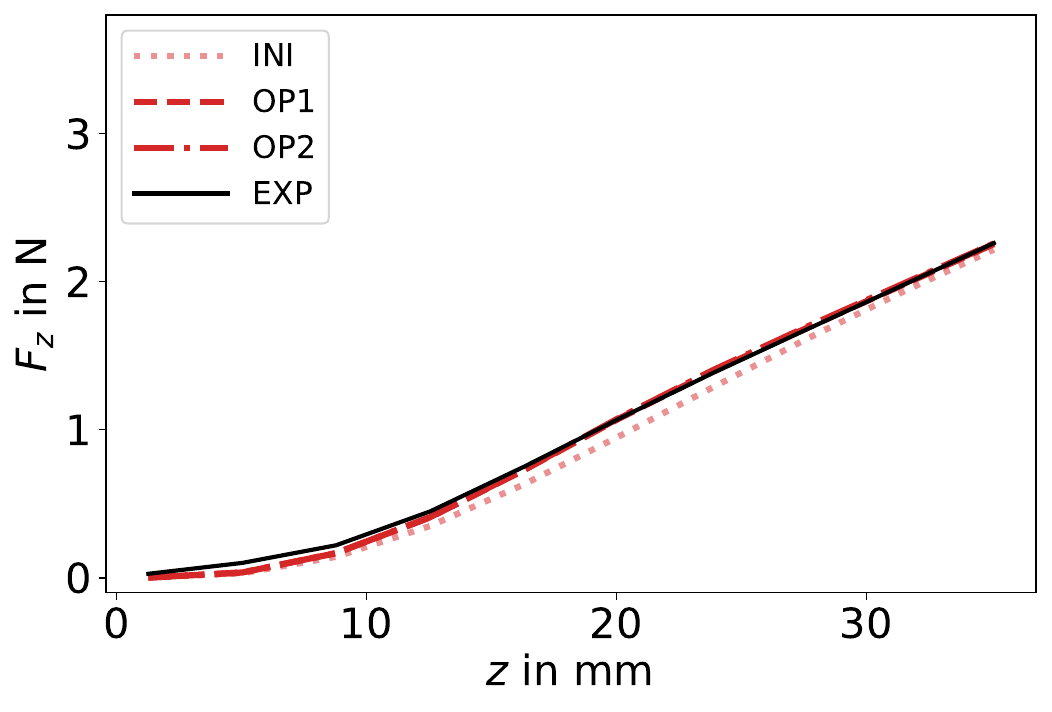}
           & \includegraphics[width=0.4\textwidth]{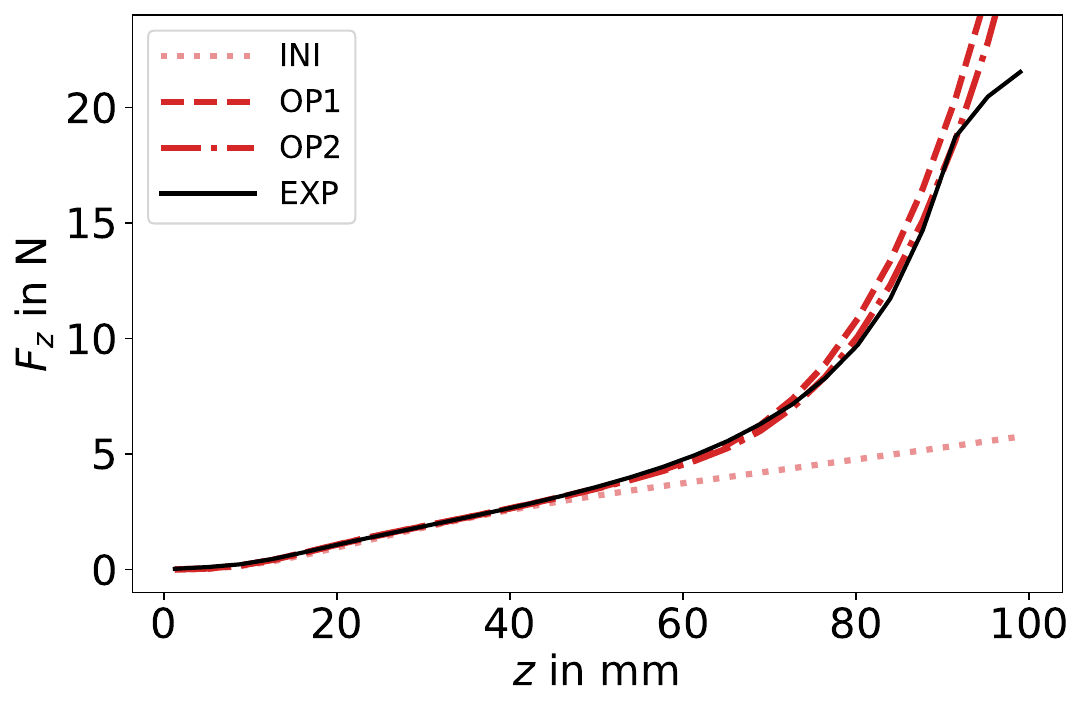} \\
{O}    & \includegraphics[width=0.4\textwidth]{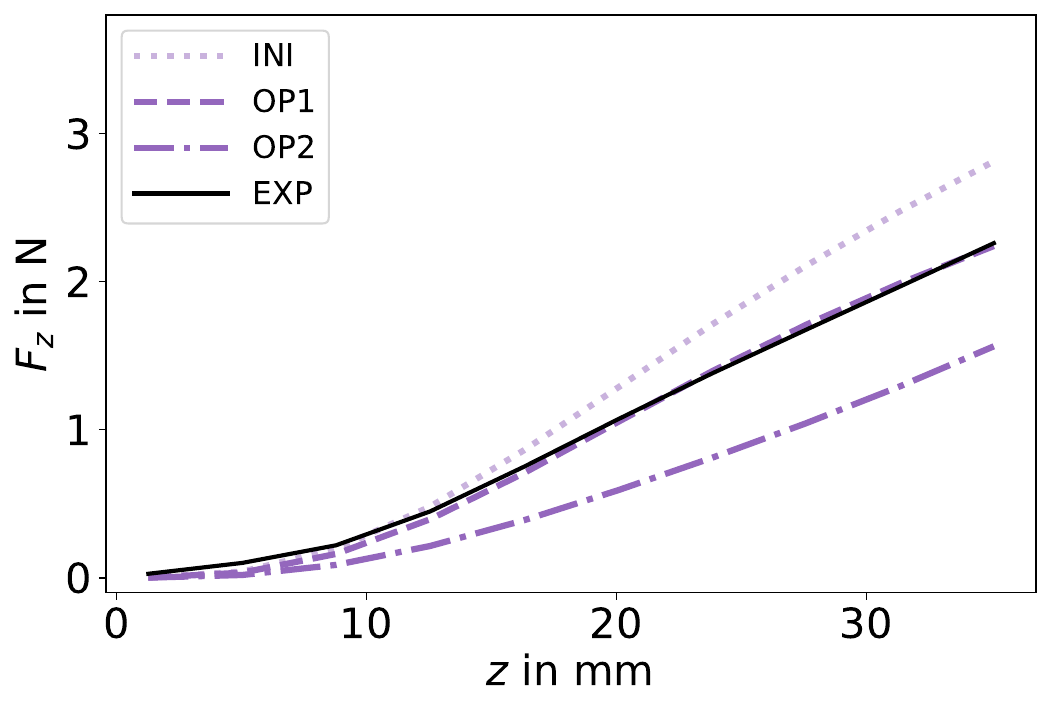}
           & \includegraphics[width=0.4\textwidth]{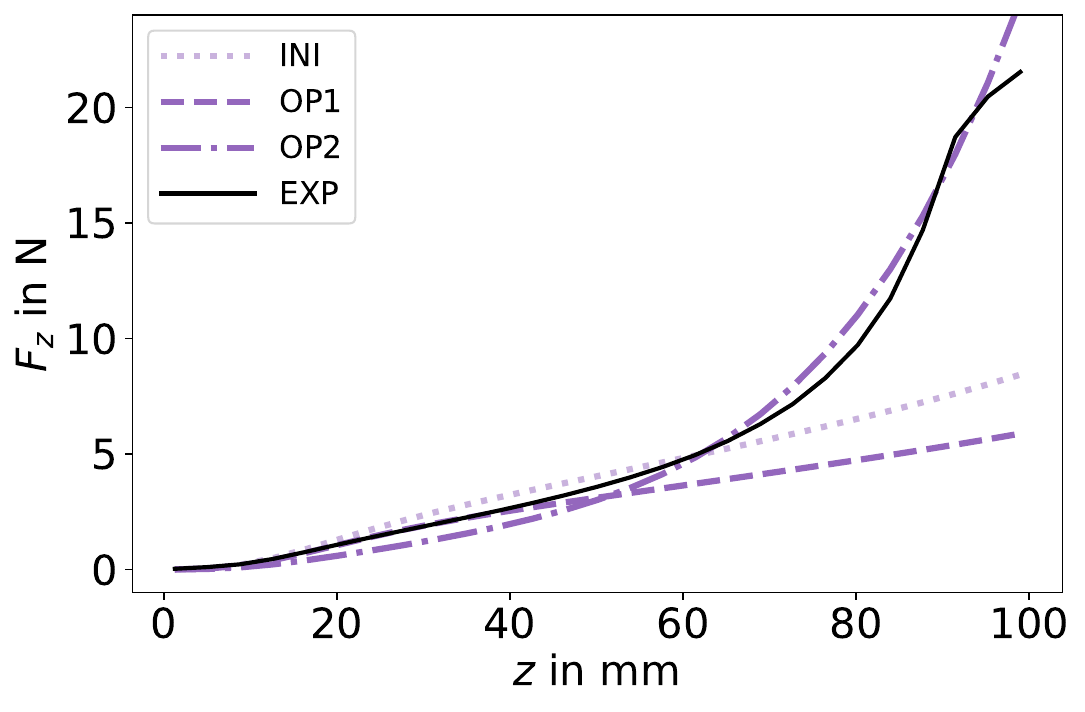} \\
\end{tabular}
\caption{Force-deflection diagrams for different model parameters}
\label{fig:optres}
\end{figure}

The resulting force-deflection diagrams for all five models are shown in Figure~\ref{fig:optres}. 

\begin{figure}[H]
\centering
\rotatebox{90}{Neo-Hookean } \includegraphics[width=.26\textwidth]{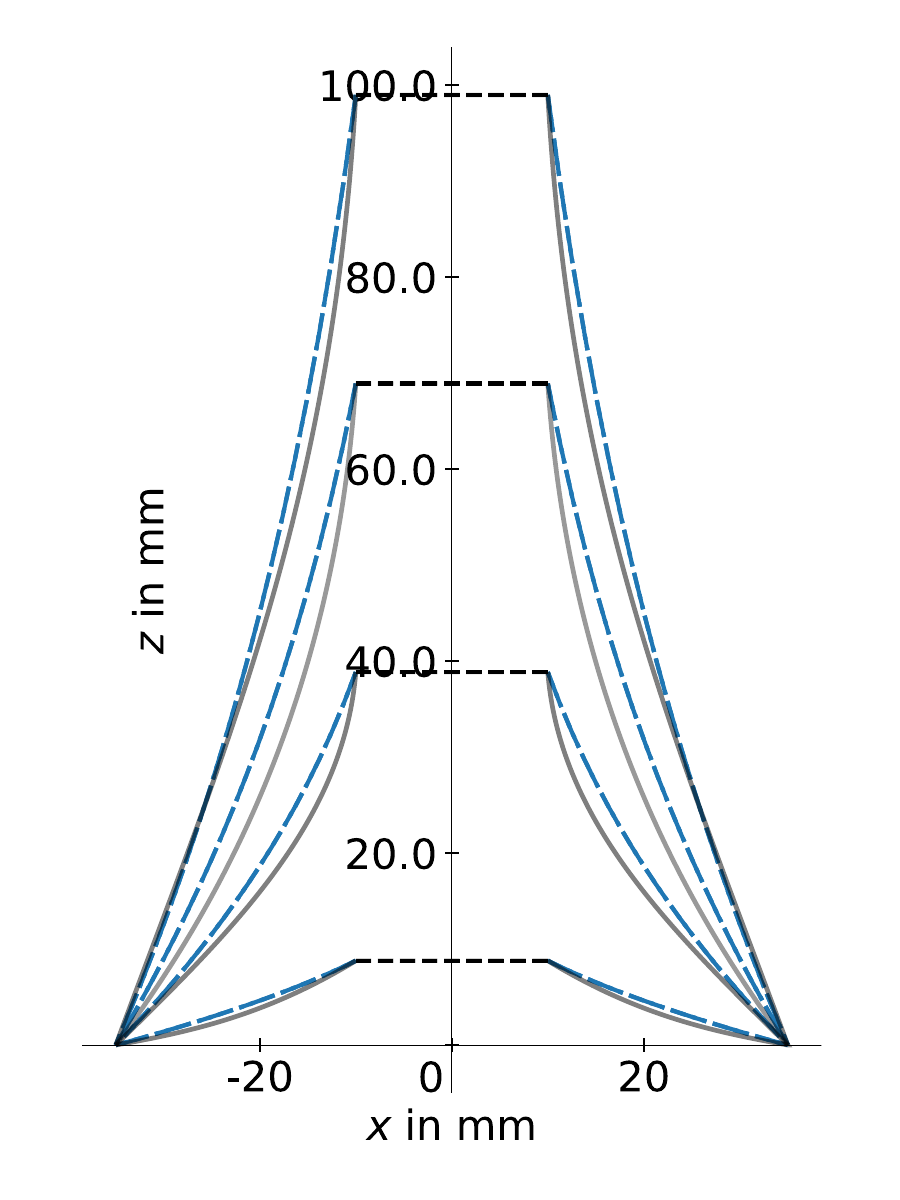}\hfill
\rotatebox{90}{Mooney} \includegraphics[width=.26\textwidth]{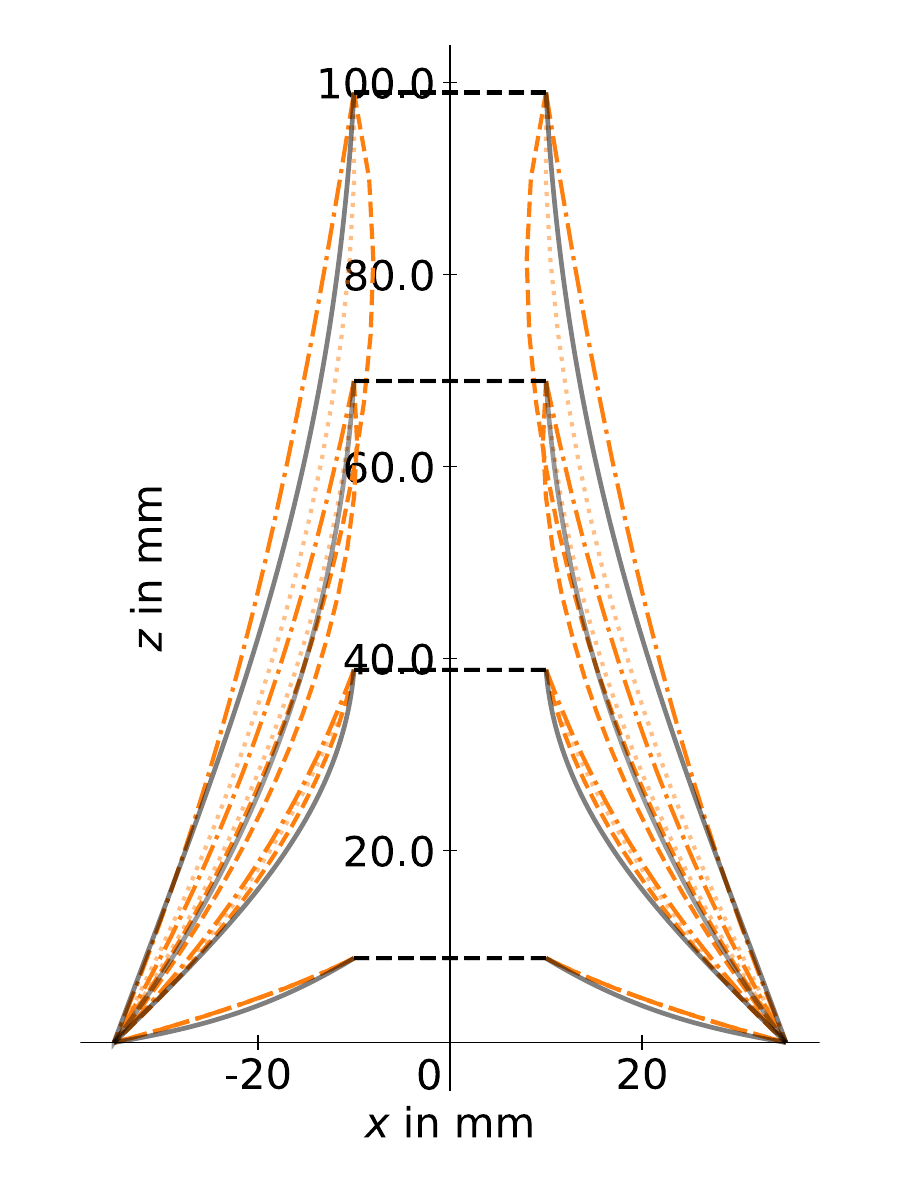}\hfill
\rotatebox{90}{Gent}\includegraphics[width=.26\textwidth]{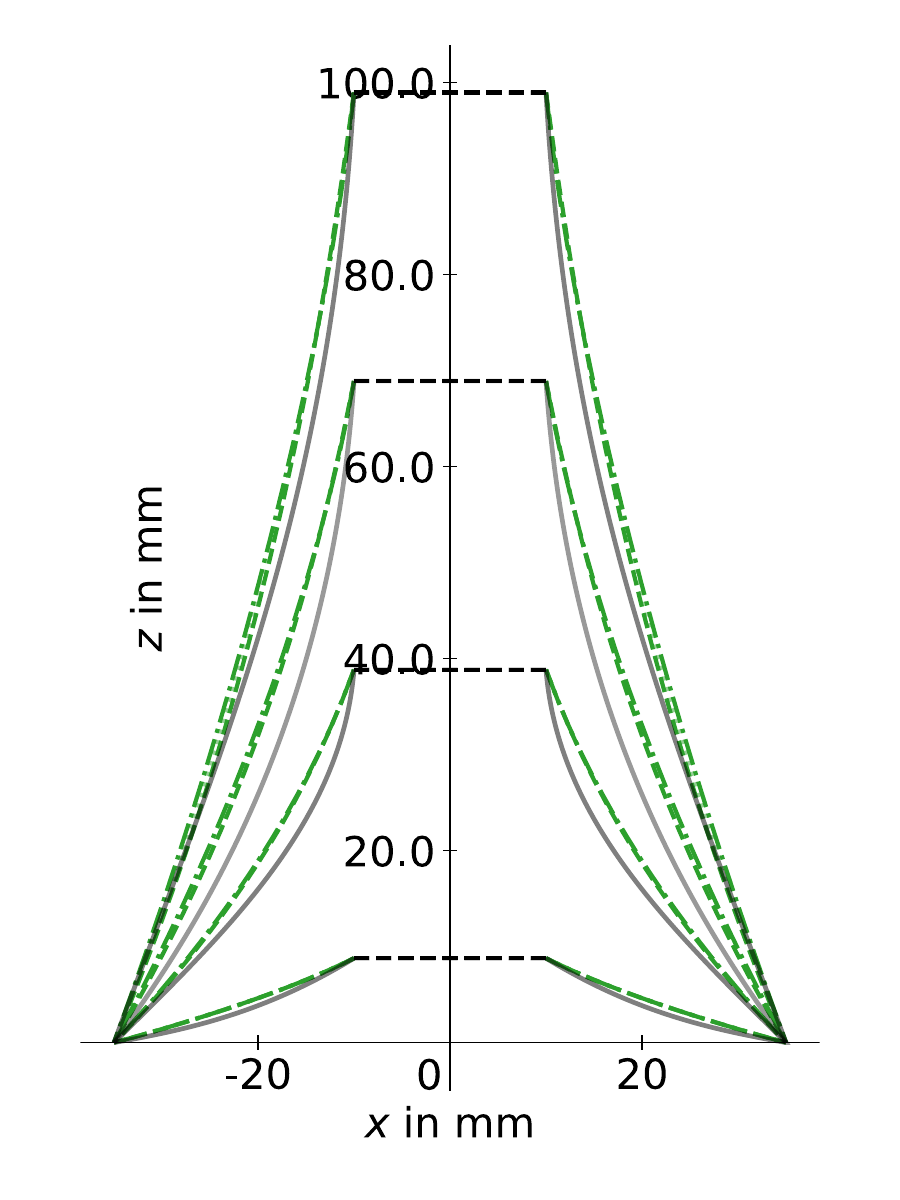}\hfill 
\\
\rotatebox{90}{Yeoh}\includegraphics[width=.26\textwidth]{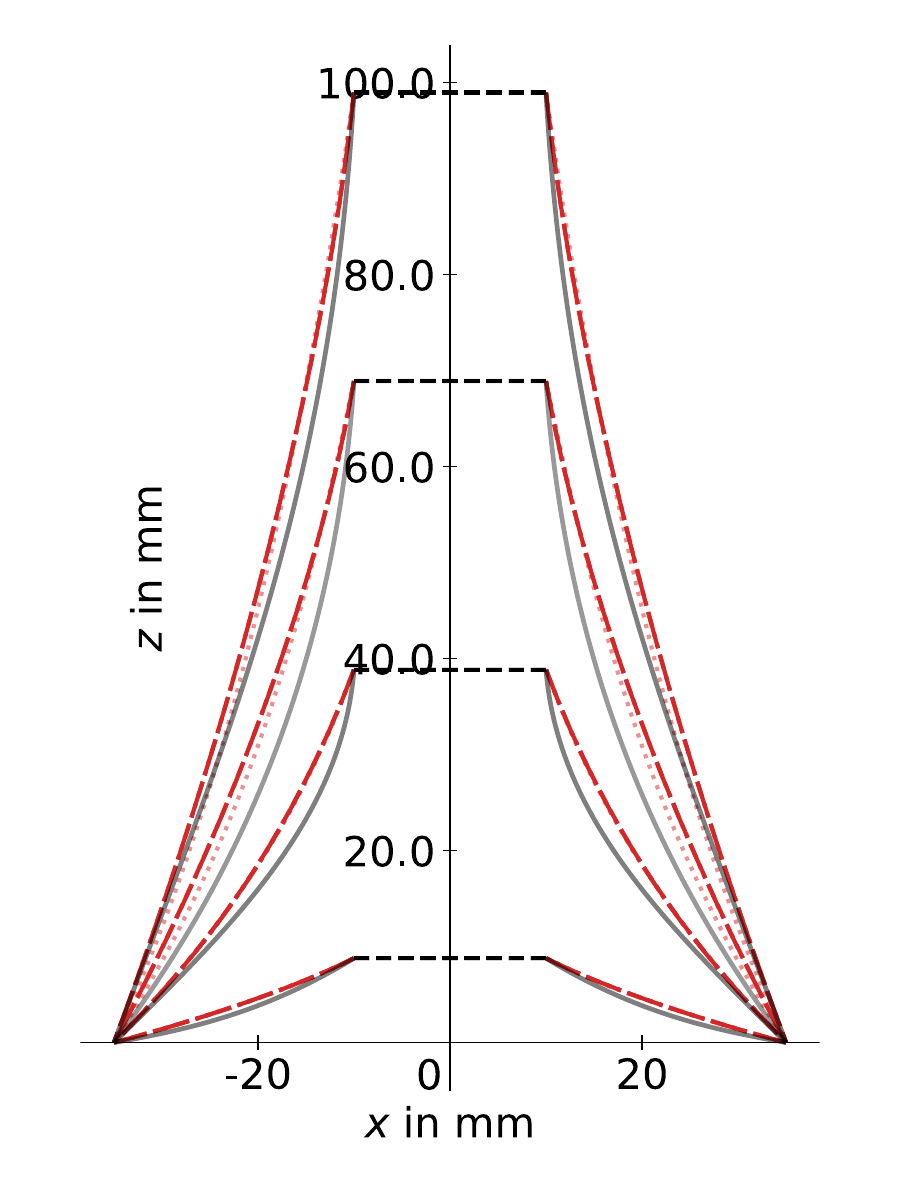}
\rotatebox{90}{Ogden}\includegraphics[width=.26\textwidth]{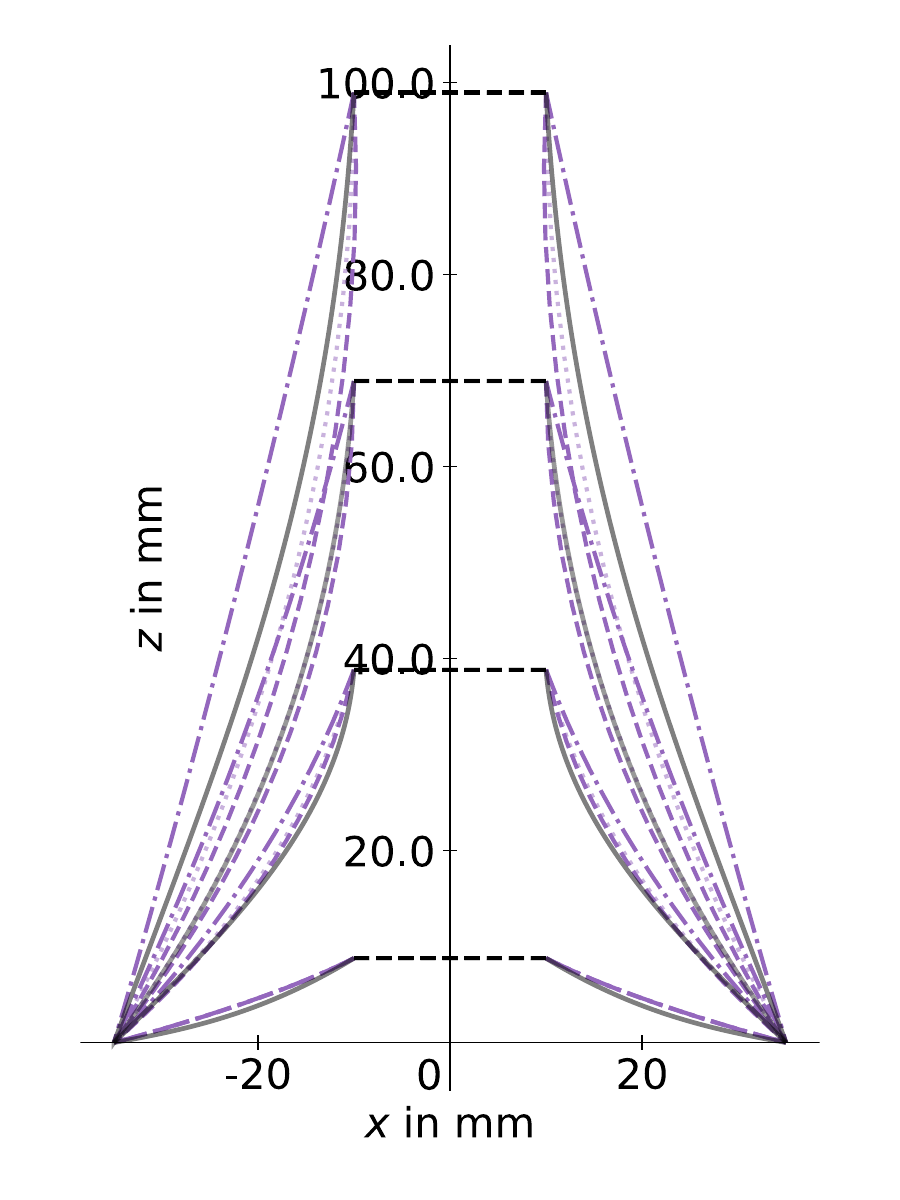}\hfill
\caption{Comparison of profiles}
\label{fig:plots_profiles}
\end{figure}

Figure~\ref{fig:plots_profiles} shows the resulting deformed profiles at different deflection levels, compared to the experimental profiles. 
In Figure~\ref{fig:plots_profiles}, the same line style coding as in Figure~\ref{fig:optres} is used to distinguish between initial parameters, optimization results, and experimental data. The legend is omitted for clarity.

\FloatBarrier
\clearpage

\subsection{Observations on the Optimization results}

The optimization of material model parameters reveals distinct behaviors across different models, with notable differences between OP1 and OP2.

\subsubsection*{Neo-Hookean Model}
The Neo-Hookean (NH) model performs adequately for OP1 but is not suitable for OP2, as evident from the significant increase in $ \text{Loss}_1 $ under OP2. 
The shape of the profiles is the same for different values for the parameter $C_1$, since it appears in the governing equations only as constant factor. 
The effect of parameter optimization only manifests in stress magnitudes.

\subsubsection*{Mooney Model}
The Mooney model demonstrates strong sensitivity to the parameter $C_2$. 
A sufficient negative $C_2$ leads to an outward bulge in the profiles, indicating an unphysical solution, this can be seen also in the nonconvex behavior of the energy function. 
Conversely, when $C_2$ is positive and of similar magnitude to $C_1$, the profiles exhibit pronounced necking, where the hoop stretch is significantly below 1. 
This suggests that the choice of $C_2$ has a critical influence on shape formation. 

For OP1, there is a minimal region in the loss function when $C_1 + C_2 \approx \SI{7.8}{\newton\per\metre}$ . 
When $C_2$ is small, the model behaves similarly to the Neo-Hookean model. 
However, Mooney proves unsuitable for OP2, similar to the Neo-Hookean model.

\subsubsection*{Gent Model}
The Gent model shows different behavior between OP1 and OP2. 
In OP1, $C_2$ can become arbitrarily large, reducing the loss while bringing the solution closer to the Neo-Hookean response. 
The loss landscape has a minimum that continues to decrease with increasing $C_2$, albeit slightly. 

For OP2, however, the optimal $C_2$ remains relatively small and significantly different from OP1, suggesting that limited chain extensibility plays an essential role at extreme strains. 
Changes in $C_2$ have only a minor effect on the profile shape, indicating that Gent's flexibility primarily affects stress distribution rather than deformation patterns.

\subsubsection*{Yeoh Model}
Among all models, Yeoh provides the best fit for the training data in both OP1 and OP2. 
The profile shapes remain largely unchanged across different optimizations, maintaining a consistent deviation from the experimental data. 
This suggests that Yeoh is robust but may not fully capture specific strain-dependent features.

\subsubsection*{Ogden Model}
Despite its theoretical flexibility, Ogden performs worse than expected, particularly when compared to the Yeoh model, which achieves lower loss values. 
This contradicts initial expectations that Ogden would outperform Yeoh due to its more complex material representation.
The Ogden model exhibits strong sensitivity to parameter choices, leading to significant variations in shape. 
Due to the high number of parameters, finding an optimal fit is challenging, and the optimization frequently results in unphysical solutions. 

\section{Conclusions}

This study provides a systematic examination of hyperelastic membranes under large deformations, emphasizing both the theoretical foundations and their practical implications. 
The formulation of membrane theory purely within a two-dimensional geometric framework highlights the mathematical beauty of the approach. 
The governing equilibrium equations take an elegant, coordinate-independent form, directly analogous to three-dimensional elasticity: the divergence of the stress tensor must vanish. 
However, the appropriate divergence operator in this setting is the one acting on the surface-defined first Piola-Kirchhoff stress tensor. 
In this work, all quantities are formulated intrinsically within the geometry of the surface itself, ensuring a purely two-dimensional description. 
This differential-geometric perspective clarifies the structure of the governing equations and reinforces the deep connection between elasticity theory and geometry.

From an applied perspective, the comparison with experimental data shows that while numerical predictions capture key aspects of the observed deformations, further refinements could be done. 
The optimization process is sometimes unstable, either failing abruptly or converging to unrealistic solutions. 
More robust strategies, such as adjusting parameter scaling, refining search spaces, or implementing fallback mechanisms, could improve convergence and stability. 
Additionally, optimizing model parameters with respect to both force-deflection curves and profile shapes could yield better agreement with experimental results.

Further experimental validation would help assess model reliability across different loading conditions. However, no additional data are currently available to us, and the present study is based on a limited experimental dataset obtained in an earlier phase of the project.
Beyond additional experiments, alternative modeling approaches could be explored. In particular, to account for the observed dissipation during unloading, one might consider extending the formulation beyond purely hyperelastic models—for example, by introducing internal variables or incorporating viscoelastic or rate-dependent effects.
Expanding the range of hyperelastic energy functions, incorporating data-driven constitutive models, or considering compressible hyperelastic models may further extend the applicability of the framework.

These findings highlight the interplay between theoretical modeling, numerical implementation, and experimental validation. 
While the current results demonstrate the feasibility of using hyperelastic models for membrane deformations, further refinements in parameter selection, optimization strategies, and constitutive modeling could significantly enhance predictive accuracy and robustness.

\section*{Acknowledgements}

We thank Christian Bär for many valuable discussions.

\section*{Statements and Declarations}

\subsection*{Funding}
The authors received no funding for this work.

\subsection*{Competing Interests}
The authors declare that they have no competing interests.

\subsection*{Ethics Approval and Consent to Participate}
Not applicable.

\subsection*{Consent for Publication}
Not applicable.

\subsection*{Data Availability}
The data sets generated during the current study are available from the corresponding author on reasonable request.

\subsection*{Code Availability}
The code used in this study is available from the corresponding author on reasonable request.

\subsection*{Author Contributions}
Grabs conducted the theoretical analysis, interpreted the experimental results, performed the simulations and wrote the manuscript. 
Wirges contributed to the experimental setup and data acquisition. 
Both authors discussed the results and approved the final manuscript.

\begin{appendices}

\section{Material models and initial parameters}\label{secA1}

We review the strain energy functions of the selected hyperelastic models in order to establish a basis for the subsequent parameter identification. 
Naturally, there exists a vast body of literature in which these models are developed, extended, tested, or applied to various materials and settings. 
The hyperelastic models are presented for example in Chapter~5 of the book~\cite{Bergstrom}, particularly in Section~5.3 on isotropic hyperelasticity. 
Concise summaries of the same models can also be found in the web articles by WELSIM \cite{NeoHookWelsim,MooneyWelsim,GentWelsim,YeohWelsim,OgdenWelsim}, with a focus on their application in finite element analysis. 

\vspace{0.5em}

Our focus here lies on preparing suitable initial parameter values for the optimization process. 
These initial values should be consistent with the material behavior observed in the experiments and provide a reasonable starting point for numerical fitting.

\begin{table}[h]
\centering
\begin{threeparttable}
    \caption{Stored energy density functions with consistency conditions}
  \label{tab:SEDcc}

  \begin{tabular}{lll}
    \toprule
  model &  $\tilde{W}=$ & consistency cond. \\
    \midrule
 Neo-Hookean & $C_{10}(I_1-3)$ & $G=2C_{10}$  \\
&& \\
 Mooney-Rivlin &  $C_{10}(I_1-3)+C_{01}(I_2-3)$ & $G=2(C_{10}+C_{01})$ \\
&& \\
Gent & $-\frac{\mu J_m}{2}\ln(1-\frac{I_1-3}{J_m}))$ &  $G=\mu$\\
&&\\
Yeoh, 3rd order & $\sum_{i=1}^3C_{i0}(I_1-3)^i$ &  $G=2C_{10}$ \\
&&\\
Ogden, 3rd order & $\sum_{i=1}^3 \frac{\mu_i}{\alpha_i}(I_1(\alpha_i)-3)$ &  $G=\tfrac{1}{2}\sum_i \mu_i \alpha_i $\\
    \bottomrule
  \end{tabular}
  \begin{tablenotes}
  \item Here, as before, $I_1=\lambda_1^2+\lambda_2^2+1/(\lambda_1^2\lambda_2^2)$, $I_2 =\lambda_{1}^{2} \lambda_{2}^{2} - 3 + \frac{\lambda_{1}^{2} + \lambda_{2}^{2}}{\lambda_{1}^{2} \lambda_{2}^{2}}$ and $I_1(\alpha_i)= \lambda_1^{\alpha_i}+\lambda_2^{\alpha_i}+1/\left(\lambda_1\cdot \lambda_2\right)^{\alpha_i}$. 
  \item We make usage of the incompressibility constraint $\lambda_1\lambda_2\lambda_3=1$ and set $\W(\lambda_1,\lambda_2)=W^{3D}(\lambda_1,\lambda_2,\frac{1}{\lambda_{1} \lambda_{2}})$.
  \end{tablenotes}
  \end{threeparttable}
\end{table}

Table~\ref{tab:SEDcc} summarizes the strain energy functions of the models considered, along with their consistency conditions.

These consistency conditions—ensuring agreement with linear elasticity (see equations 6.1.88 and 7.2.6 in~\cite{Ogden})—require that
\begin{equation*}
\partial_{1,1}^{2}\tilde{W}(1,1) = 2\partial_{1,2}^{2}\tilde{W}(1,1) = 4G,
\end{equation*}
where $G$ is the \emph{shear modulus} of the homogeneous, isotropic material under consideration.
The shear modulus provides a first-order measure of the material's stress response to shear deformations and can be determined through standardized mechanical tests (see Section~3.8 in~\cite{CiarletI}). 

For the ELASTOSIL\textregistered{} Film 2030 from Wacker, which was used in our experiments, we consulted the literature to determine a baseline shear modulus for parameter initialization.

According to~\cite{AZ}, the shear modulus of silicone rubber ranges from $ G = \SI{0.3}{\mega\pascal} $ to $ G = \SI{20}{\mega\pascal}$, depending on the exact chemical composition. 
Variations arise from differences in the polymer chain structure, the curing system used for cross-linking, and the type and quantity of fillers. 
For details on the manufacturing process of ELASTOSIL, see~\cite{wacker2}; according to~\cite{wacker1}, it is a platinum-cured silicone rubber.

The shear modulus $G$ is related to other elastic moduli for measuring the stiffness of materials, for instance for incompressible materials, 
\begin{equation} \label{eq:EG}
3G=E,
\end{equation}
 where $E$ is Young's modulus, describing the stress response of a material when undergoing uniaxial extension or compression.

Table~\ref{tab:GEF} lists two sources providing information on the Young’s modulus of Elastosil, from which the shear modulus is computed by~\eqref{eq:EG}.
 
\begin{table}[h]
\centering
\begin{threeparttable}
\caption{Elastic moduli of Elastosil film}
  \label{tab:GEF}
  \sisetup{table-format=1.2} 
  \begin{tabular}{lllS[table-format=1.2]S[table-format=1.2]}
    \toprule
    ref. & tested material & {thickness } & {Young's modulus} & {Shear modulus } \\
         &                & {$d$ in \si{\micro\meter}} &  {$E$ in \si{\mega\pascal}}    & {$G$ in \si{\mega\pascal}} \\
    \midrule
   \cite{Mazurek}     & Elastosil film & 100, 200 & 1.18 & 0.39 \\
   \cite{wacker2}\tnote{1} & Elastosil film &  20      & 0.50 & 0.17 \\
    \bottomrule
  \end{tabular}
  \begin{tablenotes}
  \item[1] In \cite{wacker2} we find a stress-strain curve that allows to calculate Young's modulus, we assume uniaxial extension test.
  \end{tablenotes}
  \end{threeparttable}
\end{table}

\FloatBarrier

Apparently, the thickness of the film has some influence on the elastic properties, thinner films seem to be softer.

For the simulations, we chose to model the deformation of the \SI{50}{\micro\metre} film, so we expect the shear modulus to be roughly between \SI{0.1}{\mega\pascal} and \SI{0.5}{\mega\pascal}. 
We use the average as baseline shear modulus $G_{\mathrm{INI}}:=\SI{0.3}{\mega\pascal}$.

The following sections briefly review each of the considered hyperelastic models. 
For each model, the strain energy function is stated, and the choice of initial parameter values is explained. 
We gather values from previously published fits in the literature, these serve as starting points for the optimization. 
A summary of all chosen initial values is provided in Section~\ref{subsec:summary} at the end of this section.

\FloatBarrier

\subsection{Neo-Hookean model}

The incompressible Neo-Hookean stored energy density function is 
\begin{align*}
\W(\lambda_1,\lambda_2)& =C_{10}(I_1-3) \\
&=C_{10} \left(\lambda_{1}^{2} + \lambda_{2}^{2} + \frac{1}{\lambda_{1}^{2} \lambda_{2}^{2}}- 3\right)
\end{align*}

This model is a simple hyperelastic material model, characterized by having only a single material parameter. 
While it provides a straightforward approach to describing material behavior, it is not highly accurate for predicting large-strain deformations. 
Its applicability is best suited for materials subjected to moderate deformations, particularly in cases of uniaxial tension ranging from 30\% to 40\% and pure shear deformation between 80\% and 90\%. 
Despite its limitations in capturing extreme deformations, the model is advantageous due to its good analytical properties, making it convenient for theoretical and computational applications.

Since $2C_{10}=G$ (consistency condition), our initial choice is $C_{10}=0.5G_{\mathrm{INI}}=\SI{0.15}{\mega\pascal}$. 

\begin{figure}[h]
    \centering
   \includegraphics[width=0.9\textwidth]{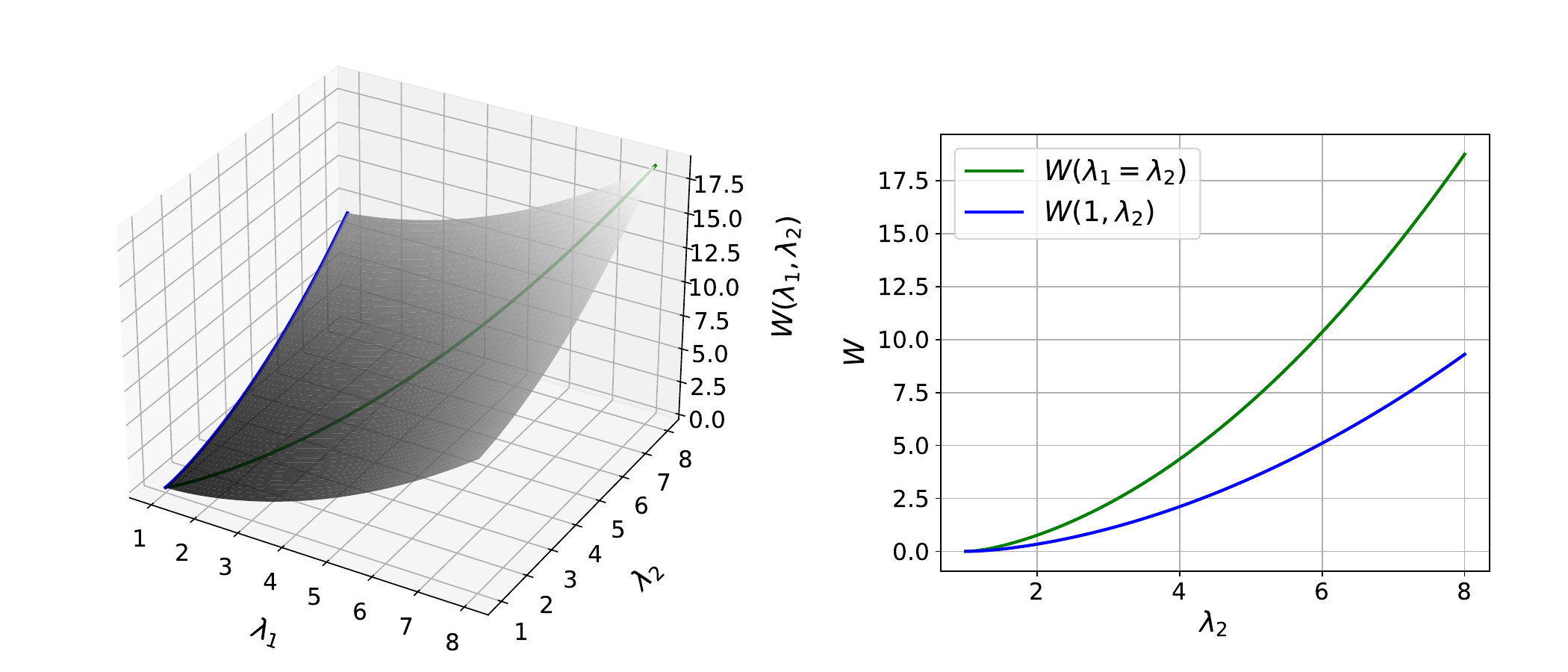}
    \caption{Neo-Hookean energy density for $C_{10}=0.15$ \si{\mega\pascal}, $(\lambda_1,\lambda_2)\in [1,8]^2$ }
    \label{fig:WNHplot}
\end{figure}

The Neo-Hookean energy density is convex for $ C_{10} > 0 $ for positive principal stretches, as can be seen in Figure~\ref{fig:WNHplot}. 
Therefore, we expect the boundary value problem to be well-posed, ensuring both existence and uniqueness of solutions.

\FloatBarrier

\subsection{Mooney-Rivlin model}

The two parameter Mooney-Rivlin energy function is 

\begin{align*}
W(\lambda_1,\lambda_2) &=C_{10}(I_1-3)+C_{01}(I_2-3) \\
&=C_{10} \left(\lambda_{1}^{2} + \lambda_{2}^{2} - 3 + \frac{1}{\lambda_{1}^{2} \lambda_{2}^{2}}\right) + C_{01} \left(\lambda_{1}^{2} \lambda_{2}^{2} - 3 + \frac{\lambda_{1}^{2} + \lambda_{2}^{2}}{\lambda_{1}^{2} \lambda_{2}^{2}}\right)
\end{align*}

This model generalizes the Neo-Hookean formulation, as it reduces to the Neo-Hookean energy when $ C_{01} = 0 $. 
The introduction of a second material parameter improves accuracy in describing uniaxial tension. 
However, despite this refinement, the model remains insufficient for accurately capturing multiaxial stress states. 

A key limitation is that material parameters derived from a specific deformation experiment do not necessarily generalize well to other types of deformation, reducing predictive reliability. 
Additionally, the model is not suitable for deformations exceeding 150\%. 
Another drawback is its analytic properties, which become problematic for $ C_{01} < 0 $, a parameter choice that is commonly encountered in practice.

\begin{table}[h]
  \centering
  \sisetup{table-format=1.2} 
  \begin{tabular}{lp{1.5cm}p{1.5cm}p{1.7cm}S[table-format=1.5]S[table-format=1.5]}
    \toprule
    ref. &  test & max. strain & $C_{10}$ & $C_{01}$  \\
       &      &  in prozent & {in \si{\mega\pascal}}    & {in \si{\mega\pascal}} \\
    \midrule
   ~\cite{Comp}      & simple tension & 220 &  0.96496 &  -0.95833\\  
   ~\cite{Putra}      &  biaxial extension & 100  &  0.1147 & -0.0161  \\ 
   ~\cite{GRNN} &  uniaxial tensile test & 60 & 0.2393 & 0.1134 \\ 
    \bottomrule
  \end{tabular}
  \caption{Different parameter values for 2 parameter Mooney-Rivlin model}
  \label{tab:mooney parameters}
\end{table}

Table~\ref{tab:mooney parameters} shows results of different methods of parameter fitting for different, unspecified silicone rubbers in different loading situations. 
It can be seen, that the results differ quantitatively and qualitatively. 
In~\cite{MooneyWelsim} it says, that for most rubber materials, the ratio $C_{10}/C_{01}$ lies between 0.1 and 0.2. 
When looking at table~\ref{tab:mooney parameters}, this seems not to be good rule of thumb.

\begin{figure}[h]
    \centering
   \includegraphics[width=0.9\textwidth]{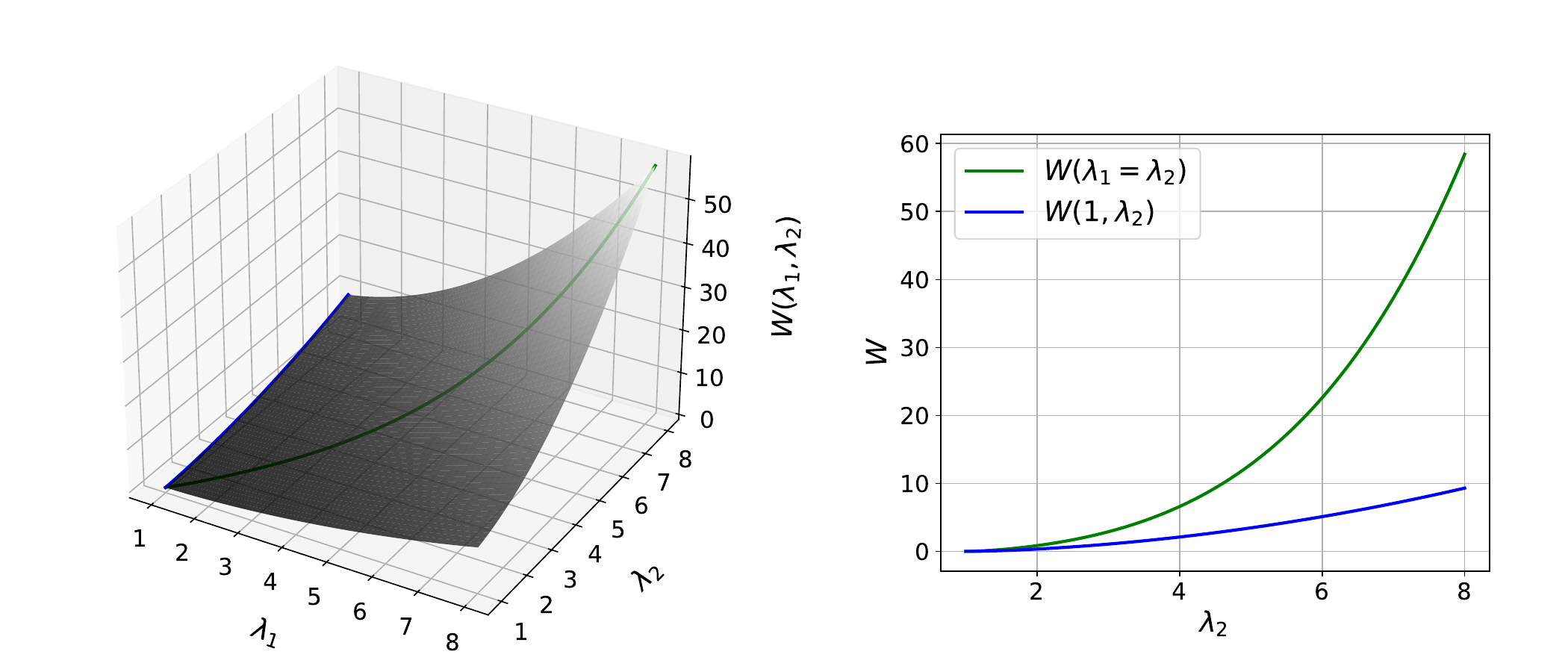}
    \caption{Mooney-Rivlin energy density for $C_{10}=0.14$ MPa, $C_{01}=0.01$ MPa, $(\lambda_1,\lambda_2)\in [1,8]^2$ }
    \label{fig:WMplot1}
\end{figure}

We chose values, such that $2(C_{10}+C_{01})=G_{\mathrm{INI}}$ and both parameters are positive, namely $C_{10}=\SI{0.14}{\mega\pascal}$, $C_{01}=\SI{0.01}{\mega\pascal}$.
Figure~\ref{fig:WMplot1} shows that the energy density for our initial parameter choice is a convex function.

\begin{figure}[h]
    \centering
   \includegraphics[width=0.9\textwidth]{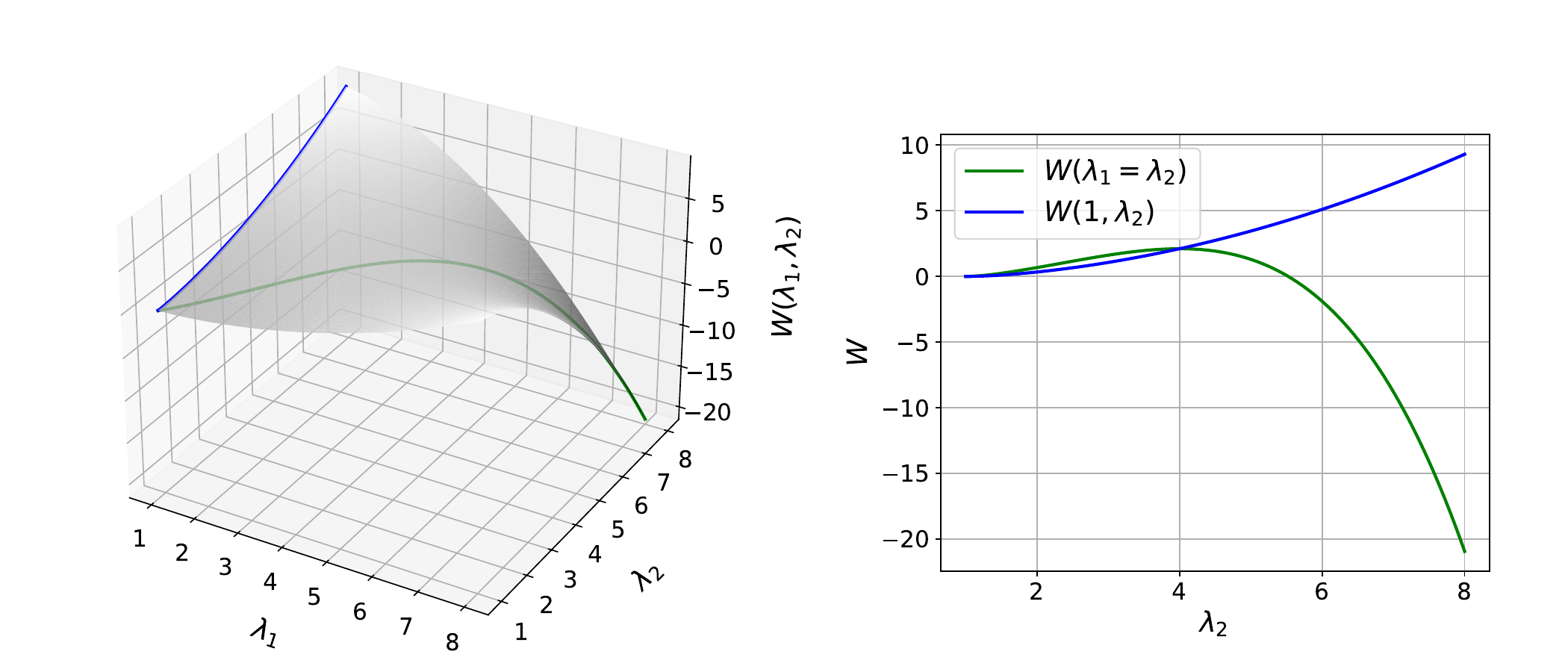}
    \caption{Mooney-Rivlin energy density for $C_{10}=0.16$ MPa, $C_{01}=-0.01$ MPa, $(\lambda_1,\lambda_2)\in [1,8]^2$ }
    \label{fig:WMplot2}
\end{figure}

Figure~\ref{fig:WMplot2} in contrast shows, that for $C_{10}=\SI{0.16}{\mega\pascal}$, $C_{01}=\SI{-0.01}{\mega\pascal}$ the Mooney-Rivlin energy density is not convex.  
However it is convex near $\lambda_1=1$ or $\lambda_2=1$.
In these regions (corresponding to extension in only one principal direction) these parameter choices would would remain acceptable.

But for biaxial deformation modes, convexity of the energy function is violated, non-physical solutions might appear, as well as non-solvability of the equations.

\FloatBarrier

\subsection{Gent model}

The Gent stored energy density function is 

\begin{align*}
\W(\lambda_1,\lambda_2)&=-\frac{\mu J_m}{2}\ln(1-\frac{I_1-3}{J_m})) \\
&=-\frac{\mu J_m}{2} \ln{\left(1 - \frac{\lambda_{1}^{2} + \lambda_{2}^{2} + \frac{1}{\lambda_{1}^{2} \lambda_{2}^{2}} - 3 }{J_m} \right)}
\end{align*}

This model is based on the concept of limiting chain extensibility and serves as an extension of the Neo-Hookean formulation, aiming to provide a more accurate representation of elastomer-like materials under large deformations. 

The energy density has a singularity when the first invariant of the left Cauchy-Green deformation tensor approaches its limiting value, i.e., when $ I_1 - 3 \to J_m $. 
Consequently, the range of validity of the model strongly depends on the choice of $ J_m $, which must be selected carefully to ensure meaningful predictions.

For rubber, typical values for the dimensionless parameter $J_m$ for simple extension range from 30 to
100. (\cite{Horgan})

We chose $J_m=50$ and for consistency $\mu=G_{\mathrm{INI}}=\SI{0.3}{\mega\pascal}$.

\begin{figure}[H]
    \centering
   \includegraphics[width=0.9\textwidth]{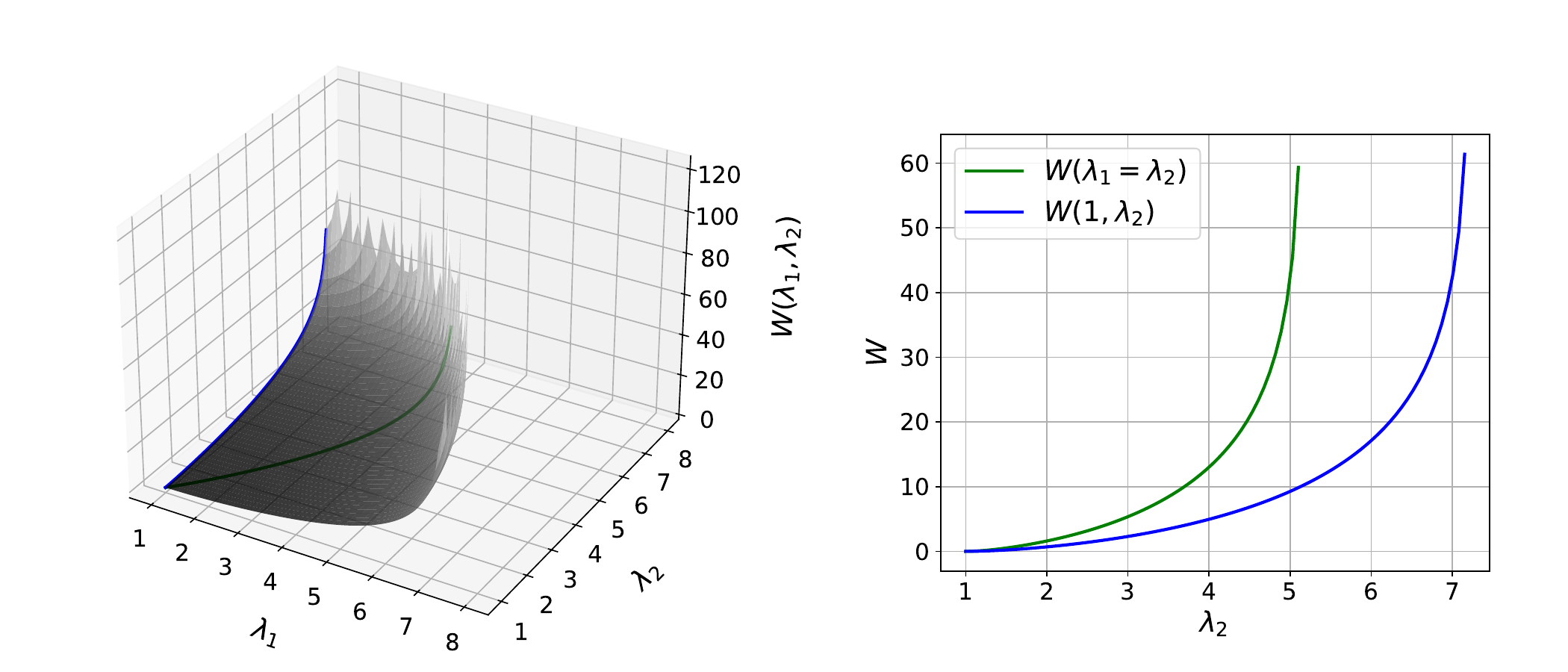}
    \caption{Gent membrane energy for $\mu=\SI{0.3}{\mega\pascal}$, $J_m=50$, $(\lambda_1,\lambda_2)\in [1,8]^2$}
    \label{fig:WGplot}
\end{figure}

The Gent stored energy density function is convex on $\{(\lambda_1,\lambda_2)\mid \lambda_{1}^{2} + \lambda_{2}^{2} + \frac{1}{\lambda_{1}^{2} \lambda_{2}^{2}} < J_m+3 \}$. 
In figure~\ref{fig:WGplot} it can be seen, that $J_m=50$ can be used to model uniaxial deformations up to 700\% strain or biaxial deformations up to 500\% strain.

\subsection{Yeoh model}

The 3rd order Yeoh stored energy density is

\begin{align*}
\tilde{W}(\lambda_1,\lambda_2)&=\sum_{i=1}^3C_{i0}(I_1-3)^i \\
&= \sum_{i=1}^3 C_{i0} \left(\lambda_{1}^{2} + \lambda_{2}^{2} - 3 + \frac{1}{\lambda_{1}^{2} \lambda_{2}^{2}}\right)^i
\end{align*}

This model provides more accurate predictions than the Neo-Hookean formulation by incorporating higher-order terms of $ I_1 $. 
Unlike the Mooney-Rivlin model, it does not depend on $ I_2 $, which helps to avoid certain stability issues associated with that formulation.  

The Yeoh model improves upon the Neo-Hookean model's predictive capabilities across different loading modes, particularly in the regime of large deformations. 
However, convexity must be carefully examined within the relevant deformation range, as analytical issues may arise when $ C_{20} < 0 $ or $ C_{30} < 0 $.

\begin{table}[h]
  \centering
  \begin{tabular}{lp{1.8cm}p{1.5cm}p{1.7cm} p{1.7cm} p{1.7cm}}
    \toprule
    ref. &   test & max. strain & $C_{10}$ & $C_{20}$ & $C_{30}$ \\
       &                  &  in \% & {in \si{\mega\pascal}}    & {in \si{\mega\pascal}} & {in \si{\mega\pascal}} \\
    \midrule
   ~\cite{FEMYeohOgden} &  uniaxial compression & 50  & 0.6465  & -0.3968 & 0.2040 \\
   ~\cite{Comp} & simple tension & 220  & 0.24162 & 0.19977 & -0.00541 \\
   ~\cite{YeohWelsim} & (unknown) & (unknown) & 0.57382&  -0.0747& 0.01132 \\
   ~\cite{Bergstrom} & equibiaxial/ shear/ uniaxial & 350/ 400/ 600  & 0.173146 & -0.00074 & 0.000034\\
    \bottomrule
  \end{tabular}
  \caption{Different parameter values for 3rd order Yeoh model}
  \label{tab:yeoh_parameters}
\end{table}

\FloatBarrier

Table~\ref{tab:yeoh_parameters} shows results of different methods of parameter fitting for different rubbers (silicon rubber, carbon black filled natural rubber, vulcanized natural rubber) in different loading situations.
In~\cite{Bergstrom} a rule of thumb is stated to be: select $C_{10}>0$, $C_{20} \approx -0.01C_{10}$ and $C_{30} \approx -0.01C_{20}$.
Note that it is not the case in any of the examples, but we used this for our initial guess.

\begin{figure}[H]
    \centering
   \includegraphics[width=0.9\textwidth]{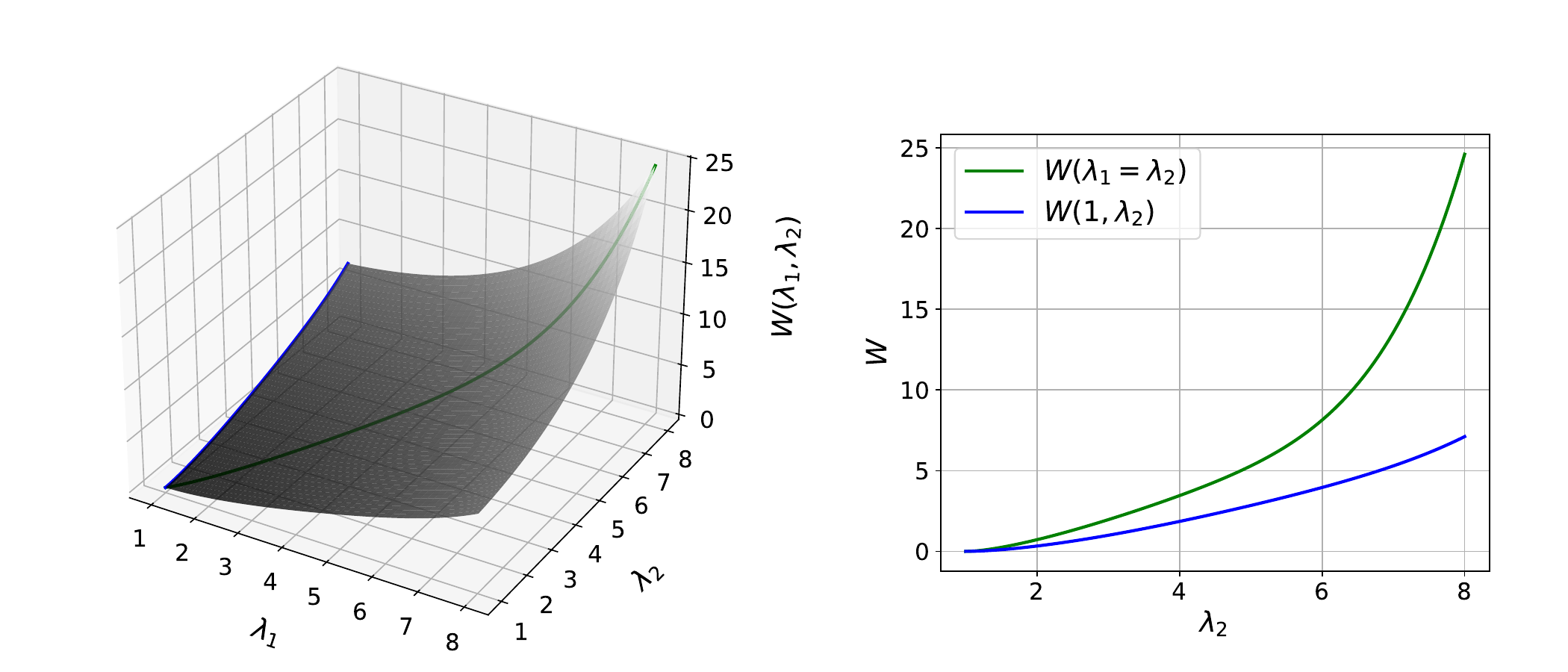}
    \caption{Yeoh membrane energy for $C_{10}=\SI{0.15}{\mega\pascal}$, $C_{20}=\SI{-0.0015}{\mega\pascal}$ and  $C_{30} =\SI{ 0.000015}{\mega\pascal}$, $(\lambda_1,\lambda_2)\in [1,8]^2$} 
    \label{fig:WYplot}
\end{figure}

So, we choose $C_{10}=\SI{0.15}{\mega\pascal}$ (for consistency), $C_{20}=\SI{-0.0015}{\mega\pascal}$ and  $C_{30} = \SI{0.000015}{\mega\pascal}$, see Figure~\ref{fig:WYplot}.

For this choice of parameters, the Yeoh energy density is a convex function on $[1,8]^2$. 

\begin{figure}[H]
    \centering
   \includegraphics[width=0.9\textwidth]{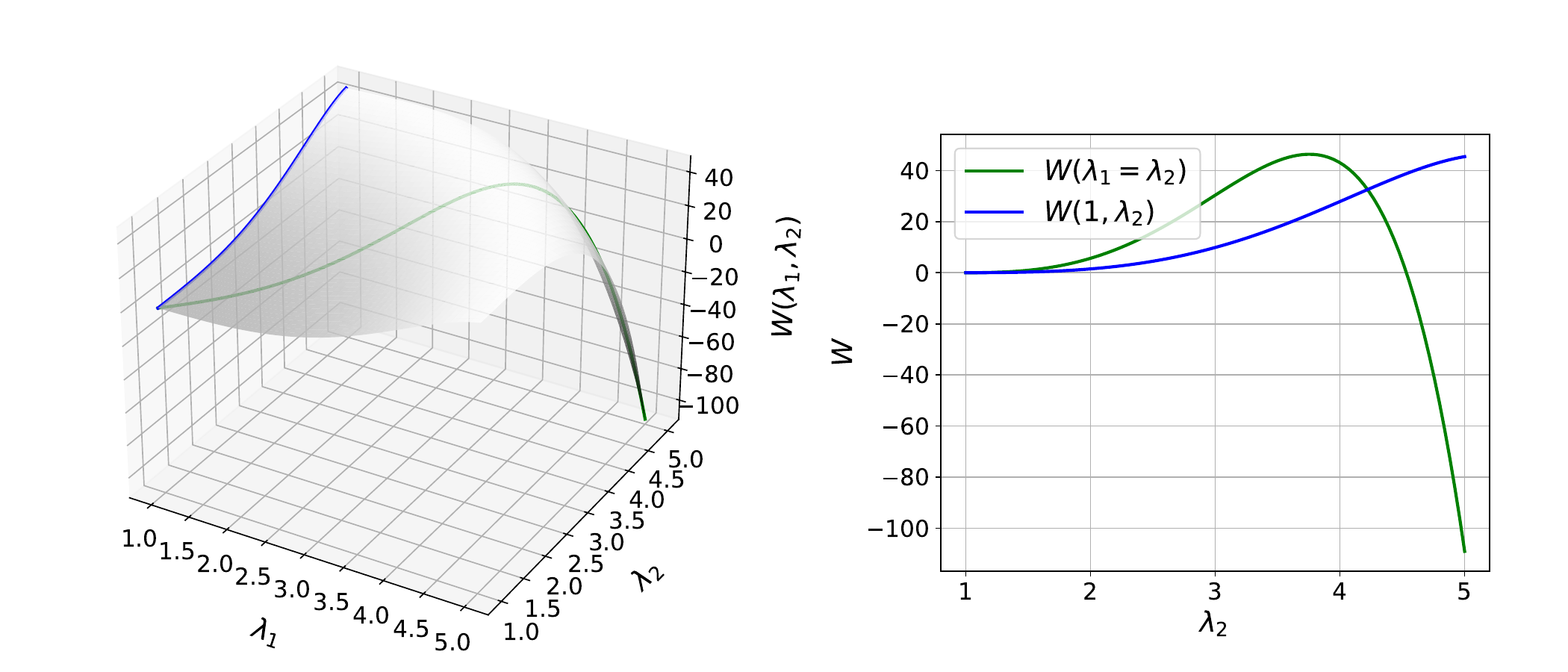}
    \caption{Yeoh membrane energy for $C_{10}=\SI{0.24162}{\mega\pascal}$, $C_{20}=\SI{0.19977}{\mega\pascal}$ and  $C_{30} =\SI{-0.00541}{\mega\pascal}$, $(\lambda_1,\lambda_2)\in [1,5]^2$} 
    \label{fig:WYplot2}
\end{figure}

But for example, the values from~\cite{Comp} are not suitable to model larger deformations, as can be seen in Figure~\ref{fig:WYplot2}.

\subsection{Ogden model}

The order three Ogden model is 

\begin{align*}
\tilde{W}(\lambda_1,\lambda_2)&= \sum_{i=1}^3 \frac{\mu_i}{\alpha_i}(I_1(\alpha_i)-3) \\
&= \sum_{i=1}^3 \frac{\mu_i}{\alpha_{i}} \left(\lambda_{1}^{\alpha_{i}} + \lambda_{2}^{\alpha_{i}} + \left(\lambda_{1} \lambda_{2}\right)^{- \alpha_{i}}- 3 \right)
\end{align*}

This model's general form makes it a powerful tool for describing hyperelastic behavior. 
However, this flexibility also introduces challenges, as selecting an appropriate set of material parameters that ensures stable predictions across general deformation states can be difficult.  

Due to its versatility, the model can be applied to a wide range of hyperelastic constitutive relations and is capable of providing accurate results over the entire strain range. 
Additionally, it effectively captures the rapid stiffness increase observed in the late stages of deformation, making it suitable for materials that exhibit strong strain stiffening.

Table~\ref{tab:ogden_parameters} provides on overview on example sets of material parameters that we gathered from the literature.

\begin{table}[h]
  \caption{Different parameter sets for 3rd order Ogden model}
  \label{tab:ogden_parameters}
  \centering
  \begin{tabular}{p{1.5cm}p{2.0cm}p{1.5cm} lll c}
    \toprule
    ref. &  test & max. strain & \makecell{$\alpha_1$ [--]\\ $\mu_1$ [\si{\mega\pascal}]} & \makecell{$\alpha_2$ [--]\\ $\mu_2$ [\si{\mega\pascal}] } & \makecell{$\alpha_3$ [--] \\ $\mu_3$ [\si{\mega\pascal}] } & set \\
    \midrule
   \cite{FEMYeohOgden} &  uniaxial compression & 50  & \makecell{2.0555 \\ 0.76332}  & \makecell{20.3664 \\ 0.36977}   & \makecell{-10.1800 \\ 0.37006\footnote{ABAQUS different convention...}} & (1) \\
    \cmidrule{4-6} 
   \cite{Comp} & simple tension & 220  & \makecell{1.30073 \\ -0.48953} & \makecell{2.95646 \\ 0.73743} & \makecell{1.35266 \\ -0.60229} & (2)\\
     \cmidrule{4-6}
    \cite{GRNN} & uniaxial compression & 60 & \makecell{-2.3069 \\-3.9516 } & \makecell{-1.3507 \\ -0.3780} & \makecell{-3.8342 \\ 5.4001} & (3)\\
      \cmidrule{4-6}
   \cite{Ogden} (7.2.28) & biaxial & 250 & \makecell{1.3 \\ 0.69} & \makecell{4.0 \\ 0.01} & \makecell{-2.0 \\ -0.0122} & (4)\\  
     \cmidrule{4-6}
   \cite{Ogden} (7.2.31) & uniaxial/ biaxial/ shear & 650/ 350/ 400 & \makecell{1.3 \\  0.618} & \makecell{5.0 \\ 0.0012} & \makecell{-2.0 \\ -0.01} & (5)\\  
    \cmidrule{4-6}
   ~\cite{Yama} & biaxial deformations & 200 & \makecell{2.11 \\ 0.355} & \makecell{2.58 \\ 0.0005} & \makecell{-3.67 \\ -0.00413} & (6) \\
   \bottomrule
  \end{tabular}

\end{table}

We observed the following:
Only the parameter sets (2),(4),(5) and (6) yield convex functions on $[1,8]^2$.
Parameter set (2) and (6) were optimized for silicone rubber, whereas especially parameter set (5) was optimized to model larger strains and different types of deformation, but for natural rubber. 
Parameter set (2) uses exponents $\alpha_i$ that all are between $1$ and $3$, which seems not so versatile. 
 
\begin{figure}[H]
    \centering
   \includegraphics[width=0.9\textwidth]{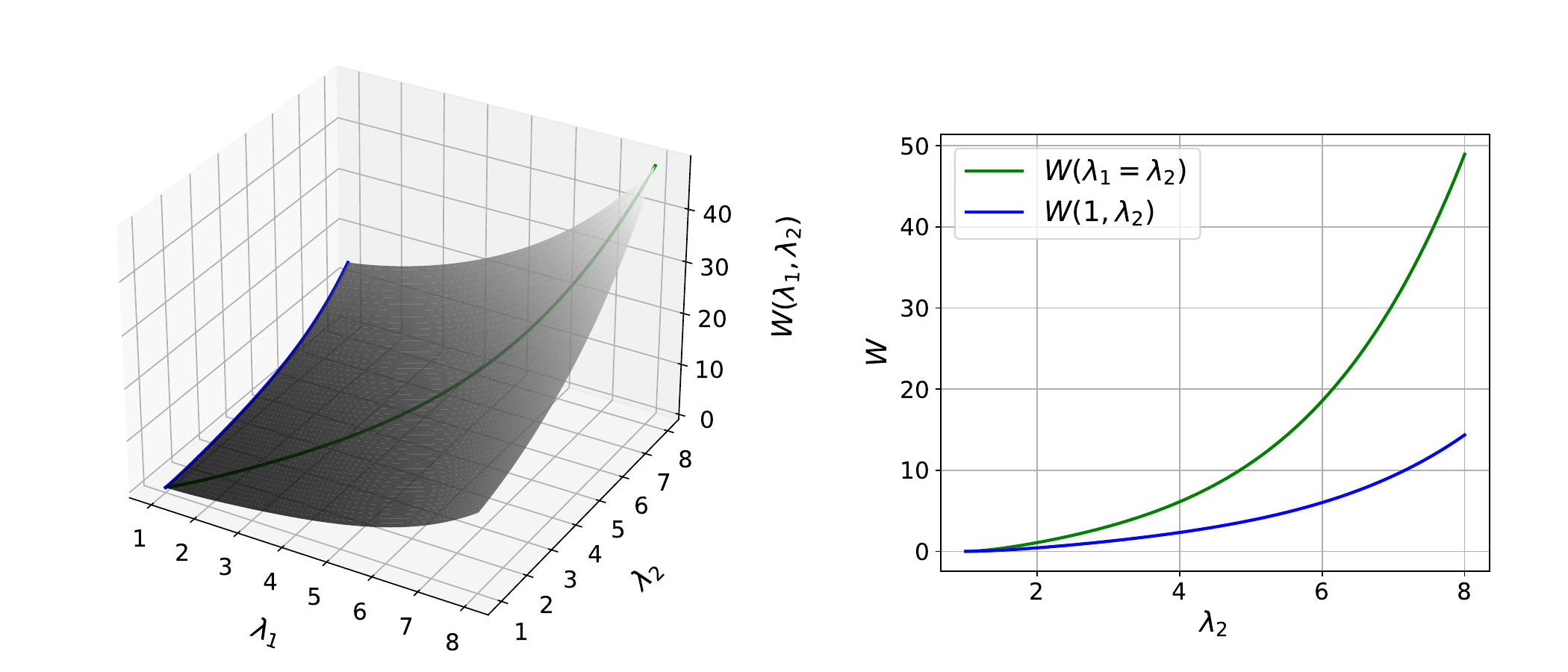}
    \caption{Ogden membrane energy for parameter set (5)} 
    \label{fig:WOplot}
\end{figure}

A choice had to be made, and we simply chose parameter set (5) as initial values, the corresponding energy function can be seen in Figure~\ref{fig:WOplot}.

\subsection{Summary initial parameters}\label{subsec:summary}

We summarize our choice of initial material parameters in Table~\ref{tab:ini}.

\begin{table}[h]
  \centering
  \begin{tabular}{llc}
    \toprule
    model &  $\tilde{W}=$ & initial parameters \\
    \midrule
    Neo-Hookean & $C_{10}(I_1-3)$ & \makecell{$C_{10} = \SI{0.15}{\mega\pascal}$} \\
    && \\
    Mooney-Rivlin &  $C_{10}(I_1-3)+C_{01}(I_2-3)$ & \makecell{$C_{10} = \SI{0.14}{\mega\pascal}$ \\ $C_{01} = \SI{0.1}{\mega\pascal}$} \\
    && \\
    Gent & $-\frac{\mu J_m}{2}\ln\left(1 - \frac{I_1 - 3}{J_m}\right)$ &  \makecell{$\mu = \SI{0.3}{\mega\pascal}$ \\ $J_m = 50$} \\
    && \\
    Yeoh, 3rd order & $\sum_{i=1}^3 C_{i0}(I_1 - 3)^i$ &  \makecell{
      $C_{10} = \SI{0.15}{\mega\pascal}$ \\
       $C_{20} = \SI{-0.0015}{\mega\pascal}$ \\
       $C_{30} = \SI{0.000015}{\mega\pascal}$} \\
    && \\
    Ogden, 3rd order & $\sum_{i=1}^3 \frac{\mu_i}{\alpha_i} \left(I_1(\alpha_i) - 3\right)$ &  
    \makecell{
      $\alpha_1 = 1.3,\ \mu_1 = \SI{0.618}{\mega\pascal}$ \\
      $\alpha_2 = 5.0,\ \mu_2 = \SI{0.0012}{\mega\pascal}$ \\
      $\alpha_3 = -2.0,\ \mu_3 = \SI{-0.01}{\mega\pascal}$} \\
    \bottomrule
  \end{tabular}
  \caption{Stored energy density functions with initial parameter choice}
  \label{tab:ini}
\end{table}

The membrane energies~\eqref{eq:defWN} to~\eqref{eq:defWO} are related to the stored energy density functions $\tilde{W}$ via $W(\lambda_1,\lambda_2)=d\cdot \tilde{W}\left(\lambda_1,\lambda_2,(\lambda_1\lambda_2)^{-1}\right)$ where $d$ is the thickness of the membrane, we have $d=\SI{50}{\micro\metre}$.

In Table~\ref{tab:rel}, we state the induced relation of the parameters from Table~\ref{tab:ini} and the parameters for ~\eqref{eq:defWN} to~\eqref{eq:defWO}, as well as the final initial parameters for the membrane energies.

\begin{table}[h]
  \centering
  \begin{tabular}{lllc}
    \toprule
    eqn. & model & relation parameters & initial parameters \\
    \midrule
    \eqref{eq:defWN} & Neo-Hookean  
      & $C_1 = d \cdot C_{10}$ 
      & \makecell{$C_1 = \SI{7.5}{\newton\per\metre}$} \\
    &&& \\
    \eqref{eq:defWM} & Mooney  
      & $C_1 = d \cdot C_{10},\ C_2 = d \cdot C_{01}$ 
      & \makecell{$C_1 = \SI{7.0}{\newton\per\metre}$ \\
                 $C_2 = \SI{0.5}{\newton\per\metre}$} \\
    &&& \\
    \eqref{eq:defWG} & Gent  
      & $C_1 = \frac{d \cdot \mu}{2},\ C_2 = J_m$ 
      & \makecell{$C_1 = \SI{7.5}{\newton\per\metre}$ \\
                 $C_2 = 50$} \\
    &&& \\
    \eqref{eq:defWY} & Yeoh 
      & $C_i = d \cdot C_{i0}$ 
      & \makecell{$C_1 = \SI{7.5}{\newton\per\metre}$ \\
                 $C_2 = \SI{-0.075}{\newton\per\metre}$ \\
                 $C_3 = \SI{0.00075}{\newton\per\metre}$} \\
    &&& \\
    \eqref{eq:defWO} & Ogden 
      & $C_i = d \cdot \mu_i,\ a_i = \alpha_i$ 
      & \makecell{
          $a_1 = 1.3,\ C_1 = \SI{30.9}{\newton\per\metre}$ \\
          $a_2 = 5.0,\ C_2 = \SI{0.06}{\newton\per\metre}$ \\
          $a_3 = -2.0,\ C_3 = \SI{-0.5}{\newton\per\metre}$} \\
    \bottomrule
  \end{tabular}
  \caption{Relation between 2D and 3D parameters, initial values}
  \label{tab:rel}
\end{table}

\FloatBarrier

\end{appendices}

\printbibliography

@book{carpi2016,
  author    = {Federico Carpi},
  title     = {Electromechanically Active Polymers: A Concise Reference},
  publisher = {Springer International Publishing},
  address   = {Cham},
  year      = {2016},
  series    = {Polymers and Polymeric Composites: A Reference Series},
  isbn      = {978-3-319-31528-7},
  doi       = {10.1007/978-3-319-31530-0}
}

@book{GreenAdkins1960,
  author    = {A.E. Green and J.E. Adkins},
  title     = {Large Elastic Deformations and Non-linear Continuum Mechanics},
  publisher = {Clarendon Press},
  address   = {Oxford},
  year      = {1960},
  isbn      = {9780198503675}
}

@book{LeeSM,
  author    = {John M. Lee},
  title     = {Introduction to Smooth Manifolds},
  series    = {Graduate Texts in Mathematics},
  volume    = {218},
  publisher = {Springer},
  address   = {New York},
  year      = {2003},
  isbn      = {978-0-387-95448-6},
  doi       = {10.1007/978-0-387-21752-9}
}

@book{LeeRM,
  author    = {John M. Lee},
  title     = {Introduction to Riemannian Manifolds},
  edition   = {2},
  series    = {Graduate Texts in Mathematics},
  volume    = {176},
  publisher = {Springer},
  address   = {Cham},
  year      = {2019},
  isbn      = {978-3-319-91754-2},
  doi       = {10.1007/978-3-319-91755-9}
}

@book{MH,
  author    = {Jerrold E. Marsden and Thomas J.R. Hughes},
  title     = {Mathematical Foundations of Elasticity},
  publisher = {Dover Publications},
  address   = {New York},
  year      = {1994},
  series    = {Dover Civil and Mechanical Engineering Series},
  isbn      = {978-0-486-67865-8}
}

@book{CiarletI,
  author    = {P.G. Ciarlet},
  title     = {Mathematical Elasticity, Volume I: Three-dimensional Elasticity},
  year      = {1988},
  publisher = {Elsevier},
  address   = {Amsterdam},
  series    = {Studies in Mathematics and its Applications},
  volume    = {20},
  isbn      = {0-444-70259-8}
}

@book{LibaiSimmonds1998,
  author    = {A. Libai and J.G. Simmonds},
  title     = {The Nonlinear Theory of Elastic Shells},
  edition   = {2},
  publisher = {Cambridge University Press},
  address   = {Cambridge},
  year      = {1998},
  isbn      = {978-0-521-59823-0}
}

@book{GreenZerna1954,
  author    = {A.E. Green and W. Zerna},
  title     = {Theoretical Elasticity},
  publisher = {Clarendon Press},
  address   = {Oxford},
  year      = {1954},
  isbn      = {9780198513797}
}

@book{Reddy,
  author    = {J.N. Reddy},
  title     = {Theory and Analysis of Elastic Plates and Shells},
  edition   = {2},
  publisher = {Taylor \& Francis},
  address   = {Boca Raton},
  year      = {2006},
  isbn      = {978-0-8493-8415-8}
}

@article{MellyLiuReview,
  author  = {Stephen K. Melly and Liwu Liu and Yanju Liu and Jinsong Leng},
  title   = {A Review on Material Models for Isotropic Hyperelasticity},
  journal = {International Journal of Mechanical System Dynamics},
  volume  = {1},
  number  = {1},
  pages   = {71--88},
  year    = {2021},
  doi     = {10.1002/msd2.12013}
}

@article{YangFeng1970,
  author  = {W.H. Yang and W.W. Feng},
  title   = {On Axisymmetrical Deformations of Nonlinear Membranes},
  journal = {J. Appl. Mech.},
  volume  = {37},
  number  = {4},
  pages   = {1002--1011},
  year    = {1970},
  doi     = {10.1115/1.3408651}
}

@article{LeDretRaoult1995,
  author  = {Hervé {Le Dret} and Annie Raoult},
  title   = {The Nonlinear Membrane Model as Variational Limit of Nonlinear Three-Dimensional Elasticity},
  journal = {J. Math. Pures Appl.},
  volume  = {74},
  pages   = {549--578},
  year    = {1995}
}

@article{FultonSimmonds1986,
  author  = {J.P. Fulton and J.G. Simmonds},
  title   = {Large Deformations under Vertical Edge Loads of Annular Membranes with Various Strain Energy Densities},
  journal = {Int. J. Non-Linear Mech.},
  volume  = {21},
  number  = {4},
  pages   = {257--267},
  year    = {1986},
  doi     = {10.1016/0020-7462(86)90033-8}
}

@online{ZR,
  author    = {{ZwickRoell GmbH \& Co. KG}},
  title     = {Tensile Testing Machines and Testers},
  year      = {2023},
  url       = {https://www.zwickroell.com/products/static-materials-testing-machines/universal-testing-machines-for-static-applications/tensile-tester/},
  note      = {Accessed 2 November 2023}
}

@online{WA,
  author    = {{Wacker Chemie AG}},
  title     = {ELASTOSIL\textregistered Film 2030},
  year      = {2023},
  url       = {https://www.wacker.com/h/en-us/silicone-rubber/silicone-films/elastosil-film-2030/p/000038005},
  note      = {Accessed 2 November 2023}
}

@book{Ogden,
  author    = {R.W. Ogden},
  title     = {Non-Linear Elastic Deformations},
  series    = {Ellis Horwood Series in Mathematics and its Applications},
  publisher = {Ellis Horwood},
  address   = {Chichester},
  year      = {1984},
  isbn      = {978-0-85312-273-9}
}

@online{TensorFlow,
  author    = {{Google Brain Team}},
  title     = {TensorFlow: Open Source Machine Learning Framework},
  year      = {2025},
  url       = {https://www.tensorflow.org},
  note      = {Accessed 19 March 2025}
}

@book{Bergstrom,
  author    = {Jörgen Bergström},
  title     = {Mechanics of Solid Polymers},
  series    = {PDL Handbook Series},
  publisher = {Elsevier},
  address   = {Oxford},
  year      = {2015},
  isbn      = {978-0-323-31150-2},
  doi       = {10.1016/C2013-0-15493-1}
}

@article{Putra,
  author  = {Ketut Putra and Jeffrey Plott and Albert Shih},
  title   = {Biaxial {Mooney--Rivlin} Coefficient of Silicone Sheet by Additive Manufacturing},
  journal = {Procedia CIRP},
  volume  = {65},
  pages   = {189--195},
  year    = {2017},
  doi     = {10.1016/j.procir.2017.04.049}
}

@article{GRNN,
  author  = {Junling Hou and Xuan Lu and Kaining Zhang and Yidong Jing and Zhenhe Zhang and Jungfeng You and Qun Li},
  title   = {Parameters Identification of Rubber-like Hyperelastic Material Based on General Regression Neural Network},
  journal = {Materials},
  volume  = {15},
  number  = {11},
  year    = {2022},
  pages   = {3776},
  doi     = {10.3390/ma15113776}
}

@article{Mazurek,
  author  = {Justina Vaicekauskaite and Piotr Mazurek and Sindhu Vudayagiri and Anne Ladegaard Skov},
  title   = {Mapping the Mechanical and Electrical Properties of Commercial Silicone Elastomer Formulations for Stretchable Transducers},
  journal = {J. Mater. Chem. C},
  volume  = {8},
  pages   = {1273--1279},
  year    = {2020},
  doi     = {10.1039/C9TC05072H}
}

@article{FEMYeohOgden,
  author  = {Qi Zhang and Guoying Meng and Haixu Geng and Shuangfu Suo and Jinsen Zhang},
  title   = {Finite Element Analysis of Silicone Rubber Based on {Yeoh} Constitutive Model and {Ogden} Constitutive Model},
  journal = {IOP Conf. Ser.: Earth Environ. Sci.},
  volume  = {714},
  number  = {3},
  pages   = {032078},
  year    = {2021},
  doi     = {10.1088/1755-1315/714/3/032078}
}

@article{Comp,
  author  = {P.A.L.S. Martins and R.M. Natal Jorge and A.J.M. Ferreira},
  title   = {A Comparative Study of Several Material Models for Prediction of Hyperelastic Properties: Application to Silicone-Rubber and Soft Tissues},
  journal = {Strain},
  volume  = {42},
  number  = {3},
  pages   = {135--147},
  year    = {2006},
  doi     = {10.1111/j.1475-1305.2006.00257.x}
}

@article{Horgan,
  author  = {Cornelius O. Horgan},
  title   = {The Remarkable {Gent} Constitutive Model for Hyperelastic Materials},
  journal = {Int. J. Non-Linear Mech.},
  volume  = {68},
  pages   = {9--16},
  year    = {2015},
  note    = {Mechanics of Rubber -- in Memory of Alan Gent},
  doi     = {10.1016/j.ijnonlinmec.2014.05.010}
}

@article{Yama,
  author  = {Yoshihiro Yamashita and Hideyuki Uematsu and Shuichi Tanoue},
  title   = {Calculation of Strain Energy Density Function Using {Ogden} Model and {Mooney--Rivlin} Model Based on Biaxial Elongation Experiments of Silicone Rubber},
  journal = {Polymers},
  volume  = {15},
  number  = {10},
  pages   = {2266},
  year    = {2023},
  doi     = {10.3390/polym15102266}
}

@online{AZ,
  author    = {{AZoM.com}},
  title     = {Property Data: Silicone Rubber},
  year      = {2023},
  url       = {https://www.azom.com/properties.aspx?ArticleID=920},
  note      = {Accessed 2 November 2023}
}

@online{NeoHookWelsim,
  author    = {{WELSIM}},
  title     = {Neo-Hookean Hyperelastic Model for Nonlinear Finite Element Analysis},
  year      = {2025},
  url       = {https://medium.com/@getwelsim/neo-hookean-hyperelastic-model-for-nonlinear-finite-element-analysis-16ac996aa507},
  note      = {Accessed 4 March 2025}
}

@online{MooneyWelsim,
  author    = {{WELSIM}},
  title     = {Mooney--Rivlin Hyperelastic Model for Nonlinear Finite Element Analysis},
  year      = {2025},
  url       = {https://getwelsim.medium.com/mooney-rivlin-hyperelastic-model-for-nonlinear-finite-element-analysis-b0a9a0459e98},
  note      = {Accessed 4 March 2025}
}

@online{GentWelsim,
  author    = {{WELSIM}},
  title     = {Gent Hyperelastic Model for Nonlinear Finite Element Analysis},
  year      = {2025},
  url       = {https://getwelsim.medium.com/gent-hyperelastic-model-for-nonlinear-finite-element-analysis-6f6bd18b0d8b},
  note      = {Accessed 4 March 2025}
}

@online{YeohWelsim,
  author    = {{WELSIM}},
  title     = {Yeoh Hyperelastic Model for Nonlinear Finite Element Analysis},
  year      = {2025},
  url       = {https://getwelsim.medium.com/yeoh-hyperelastic-model-for-nonlinear-finite-element-analysis-6b45e59d2634},
  note      = {Accessed 4 March 2025}
}

@online{OgdenWelsim,
  author    = {{WELSIM}},
  title     = {Ogden Hyperelastic Model for Nonlinear Finite Element Analysis},
  year      = {2025},
  url       = {https://getwelsim.medium.com/ogden-hyperelastic-model-for-nonlinear-finite-element-analysis-df9518de3b48},
  note      = {Accessed 4 March 2025}
}

@online{wacker1,
  author    = {{Wacker Chemie AG}},
  title     = {Stress--Strain Curve for ELASTOSIL},
  year      = {2025},
  url       = {https://www.wacker.com/h/medias/7091-EN.pdf},
  note      = {Accessed 4 March 2025}
}

@online{wacker2,
  author    = {{Wacker Chemie AG}},
  title     = {Manufacturing Process of ELASTOSIL},
  year      = {2025},
  url       = {https://www.immortal-green.com/upload/WS-R.pdf},
  note      = {Accessed 4 March 2025}
}

\end{document}